\documentclass[onecolumn,12pt]{revtex4-1}

\usepackage[english]{babel}
\usepackage{color}
\usepackage{amsmath}
\usepackage{physics}
\usepackage{xr-hyper}
\usepackage[hidelinks]{hyperref}
\usepackage{diagbox}
\usepackage{multirow}
\usepackage{amsfonts}
\usepackage{amssymb}
\usepackage{latexsym}
\usepackage{graphicx}
\usepackage{adjustbox}
\usepackage{float}
\usepackage{xr}
\usepackage{subcaption}
\usepackage{caption}

\newcommand{\pa}{\partial}

\newcommand{\veps}{\varepsilon}

\newcommand{\la}{\lambda}

\newcommand{\rar}{\rightarrow}

\begin{document}

\title{Radial Anharmonic Oscillator: Perturbation Theory, New Semiclassical Expansion, Approximating Eigenfunctions.\\ II. Quartic and Sextic Anharmonicity Cases}

\author{J.C.~del~Valle}
\email{delvalle@correo.nucleares.unam.mx}

\author{A.V.~Turbiner}
\email{turbiner@nucleares.unam.mx, alexander.turbiner@stonybrook.edu}
\affiliation{Instituto de Ciencias Nucleares, Universidad Nacional Aut\'onoma de M\'exico,
A. Postal 70-543 C. P. 04510, Ciudad de M\'exico, M\'exico.}

\begin{abstract}
	
In our previous paper I (del Valle--Turbiner, 2019) it was developed the formalism to study the general $D$-dimensional radial anharmonic oscillator with potential
$V(r)= \frac{1}{g^2}\,\hat{V}(gr)$. It was based on the Perturbation Theory (PT) in powers of $g$ (weak coupling regime) and in inverse, fractional powers of $g$ (strong coupling regime) in both $r$-space and in $(gr)$-space, respectively. As the result it was introduced - the {\it Approximant} - a locally-accurate uniform compact approximation of a wave function. If taken as a trial function in variational calculations it has led to variational energies of unprecedented accuracy for cubic anharmonic oscillator. In this paper the formalism is applied to both quartic and sextic, spherically-symmetric  radial anharmonic oscillators with two term potentials $V(r)= r^2 + g^{2(m-1)}\, r^{2m}, m=2,3$, respectively. It is shown that a two-parametric Approximant for quartic oscillator and a five-parametric one for sextic oscillator for the first four eigenstates used to calculate the variational energy are accurate in 8-12 figures for any $D=1,2,3\ldots $ and $g \geq 0$, while the relative deviation of the Approximant from the exact eigenfunction is less than $10^{-6}$ for any $r \geq 0$.
\end{abstract}

\maketitle
\newpage

\section*{Introduction}

In our previous paper \cite{DelValle1}, henceforth denoted by I, we studied the energy and  wave function of the radial anharmonic potential given by
\begin{equation}
\label{potential}
    V(r)\ =\ \frac{1}{g^2}\,\hat{V}(g\,r)\  =\ \frac{1}{g^2}\,\sum_{k=2}^{m}a_k\,g^k\,r^k\ ,\ r \in [0, \infty)\ .
\end{equation}
Here $r$ is hyperradius in $D$-dimensional space, $g \geq 0$ is a coupling constant and $a_k$, $k=2,3,...,m$, are parameters. It was assumed that $a_2$ and $a_m$ are both positive and the potential $V(r) > 0$ at $r>0$ is positive as well whose possible minima are non-degenerate with minimum at $r=0$: $V(0)=0$. Hence, the minimum at origin is global. The corresponding radial Schr\"odinger operator takes the form
\begin{equation}
\label{radialop}
  \hat{h}_r\ =\ -\frac{\hbar^2}{2M}\left(\pa_r^2\ +\ \frac{D-1}{r}\pa_r\ -\ \frac{l(l+D-2)}{r^2}\right)\ +\ \frac{1}{g^2}\hat{V}(gr)\quad ,\quad
  \pa_r\equiv\frac{d}{dr}\ ,
\end{equation}
where $\hbar$ is the Planck constant and $M$ is the mass of the system. It
has an infinite discrete spectra and it does not contain non-analytic in $g$ terms.
Needless to say that at $m=3,4,\ldots$ and $g \neq 0$, the spectral problem -
the radial Schr\"odinger   equation: $\hat{h}_r\Psi=E\Psi$  - is non-solvable.
Hence, energies and radial wave functions can be found only approximately.

The general formalism to study (\ref{radialop}) was developed in I, where it was successfully applied to a particular case of cubic radial anharmonic oscillator. In order to make this paper self-contained, some results and relevant equations obtained in I will be repeated briefly here.
Our ultimate goal was to construct a \textit{locally accurate} uniform approximation of the wave function $\Psi$ for some of the low-lying states via  the variational method.  It is worth mentioning that in $D=1$ any eigenfunction is labeled by single quantum number like $\Psi_n$ with $n=0,1,\ldots$\,. It measures a number of nodes, where the eigenfunction vanishes. In $D>1$ two quantum numbers are needed to identify the state, $\Psi_{n_r,\ell}$, where $n_r,\ell=0,1,\ldots$, here $n_r$ and $\ell$ are radial quantum number and angular momentum, respectively.
We focus on the ground state function $n_r=\ell=0$, thus, we drop labels and denote it as $\Psi(r)$, in  the exponential representation
\begin{equation}
\label{Phi}
\Psi(r)\ =\ e^{-\frac{1}{\hbar}\,\Phi(r)}\ ,
\end{equation}
where  the function $\Phi$  is the \textit{phase} of the wave function. This representation allows us to transform the radial  Schr\"odinger equation into a Riccati one,
\begin{equation}
\label{riccati}
  \hbar\,\pa_r\, y\ -\ y\,\left(y \ -\ \frac{\hbar\,(D-1)}{r}\right)\ =\ 2 M\, \left(E \ -\  V \right)\quad ,\quad\ y = \pa_r\,\Phi(r)\ .
\end{equation}
There are two ways to remove the explicit appearance of the Planck constant $\hbar$ and the mass $M$ in this equation:

${\bf (i)}$ by introducing in (\ref{riccati}) the new $\hbar$-dependent variable
\begin{equation}
\label{change-v}
     v\ =\ \bigg(\frac{2M}{\hbar^2}\bigg)^{\frac{1}{4}} \,r \ ,
\end{equation}
which we call the {\it quantum} coordinate, and then changing phase and energy
\begin{equation}
\label{change-to-Y}
 y\ =\ (2M \hbar^2)^{\frac{1}{4}}\ \mathcal{Y}\quad , \quad E\ =\ \frac{\hbar}{(2M)^{\frac{1}{2}}}\,\veps\ .
\end{equation}
After that the equation (\ref{riccati}) becomes the \textit{Riccati-Bloch} (RB) equation, see I,
\begin{equation}
\label{riccati-bloch}
 \pa_v\mathcal{Y}\ -\ \mathcal{Y}\left(\mathcal{Y} - \frac{D-1}{v}\right)\ =\ \veps\left(\la\right)\ -\ \frac{1}{\la^2}\,\hat{V}\left(\la v\right)\quad ,\quad
 \pa_v \equiv \frac{d}{dv}\ ,
\end{equation}
where the \textit{effective} coupling constant is
\begin{equation}
\label{effective}
 \la\ =\ \left(\frac{\hbar^2}{2M}\right)^{\frac{1}{4}}\, g\ ,
\end{equation}
while $\veps$ plays the role of energy. The RB equation governs the dynamics in $v(r)$-space.

${\bf (ii)}$ one can introduce in (\ref{riccati}) the \textit{classical}, $\hbar$-independent  coordinate
\begin{equation}
\label{change-to-u}
    u\ =\ g\,r\ ,
\end{equation}
and define a new unknown function
\begin{equation}
\label{change-to-Z}
  \mathcal{Z}\ =\ \frac{g}{(2M)^{1/2}}\, y\ .
\end{equation}
It is easy to check that $\mathcal{Z}(u)$ obeys a non-linear differential equation
\begin{equation}
 \la^2\,\pa_u\mathcal{Z}\ -\ \mathcal{Z}\left(\mathcal{Z} - \frac{\la^2(D-1)}{u}\right)\ =\ \la^2\,\veps(\la)\ -\ \hat{V}(u) \quad , \quad \pa_u\equiv\frac{d}{du}\ ,
\label{Bloch}
\end{equation}
which was called the (radial) \textit{Generalized Bloch} (GB) equation in I.
The definition of $\veps$ and $\la$ remains the same (\ref{change-to-Y}) and (\ref{effective}) as in RB. The GB equation governs the dynamics in $(gr)$-space. Let us note that for $D=1$ the equation (\ref{Bloch}) was called in \cite{ESCOBARI,ESCOBARII,Shuryak} the \textit{(one-dimensional) GB Equation}. The classical coordinate $u$ is related with the quantum one $v$ in a remarkable easy relation
\begin{equation}
\label{u vs v}
         u\ =\  \la\,v\ ,
\end{equation}
see (\ref{change-v}), (\ref{effective}) and (\ref{change-to-u}).

For both equations (\ref{riccati-bloch}) and (\ref{Bloch}) the Perturbation Theory (PT) in powers of the effective coupling constant $\la$ can be developed; it generates the weak coupling expansion for $\veps$
\begin{equation}
 \veps(\la)\ =\ \sum_{n=0}^{\infty} \veps_n \la^{n}\
\label{eps-in-la}
\end{equation}
and for the functions $\mathcal{Y}(v)$ and $\mathcal{Z}(u)$,
\begin{equation}
 \mathcal{Y}(v)\ =\ \sum_{n=0}^{\infty}\mathcal{Y}_n(v)\la^n\ ,
\label{Y-in-la}
\end{equation}
\begin{equation}
 \mathcal{Z}(u)\ =\ \sum_{n=0}^{\infty}\mathcal{Z}_n(u)\la^n\ .
\label{Z-in-la}
\end{equation}
Due to the fact that $\veps_n$ depend on the parameters $a_k$, see (\ref{potential}), the developed PT in powers of $\la \sim (g\,\hbar^{1/2})$ (\ref{eps-in-la}) can be considered as the expansion in powers of coupling constant $g$ and at the same time as the semiclassical expansion in powers of $\hbar^{1/2}$. In a similar way due to the fact that $\mathcal{Z}_n(u)$ (\ref{Z-in-la}) does not depend on $\hbar$ the developed PT for $\mathcal{Z}(u)$ in powers of $\la \sim (g\,\hbar^{1/2})$ can be considered as the semiclassical expansion in powers of $\hbar^{1/2}$ as well as the expansion in powers of $g$.

When PT in powers of $\la$ is developed for the RB equation, it leads to the weak coupling expansion of $\mathcal{Y}(v)$ via the so-called Non-Linearization Procedure \cite{TURBINER:1984}. In turn, PT for the case of the GB equation leads to a \textit{new version} of the semiclassical expansion. A non-trivial connection between two expansions (\ref{Y-in-la}) and (\ref{Z-in-la}) has already been established in I. Expansion (\ref{Z-in-la}) can be transformed into an expansion of the phase,
\begin{equation}
\label{phase}
   \Phi(r;\la)\ =\ \sum_{n=0}^{\infty}\la^n G_n(r)\quad ,\quad G_1(r)\ =\ 0\ ,
\end{equation}
see (\ref{Phi}), where
\begin{equation}
\label{gndef}
   G_n(r)\ =\ \left(\frac{2M}{g^2}\right)^{1/2}\int^{r}Z_n(gr)\,dr\ .
\end{equation}
Keeping $g$ fixed the expansion (\ref{phase}) can be regarded as a semiclassical expansion in powers of $\hbar^{1/2}$ of the phase (\ref{Phi}) in the non-classical domain at large $r$ (beyond a turning point).

RB and GB equations can also be used to study the strong coupling regime, a domain of large $g$. In this case, a perturbative approach implemented in the RB equation leads to the strong coupling  expansion of the energy,
\begin{equation}
 E\ =\ \left(\frac{\hbar^2}{2M}\right)^{\frac{1}{m+2}} g^{2\frac{(m-2)}{m+2}}\sum_{n=0}^{\infty}\,\tilde{\veps}_n\tilde{\la}^{-n}\quad ,\quad \tilde{\la}\ =\ \left(\frac{\hbar^2\,g^4}{2M}\right)^{\frac{1}{m+2}}\ =\ \la^{\frac{4}{m+2}}\ .
\label{Scoupling}
\end{equation}
where $\tilde{\veps}_n$, $n=1,2,...$, are  coefficients. A similar expression occurs for the energy of any exited state. There is an interesting connection between the behavior of the wave function at small distances and the strong coupling regime, see I.

The analytical information on the phase (\ref{Phi}), collected from the weak and strong coupling regime,  was used to design the \textit{Approximant}: an approximation of the exact ground state $(0,0)$ wave function valid at any $D > 0$ in the form
\begin{equation}
\label{approximant}
 \Psi_{(0,0)}^{(t)}\ =\ e^{-\frac{1}{\hbar}\,\Phi_t}\ .
\end{equation}
A straightforward modification of $\Psi^{(t)}_{(0,0)}$ via multiplication on a suitable polynomial in $r$ with real roots allows to construct Approximants for excited states, therefore the Approximant for the ground state appears as the building block.  The approximate phase $\Phi_t$, called {\it Phase Approximant}, is constructed in such a way that it interpolates the expansions at small and large $r$, in the weak and strong coupling regimes. The main result obtained in I is a simple formula for the phase $\Phi_t$, which has to be applicable for the general anharmonic oscillator potential
\begin{equation}
   \frac{1}{\hbar}\Phi_t\ =\
   \frac{\tilde{a}_0\ +\ \tilde{a}_1\,g\,r\ +\ \frac{1}{g^2}{\hat V}(r\,;\ \tilde{a}_2, \dots , \tilde{a}_{m})}{\sqrt{\frac{1}{g^2\,r^{2}}\,{\hat V}(r\,;\ \tilde{b}_2, \ldots ,\tilde{b}_{m})}}\ +\
   \text{Logarithmic Terms\,($r\,;\, \{\tilde{c}\}$)} \ ,
\label{generalrecipe}
\end{equation}
cf. \cite{Turbiner2005} for $D=1$ at $m=4$, where one can put for normalization
${\tilde b}_2 = 1$. Here ${\hat V} (r; \{\tilde{a}\})$ and ${\hat V} (r; \{\tilde{b}\})$ are modified versions of the original potential (\ref{potential}):
instead of the parameters $\{a\}$, some parameters $\{\tilde{a}\}$ and $\{\tilde{b}\}$ are taken, respectively.  The insertion of logarithmic terms in the Phase Approximant $\Phi_t$ (with dependence on some extra parameters $\{\tilde{c}\}$) mimics the logarithmic terms in the exact wave function.  In order to fix the values of the free parameters in $\Phi_t$ (\ref{generalrecipe}), the function $\Psi_{0,0}^{(t)}$ (\ref{approximant}) is used as a trial function in variational calculations: we compute the parameter-dependent variational energy
\begin{equation}
\label{evar}
     E_{var}\ =\ \frac{\int_0^{\infty}\Psi^{(t)}_{0,0}\,(\hat{h}_r\,\Psi^{(t)}_{0,0})\,r^{D-1}\,dr}
     {\int_0^{\infty}(\Psi^{(t)}_{0,0})^2 \ r^{D-1}\,dr}\ ,
\end{equation}
and then minimize it with respect to parameters $\{\tilde{a},\tilde{b},\tilde{c}\}$ to obtain the upper bound of the exact energy.  Since $E_{var}$ corresponds to the first two terms in PT, namely,
\begin{equation}
  E_{var}\ \equiv\ E_{0}^{(1)}\ =\ E_0\ +\ E_1\ ,
\label{firsta}
\end{equation}
the Non-Linearization Procedure could be used to estimate its accuracy by calculating higher-order corrections $E_2$, $E_3$, ..., to $E_{var}$. Therefore, various partial sums define different approximations to the exact energy. For instance, the partial sum
\begin{equation}
 E_0^{(2)}\ =\ E_0\ +\ E_1\ +\ E_2\ ,
\label{seconda}
\end{equation}
corresponds to the second order approximation, while the variational energy itself (\ref{firsta}) is the first order one. In general, the partial sum
\begin{equation}
\label{nth}
 E^{(n)}_0\ =\ E_0\ +\ E_1\ +\ \ldots\ E_n\ ,
\end{equation}
defines the $n$th approximation.

In our previous work I, we presented results for the cubic anharmonic oscillator
when the Approximant (\ref{approximant}) is used as trial function
in variational calculations for several low-lying states. It was explicitly checked
that the relative deviation of the Approximant (\ref{approximant})
from the exact eigenfunction is less than $10^{-4}$ for all $r \in[0,\infty)$. Therefore
the Approximant (\ref{approximant}) represents a locally accurate, uniform approximation of the exact wave function. Simultaneously, the absolute accuracy in energy reaches the extremely high value $\sim 10^{-7}$ at any dimension $D$ and coupling constant $g \in[0, \infty)$! In the present paper it will be shown that in the same formalism, the Approximants leads to highly accurate results for the quartic and sextic radial anharmonic potentials.

We follow the same program as in I: as the first step we focus on the study of the ground state:
the structure of PT in powers of $\la$ and in inverse powers of $\la$ is investigated in order
to construct the Approximant for the phase of the ground state wave function. Then ``functionally-similar" Approximant is used for the phase of the trial functions of the excited states. With this knowledge we
perform variational calculations for some low-lying states imposing the orthogonality conditions with respect to the ground state and between excited states. The  accuracy of the variational energy and the quality of the Approximant are evaluated in two different manners: $(i)$ by using PT and calculating corrections to variational calculations in the framework of the Non-Linearization Procedure; $(ii)$ by using one of the most accurate numerical methods to solve the Schr\"odinger equation: the Lagrange Mesh Method (LMM) in a formulation proposed by D. Baye, see \cite{BAYE}.
Eventually, we use the Approximant to calculate the first two dominant terms of the strong coupling expansion for the ground state. The paper is divided in two large parts: Section I is dedicated to the quartic two-term radial anharmonic oscillator and Section II is devoted to the sextic two-term radial anharmonic oscillator.

In Conclusions it will be shown that choosing the parameters in the (phase) Approximant in such a way to reproduce {\it all} growing terms at large $r$ exactly and removing linear in $r$ term at the small $r$ expansion leads to the striking fact that the relative deviation
of the (phase) Approximant from the exact phase is bounded and do not exceed $\sim 10^{-2}$.
It reduces the number of free parameters to one, two, five parameters for cubic, quartic and sextic radial anharmonic oscillators, respectively, while the accuracy of the variational energy is reduced to five-six figures for any coupling constant, which is still unprecedented result.

\numberwithin{equation}{section}
\section{Quartic Anharmonic Oscillator}

The simplest, formally even $V(-r)=V(r)$, radial anharmonic oscillator potential is characterized by
quartic anharmonicity,
\begin{equation}
V(r)\ =\ r^2\ +\ g^2\,r^4\ ,
\label{potquartic}
\end{equation}
cf. (\ref{potential}) at $m=4$ with $a_3=0$ and $a_4=1$.
It is worth mentioning that many properties those the quartic anharmonic radial oscillator exhibits are typical for any (formally) even anharmonic potential, $V(r)=V(-r)$. In particular, the polynomial nature of corrections $\veps_n$ and $\mathcal{Y}_n(v)$, see (\ref{eps-in-la}) and (\ref{Y-in-la}) is one of such common properties.
The results of the forthcoming Section are obtained in similar way to those for the cubic anharmonic potential, therefore we omit some details already presented in I.

\subsection{PT in the Weak Coupling Regime}

For the quartic anharmonic oscillator the  perturbative expansions of $\veps$ and $\mathcal{Y}(v)$, derived from RB equation (\ref{riccati-bloch}),
\begin{equation}
\label{riccati-bloch-4}
 \pa_v\mathcal{Y}\ -\ \mathcal{Y}\left(\mathcal{Y} - \frac{D-1}{v}\right)\ =\
 \veps\left(\la\right)\ -\ v^2\ \ -\ \la^2\,v^4 \quad ,
 \quad \pa_v \equiv \frac{d}{dv}\ ,
\end{equation}
where $v$ and $\la$ are defined in (\ref{change-v}) and (\ref{effective}), correspondingly,
are of the form
\begin{equation}
\label{eps-in-la-4}
  \veps\ =\ \veps_0\ +\ \veps_2\,\la^2\ +\ \veps_4\,\la^4\ +\ \ldots \quad ,\quad \veps_0\ =\ D\ ,
\end{equation}
and
\begin{equation}
\label{Y-in-la-4}
 \mathcal{Y}(v)\ =\ \mathcal{Y}_0(v)\ +\ \mathcal{Y}_2(v)\,\la^2\ +\ \mathcal{Y}_4(v)\,\la^4\ +\
 \ldots\ \quad ,\quad \mathcal{Y}_0(v)\ =\ v \ ,
\end{equation}
respectively. All odd terms in $\la$ vanish in both expansions, see (\ref{eps-in-la}) and (\ref{Y-in-la}). In general, a finite number of corrections can be calculated by linear algebra means,
in particular, the first non-vanishing corrections are
\begin{equation}
\label{correction2-4}
\veps_2\ =\ \frac{1}{4}\,D\,(D+2)\quad ,\quad \mathcal{Y}_2(v)\ =\ \frac{1}{2}\,v^3\ +\ \frac{1}{4}\,(D+2)\, v\ ,
\end{equation}
while the next two corrections $\veps_{4,6}$ and $\mathcal{Y}_{4,6}(v)$ are presented in Appendix A. In principle, the algebraic procedure of finding PT corrections holds for all even anharmonic potentials $V(r)=V(-r)$, however, it is enough to have a single odd monomial term to occur in the potential this property breaks down. In this situation, the calculation of correction $\veps_n$ becomes numerical procedure like it happens for the cubic case, see the paper I. Eventually, in contrast to the cubic case, it can be shown that for all even potentials any correction $\veps_{2n}$ is a polynomial in $D$.

In general, all corrections $\mathcal{Y}_{2n}(v)$ are odd-degree polynomials in $v$ of the form,
\begin{equation}
  \mathcal{Y}_{2n}(v)\ =\ v\,\sum_{k=0}^{n}c_{2k}^{(2n)}v^{2(n-k)}\ ,
\label{Y2n}
\end{equation}
where any coefficient $c_{2k}^{(2n)}$ is a polynomial in $D$ of degree $k$,
\begin{equation}
 c_{2k}^{(2n)}\ =\ P_{k}^{(2n)} (D)\ \quad ,\quad c_{2n}^{(2n)}\ =\ \frac{\veps_{2n}}{D}\ .
\label{Y2n-c}
\end{equation}

Due to invariance $v \rar -v$ of the original equation (\ref{riccati-bloch-4}), it is convenient to simplify it by introducing a new unknown function and changing $v$-variable,
\[
    \mathcal{Y}\ =\ v\, \mathcal{\tilde Y}\quad \mbox{and}\quad {\rm v}\ =\ v^2\ .
\]
As a result, (\ref{riccati-bloch-4}) is reduced to
\begin{equation}
\label{riccati-bloch-4-tilde}
 2 {\rm v} \pa_{\rm v} \mathcal{\tilde Y}\ -\ \mathcal{\tilde Y}\left({\rm v} \mathcal{\tilde Y} - D\right)\ =\
 \veps\left(\la\right)\ -\ {\rm v}\ \ -\ \la^2\,{\rm v}^2 \quad ,
 \quad \pa_v \equiv \frac{d}{dv}\ .
\end{equation}
This is a convenient form of the RB equation to carry out the PT consideration. In particular, the first correction (\ref{correction2-4}) to $\mathcal{\tilde Y}$ becomes a linear function,
\[
  \mathcal{\tilde Y}_2\ =\ \frac{1}{4}\,\left[2{\rm v}\ +\ (D+2)\right]\ ,
\]
and, in general, $\mathcal{\tilde Y}_{2n}$ is a polynomial in ${\rm v}$ of degree $n$, see (\ref{Y2n}).
Corrections $\mathcal{\tilde Y}_{4,6}$ are presented in Appendix A.

The energy corrections $\veps_{2n}$ are of the form \cite{DOLGOVPOPOV1978}
\begin{equation}
     \veps_{2n}(D)\ =\ D\,(D+2)\,R_{n-1}(D)\ ,
\label{factorizationq}
\end{equation}
where $R_{n-1}(D)$ is a polynomial  of degree $(n-1)$ in $D$, in particular, $R_0=\frac{1}{4}$.

From (\ref{factorizationq}) one can see that any energy correction $\veps_{2n}$ vanishes when $D=0$, the property which holds for any anharmonic oscillator, see I. Consequently, their formal sum  results in $\veps(D=0)=0$ and  ultimately  in $E(D=0)=0$.  Thus, at $D=0$ the radial Schr\"odinger equation is reduced to
\begin{equation}
   -\frac{\hbar^2}{2M}\left(\frac{d^2\Psi(r)}{dr^2}\ -\ \frac{1}{r}\,\frac{d\Psi(r)}{dr}\right)\
   +\ (r^2+g^2\,r^4)\,\Psi(r)\ =\ 0\ .
\label{D=0-q}
\end{equation}
Needless to say, this equation defines the zero mode of the Schr\"odinger operator.  It can be solved exactly in terms of  Airy functions \cite{DOLGOVPOPOV1979},
\begin{equation}
\label{D=0-q-psi}
   \Psi\ =\ C_1\,\text{Ai}\left(\frac{1+(\la v)^2}{\la^{4/3}}\right)\  +\ C_2\,\text{Bi}\left(\frac{1+(\la v)^2}{\la^{4/3}}\right)\ ,
\end{equation}
for definition of $\la$ and $v$ see (\ref{change-v}) and (\ref{effective}), respectively.
However, this linear combination can not be made normalizable at $D=0$ by any choice of constants $C_1$ and $C_2$. Hence, the original assumption $E=0$ is incorrect; it opens the possibility for non-perturbative contributions at $D=0$ in order to have $E \neq 0$. Interestingly,
at the non-physical dimension $D=-2$, all corrections $\veps_{2n}$ with $n > 1$ also vanish,
thus, the formal sum of corrections results is  $\veps=-2$, see (\ref{factorizationq}).
In this case, no exact solution for the corresponding radial Schr\"odinger equation is found.
It is not clear whether the Schr\"odinger equation has the solution in the Hilbert space at
$\veps=-2$.

\subsection{Generating Functions}

As it was mentioned above, one can determine the coefficients in the polynomial correction $\mathcal{Y}_{2n}(v)$, i.e. $c_{2k}^{(2n)}, k=0,1,\ldots, n$, see (\ref{Y2n}), by algebraic means.
However, as has discussed in I, more efficient procedure to calculate them is through constructing their generating functions in $(u=gr)$-space.
It was shown that the correction $\mathcal{Z}_{2n}(u)$ is, in fact, a generating function of the coefficients $c_{2k}^{(2n)}, k=0,1,2,\ldots$, see below.

For quartic anharmonic oscillator the function $\mathcal{Z}(u)$, derived from GB equation
\begin{equation}
 \la^2\,\pa_u\mathcal{Z}\ -\ \mathcal{Z}\left(\mathcal{Z} - \frac{\la^2(D-1)}{u}\right)\ =\ \la^2\,\veps(\la)\ -\ u^2 \ -\ u^4 \quad , \quad \pa_u\equiv\frac{d}{du}\ ,
\label{Bloch-4}
\end{equation}
cf. (\ref{Bloch}), can be written as an expansion in terms of generating functions, namely,
\begin{equation}
\label{expansionZq}
   \mathcal{Z}(u)\ =\ \mathcal{Z}_0(u)\ +\ \mathcal{Z}_2(u)\,\la^2\ +\ \mathcal{Z}_4(u)\,\la^4\ +\ \ldots\ ,
\end{equation}
where each coefficient
\begin{equation}
   \mathcal{Z}_{2k}(u)\ =\ u\, \sum_{n=k}^{\infty}c_{2k}^{(2n)}u^{2(n-k)} \quad ,\quad k\ =\ 0,1,\ldots\ ,
\end{equation}
is given by infinite series. Here the expansion of $\veps$ in powers of $\la$ is given by (\ref{eps-in-la-4}). Note that all generating functions $\mathcal{Z}_{2k+1}(u)$, $k=1,2,...$ of odd order $\la^{2k+1}$ are absent in expansion (\ref{Z-in-la}).

Due to invariance $u \rar -u$ it is convenient to simplify (\ref{Bloch-4}) by introducing
\[
    \mathcal{Z}\ =\ u \mathcal{\tilde Z}\quad \mbox{and}\quad {\rm u}\ =\ u^2\ .
\]
Finally, (\ref{Bloch-4}) is reduced to
\begin{equation}
 2 \la^2\,{\rm u}\,\pa_{\rm u} \mathcal{\tilde Z}\ -\ \mathcal{\tilde Z}\left({\rm u}
 \mathcal{\tilde Z} - {\la^2\,D}\right)\ =\ \la^2\,\veps(\la)\ -\ {\rm u} \ -\ {\rm u}^2 \quad ,
 \quad \pa_u\equiv\frac{d}{du}\ .
\label{Bloch-4-tilde}
\end{equation}
It is easy to find the first two terms of the expansion (\ref{expansionZq}) explicitly,
\begin{equation}
     \mathcal{\tilde Z}_0({\rm u})\ =\ \sqrt{1+{\rm u}}\ ,
\label{Z0quartic}
\end{equation}
\begin{equation}
\label{Z2quartic}
\mathcal{\tilde Z}_2({\rm u})\ =\ \frac{{\rm u}+D\left(1+{\rm u}-\sqrt{1+{\rm u}}\right)}{2{\rm u}(1+{\rm u})}\ .
\end{equation}
Interestingly, from the polynomial form of the coefficient $c_{2k}^{(2n)}$ in $D$, see (\ref{factorizationq}), one can find the structure of generating function in $D$,
\begin{equation}
   \mathcal{\tilde Z}_{2k}({\rm u})\ =\ \sum_{n=0}^{k}f^{(k)}_n({\rm u})\,D^n\ ,
\end{equation}
where all $f^{(k)}_n({\rm u}), \ n=0,1,\ldots k$ are real functions.
In general, $\mathcal{\tilde Z}_{2k}({\rm u})$ is a polynomial in $D$ of degree $k$.

The asymptotic behavior of the generating functions $\mathcal{Z}_{2k}(u),\ k=0,1,2,\ldots$ in expansion (\ref{expansionZq}) at large $u$ is related to the asymptotic behavior of the function $y$ at large  $r$ in a quite interesting manner.
It can be easily found that for fixed (effective) coupling constant $g (\la)$, the asymptotic expansion of $y$ at large $r$, rewritten in variable $v$, see (\ref{change-v}), has the form
\begin{equation}
  y\ =\ (2M\hbar^2)^{\frac{1}{4}}\left(\la v^2\ +\ \frac{1}{2\la}\ +\ \frac{D+1}{2}v^{-1}\ -\ \frac{4\la^2\veps+1}{8\la^3}v^{-2}\ +\ \ldots\right)\ ,\quad v\rar\infty\ .
\label{qexpansion}
\end{equation}
Note that the first three terms of the expansion are $\veps$- and $D$-independent.
On the other hand, the first three terms in the expansion of lowest generating function $(\frac{2M}{g^2})^{1/2}\mathcal{Z}_0(u)$ at large $u$ are
\begin{equation}
\label{Z0quartic_exp}
 \left(\frac{2M}{g^2}\right)^{1/2}\mathcal{Z}_0(u)\ =\ \left(\frac{2M}{g^2}\right)^{1/2}\left(u^2\ +\ \frac{1}{2}\ -\ \frac{1}{8}u^{-2}\ +\ \ldots\right)\ , \quad u\rar\infty \ ,
\end{equation}
see (\ref{change-to-Z}), and also are $\veps$- and $D$-independent. To compare the expansions (\ref{qexpansion}) and (\ref{Z0quartic_exp}), let us replace the classical coordinate $u$ by quantum $v$ (\ref{u vs v}),
\[
 u\ =\  \la\,v\ .
\]
(evidently, large $v$ implies large $u$ and vice versa as long as $\la$ is fixed). Then the expansion (\ref{Z0quartic_exp}) becomes
\begin{equation}
\left(\frac{2M}{g^2}\right)^{1/2}\mathcal{Z}_0(\la v)\ =\ (2M\hbar^2)^{\frac{1}{4}}\left(\la v^2\ +\ \frac{1}{2\la}\ -\ \frac{1}{8\la^3}v^{-2}\ +\ \ldots\right)\  ,\quad v \rar \infty\ .
\end{equation}
It reproduces exactly the first two terms in (\ref{qexpansion}) but fails to reproduce $O(v^{-1})$, this term is absent in the expansion. However, the next generating function $(\frac{2M}{g^2})^{1/2}\la^2\mathcal{Z}_2(\la v)$ at large $v$ reproduces the term  $O(v^{-1})$ exactly in the original expansion (\ref{qexpansion}),
\begin{equation}
 \left(\frac{2M}{g^2}\right)^{1/2}\la^2\mathcal{Z}_2(\lambda v)\ =\ (2M\hbar^2)^{\frac{1}{4}}\left(\frac{D+1}{2}v^{-1}\ -\
 \frac{D}{2\la}v^{-2}\ +\ \ldots\right)\ , \quad\quad v\rar\infty\ .
\end{equation}
In turn, it fails to reproduce correctly the term $O(v^{-2})$.
Thus, the expansion of the sum $(\frac{2M}{g^2})^{1/2}(\mathcal{Z}_0(\la v)+\la^2\mathcal{Z}_2(\la v))$ at large $v$ reproduces exactly the first three, $\veps$-independent terms in the expansion (\ref{qexpansion}). These three terms are responsible for normalizability of the wavefunction
at large $v$.

All higher generating functions $\mathcal{Z}_4(\la v)$, $\mathcal{Z}_6(\la v)\ldots\ $ contribute at large $v$ to the same term $O(v^{-2})$ as follows
\begin{equation}
 \left(\frac{2M}{g^2}\right)^{1/2}\la^{2n}\mathcal{Z}_{2n}(\la v)\ =\ (2M\hbar^2)^{\frac{1}{4}} \left(-\frac{\veps_{2n-2}\,\la^{2n-3} }{2} \, v^{-2}\ +\ \ldots\right)\ ,\quad v\rar\infty\ ,\ n\,>\,2\ ,
\end{equation}
where $\veps_{2n-2}$ is the energy PT correction of the order $(2n-2)$.
As a consequence, no matter how many generating functions we consider in the expansion $(\frac{2M}{g^2})^{1/2}(\mathcal{Z}_0(\la v)+\la^2\mathcal{Z}_2(\la v)+...)$, the term of order  $O(v^{-2})$  of (\ref{qexpansion}) can not be reproduced exactly.

\subsection{The Approximant and Variational Calculations }

The expansion of $\mathcal{Z}(u)$, see (\ref{expansionZq}), can be easily transformed into the expansion of the phase by making integration in $u$: any $\mathcal{Z}_{2n}(u)$ becomes the generating function $G_{2n}(u)$. Eventually, the expansion of the phase in generating functions becomes
\begin{equation}
  \Phi(u;\la)\ =\ G_0(u)\ +\ \la^2\, G_2(u)\ +\ \la^4\, G_4(u)\ +\ \ldots\ ,
\label{genexpphi-u}
\end{equation}
using (\ref{gndef}). Keeping $g$ fixed (\ref{genexpphi-u}) can be regarded as a semiclassical expansion of the phase.  For the present case - quartic anharmonic oscillator - this expansion results in integer powers of the Planck constant $\hbar$, see (\ref{effective}).

Without loss of generality we set $\hbar=1$ and $M = 1/2$, thus, it becomes $v = r$,
$\veps = E$, $\la = g$ and $\mathcal{Y} = y$, see (\ref{change-v}), (\ref{change-to-Y}) and (\ref{effective}), and it remains $u=g\,r$. The expansion of the phase (\ref{genexpphi-u}) is reduced into
\begin{equation}
  \Phi(r;\,g)\ =\ G_0(r;\,g)\ +\ g^2\, G_2(r;\,g)\ +\ g^4\, G_4(r;\,g)\ +\ \ldots\ ,
\label{genexpphi}
\end{equation}

Note that any generating function $G_{2k}(r;\,g), k=0,1,\ldots$ can be written in closed analytic form in terms of elementary functions. For example,
\begin{align}
 G_0(r;\,g) &\ =\ \frac{1}{3 g^2}\left(1+g^2r^2\right)^{3/2} ,
\label{firstr4}\\
 g^2\,G_2(r;\,g) &\ =\ \frac{1}{4}\log[1+g^2r^2]\ +\ \frac{D}{2}\log\left[1+\sqrt{1+g^2r^2}\right]\ ,
\label{secondr4}
\end{align}
while the next generating functions $G_{4,6}(r;\,g)$ are presented in Appendix A. 
The explicit expressions for the  generating functions $G_0(r;\,g)$ and $G_2(r;\,g)$ allow us to construct the Approximant $\Psi_{0,0}^{(t)}=e^{-\Phi_t}$. Following (\ref{generalrecipe}) the (phase) Approximant has the form
\begin{equation}
 \Phi_t\ =\ \dfrac{\tilde{a}_0\ +\ \tilde{a}_2\, r^2\ +\
 \tilde{a}_4\, g^2\, r^4}{\sqrt{1\ +\ \tilde{b}_4\,g^2\,r^2}}\ +\
 \dfrac{1}{4}\log\left[1\ +\ \tilde{b}_4\,g^2\, r^2\right]\ +\ \dfrac{D}{2}\log\left[1\ +\
 \sqrt{1\ +\ \tilde{b}_4\,g^2\,r^2}\right] \ ,
\label{quartictrialg}
\end{equation}
where $\tilde{a}_{0,2,4}, \tilde b_4$ are free parameters.
The logarithmic terms, added in (\ref{quartictrialg}), generate prefactor to the exponential function, in fact, they are just a certain \textit{minimal} modification of ones which occur in the second generating function $G_2(r;\,g)$ (\ref{secondr4}). As a result, the Approximant of the ground state function for arbitrary $D=1,2,3,\ldots$ is given by
\begin{equation}
\label{ApproximantQuartic}
     \Psi_{(0,0)}^{(t)}\ =\
     \frac{1}{\left(1\ +\ \tilde{b}_4\,g^2\,r^2\right)^{1/4}
     \left(1\ +\ \sqrt{1\ + \tilde{b}_4\, g^2\, r^2}\right)^{D/2}}\,
     \exp\left(-\ \dfrac{\tilde{a}_0\ +\ \tilde{a}_2\,r^2\ +\ \tilde{a}_4\,g^2\,r^4}
     {\sqrt{1\ +\ \tilde{b}_4\,g^2\,r^2}}\right)\ .
\end{equation}
This is the central formula of Section. Following the derivation we are certain it has to provide the highly accurate uniform approximation for the exact ground state eigenfunction, which will be checked and confirmed below. It must be emphasized that at $D=1$ the exponent (\ref{quartictrialg}) in formula (\ref{ApproximantQuartic}) coincides with the exponent found in \cite{Turbiner2005}, \cite{Turbiner2010} but slightly differ in logarithmic terms, hence, in the form of pre-factors in (\ref{ApproximantQuartic}) with the same asymptotic behavior at $r \rar \infty$. This is the consequence of the fact that in the time when \cite{Turbiner2005}, \cite{Turbiner2010} were written the GB equation, thus, 
the expansion in generating functions (\ref{genexpphi}) were unknown.
This difference leads to nonessential increase in accuracy in variational energy based on
(\ref{ApproximantQuartic}) with respect to ones used in \cite{Turbiner2005}, \cite{Turbiner2010}, while the local deviation from exact function remains almost the same.

\begin{figure}[h]
	\centering
	\begin{subfigure}[t]{0.47\textwidth}
		\centering
		\includegraphics[width=\linewidth]{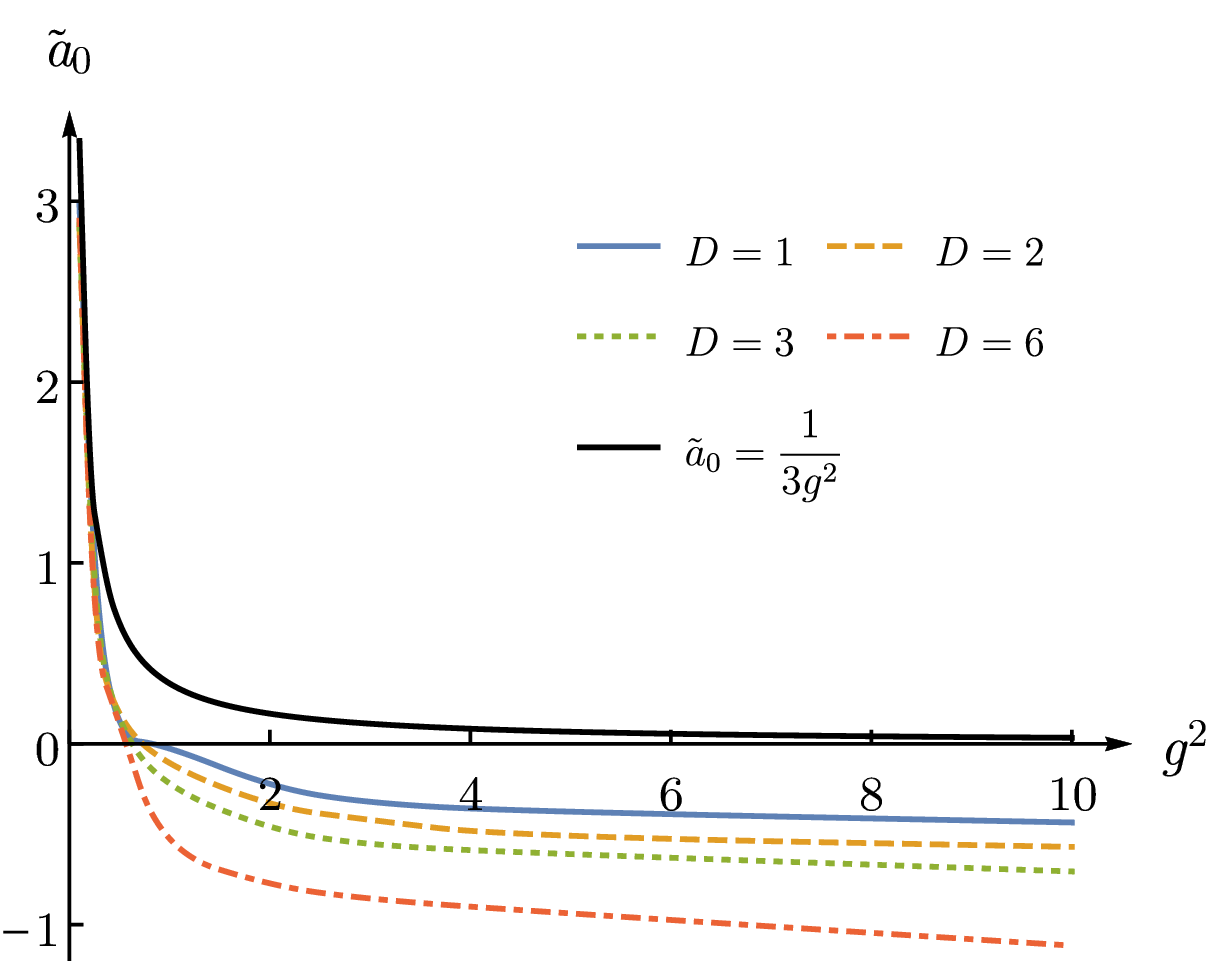}
		\caption{} \label{fig:a0}
	\end{subfigure}
	\hfill
	\begin{subfigure}[t]{0.47\textwidth}
		\centering
		\includegraphics[width=\linewidth]{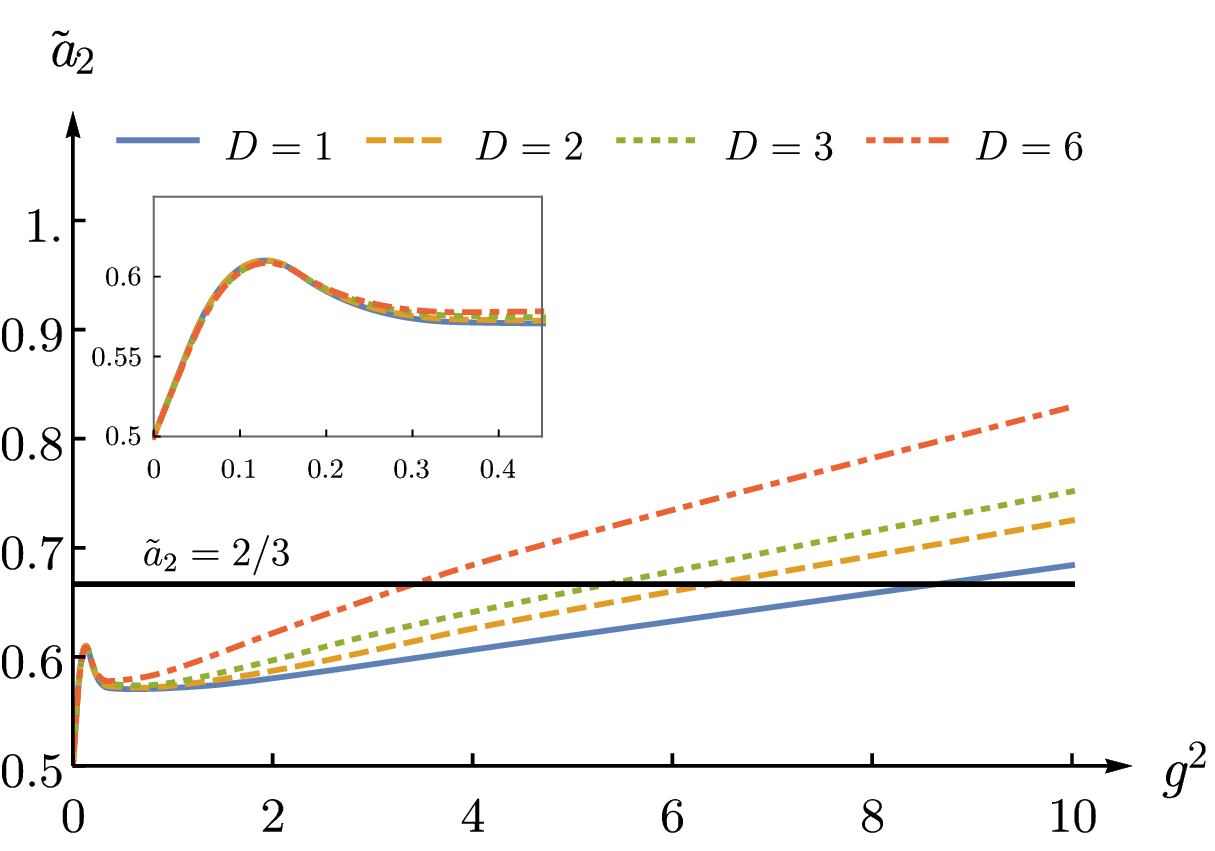}
		\caption{} \label{fig:a3}
	\end{subfigure}
	\vspace{1cm}
	\begin{subfigure}[t]{0.5\textwidth}
		\centering
		\includegraphics[width=\linewidth]{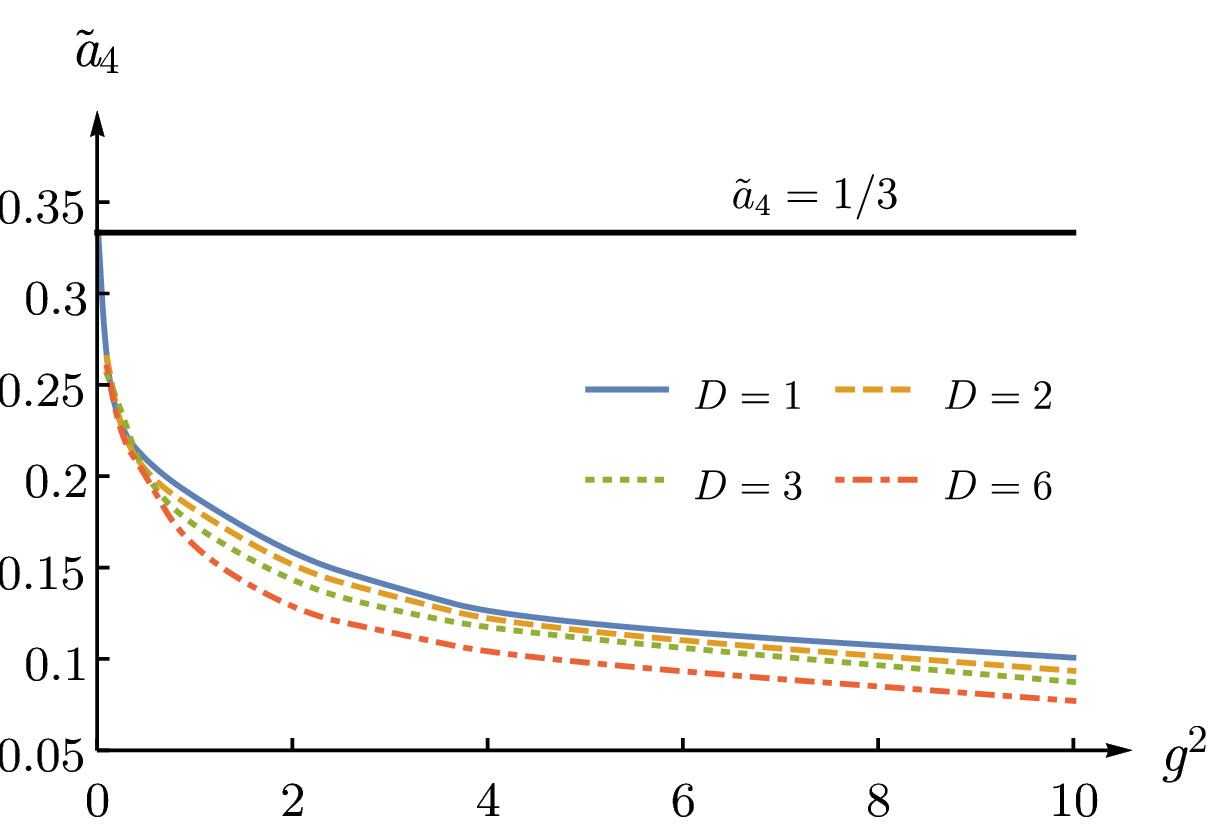}
		\caption{} \label{fig:a2}
	\end{subfigure}
  \caption{Ground state $(0,0)$: Variational parameters ${\tilde a}_0$ (a), ${\tilde a}_2$ (b) and ${\tilde a}_4$ (c) {\it vs} the coupling constant $g$ for $D=1,2,3,6$. Parameters (\ref{reproduction1}), which allow to reproduce the first two terms $G_0, G_2$ in expansion (\ref{genexpphi}) (see text), shown by solid (black) line which is horizontal for ${\tilde a}_2$ (b) and ${\tilde a}_4$ (c).}
\label{fig:varpar}
\end{figure}

\begin{figure}[h]
	\centering
	\begin{subfigure}[t]{0.47\textwidth}
		\centering
		\includegraphics[width=\linewidth]{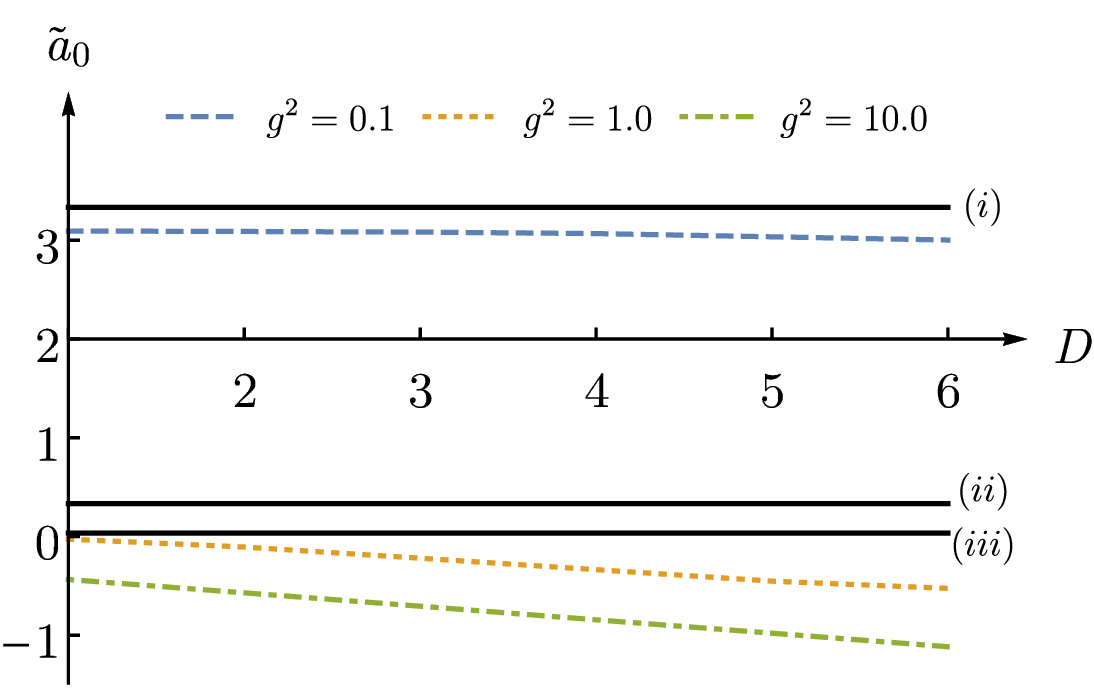}
		\caption{}
\label{fig:a0fixed}
	\end{subfigure}
	\hfill
	\begin{subfigure}[t]{0.47\textwidth}
		\centering
		\includegraphics[width=\linewidth]{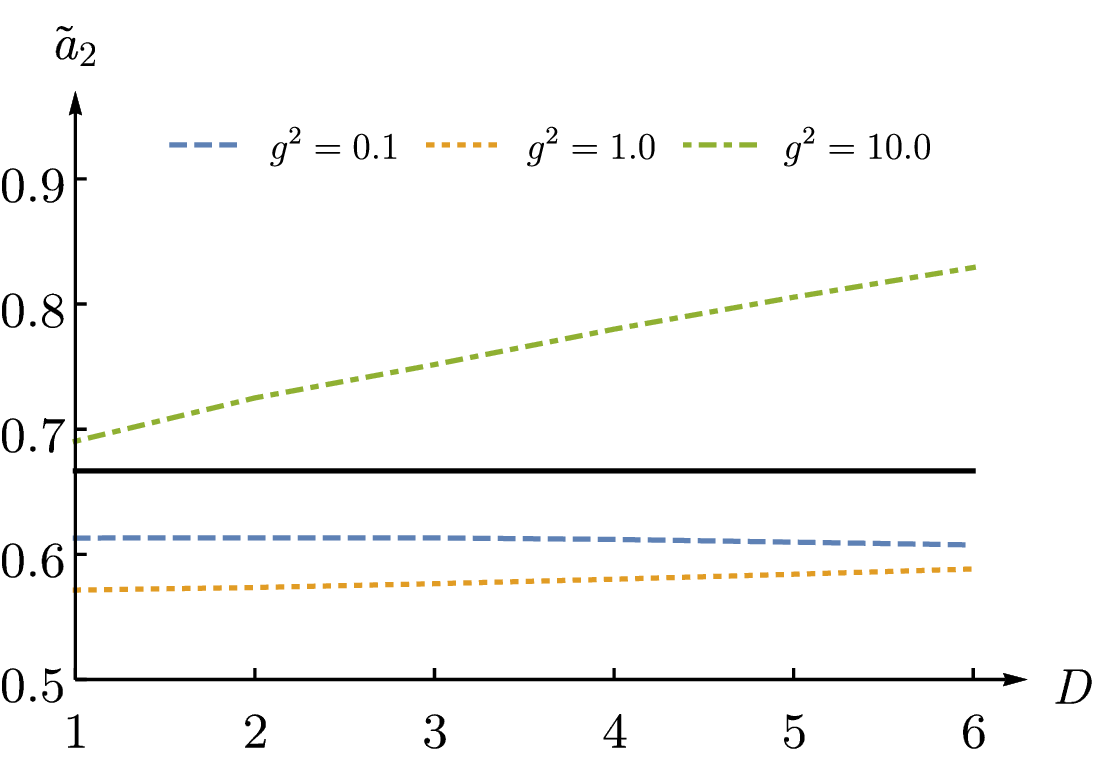}
		\caption{}
\label{fig:a2fixed}
	\end{subfigure}
	\vspace{1cm}
	\begin{subfigure}[t]{0.5\textwidth}
		\centering
		\includegraphics[width=\linewidth]{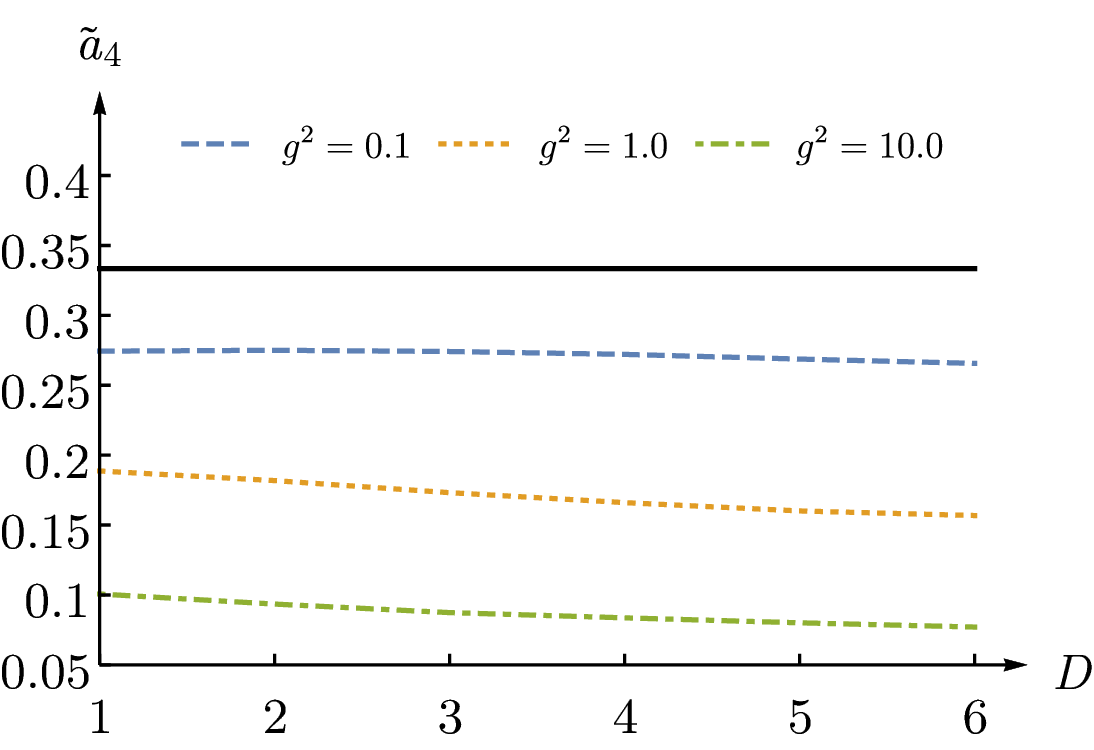}
		\caption{}
\label{fig:a4fixed}
	\end{subfigure}
	\caption{Ground state $(0,0)$: Variational parameters ${\tilde a}_0 (a), {\tilde a}_2 (b),
            {\tilde a}_4 (c)$ {\it vs} $D$ for fixed $g^2=0.1, 1, 10$. $D$-independent parameters (\ref{reproduction1}), which allow to reproduce the first two terms $G_0, G_2$ in expansion (\ref{genexpphi}) (see text), shown by solid (black) horizontal lines. In (a) the horizontal lines correspond to ${\tilde a}_0=\frac{1}{3g^2}$ at $g^2=0.1\, (i), 1.\, (ii), 10.\, (iii)$\,. }
\label{fig:varparfixed}
\end{figure}

It is easy to check that by setting the constraint
\begin{equation}
\label{a4}
    \tilde{b}_4\ =\ 9\,\tilde{a}_4^2\ ,
\end{equation}
it allows us to reproduce the dominant term in expansion (\ref{qexpansion}) exactly, hence, the asymptotic behavior of the phase at large distances. It is worth mentioning that the relaxing this constraint by keeping parameters $\tilde{a}_4$ and $\tilde{b}_4$ free demonstrates this constraint
is fulfilled with high accuracy. It justifies imposing the constraint (\ref{a4}) on variational parameters. Note that by choosing
\begin{equation}
 \tilde{a}_0\ =\ \frac{1}{3 g^2}\quad ,\quad \tilde{a}_2\ =\ \frac{2}{3}\quad ,\quad \tilde{a}_4\ =\ \frac{1}{3}\ ,
\label{reproduction1}
\end{equation}
the (phase) Approximant $\Phi_t$ reproduces exactly the first two terms in the expansion in generating functions (\ref{genexpphi}). It already leads to a highly accurate variational energies, see below Table I. However, all three parameters $\tilde{a}_0, \tilde{a}_2, \tilde{a}_4$ in (\ref{reproduction1}) are far from being optimal from the viewpoint of the variational calculations. Making minimization of the energy with respect to these parameters one can see that they appear as smooth functions in $g^2$, simultaneously being slow-changing versus $D$ for fixed $g^2$. Plots of the parameters ${\tilde a}_{0,2,4}$ {\it vs} $g^2$ for $D=1,2,3,6$ are shown in Fig. \ref{fig:varpar}, while ${\tilde a}_{0,2,4}$ {\it vs} $D$ for fixed $g^2$ are shown in Fig. \ref{fig:varparfixed}. It is worth mentioning that at small $g \lesssim 0.1$ the parameters ${\tilde a}_{0,2,4}$ are $D$-independent.

Making analysis of the parameters ${\tilde a}_{0,2,4}$ {\it vs} $g^2$ for different $D$
one can see the appearance of another, $D$-independent constraint on the parameters,
\begin{equation}
\label{a2}
    \tilde{a}_2\ \approx \ \frac{1 + 27\,\tilde{a}_4^2}{18 \tilde{a}_4}\ .
\end{equation}
It corresponds to the fact that the coefficient in front of $r$ - another growing term at $r \rar \infty$ in the trial phase (\ref{quartictrialg}) - is reproduced {\it almost exactly} in accordance to (\ref{qexpansion}). Thus it can be concluded that the trial phase (\ref{quartictrialg}) at large $r$ reproduce (almost) exactly all three growing with $r$ terms: $r^3$, $r$ and $\log r$. Eventually, if we require to reproduce all those terms exactly the Approximant in its final form will contain two free parameters $\{\tilde{a}_0, \tilde{a}_4\}$ ONLY: parameters $\{\tilde{a}_2, \tilde{b}_4\}$ obey constraints (\ref{a2}), (\ref{a4}), respectively.


As was indicated in I, the Approximant of the ground state function $\Psi_{(0,0)}^{(t)}$ is a building block to construct the Approximants of excited states. In particular, for $D=1$ in the case of the $n_r$-th excited state (at $r \geq 0$, see below) the Approximant has the form
\begin{equation}
\label{approximantcuartic1}
 \Psi_{(n_r,p)}^{(t)}\ =\
 \frac{r^p P_{n_r}(r^2)}{\left(1\ +\ \tilde{b}_4\,g^2\,r^2\right)^{1/4}\left(1\ +\
 \sqrt{1\ +\ \tilde{b}_4\, g^2\,r^2}\right)^{1/2}}\,
 \exp \left(-\ \dfrac{\tilde{a}_0\ +\ \tilde{a}_2\,r^2\ +\ \tilde{a}_4\,g^2\,r^4}
 {\sqrt{1\ +\ \tilde{b}_4\,g^2\,r^2}}\right)\ ,
\end{equation}
where $P_{n_r}(r^2)$ is a polynomial of degree $n_r$ with real coefficients and all real roots, and with $P_{n_r}(0)=1$ chosen for normalization, here $p=0,1$ and $(-)^p=\pm$ has the meaning of parity w.r.t. reflection $(r \rar -r)$. At $d=1$ there are two possible domains for the Schr\"odinger operator: $r \in [0, \infty)$ (i) and $r \in (-\infty, +\infty)$ (ii). For the first domain (i) there exist the states of positive parity $p=0$ only, we denote them $(n_r, 0)$, thus negative nodes at $r < 0$ in (\ref{approximantcuartic1}) are ignored. Hence,  the state $(n_r, 0)$ is the $n_r$-th excited state. As for the second domain (ii) there exist both states of positive and negative parity, we denote the state as $(n_r, p)$. The state $(n_r, p)$ corresponds to $(2n_r+p)$-th excited state. It is evident that the energy of the state $(n_r, 0)$ for the first domain (i) coincide with energy of the state $(n_r, 0)$ for the second domain (ii). It is easily demonstrated that energies $E_{(n_r, p)}$ obey the following inequality
\[
   E_{(n_r, 0)} < E_{(n_r, 1)} < E_{(n_r+1, 0)} \ ,
\]
for any coupling constant.
For fixed $n_r$, the $(n_r-1)$ free parameters of $P_{n_r}(r^2)$ are found by imposing the orthogonality constraints
\begin{equation}
\label{constraint1}
(\Psi_{(n_r,p)}^{(t)},\Psi_{(k_r,p)}^{(t)})\ =\ 0\quad ,\qquad k_r=0,\ldots,(n_r-1)\ .
\end{equation}

For higher dimensions, $D>1$, the Approximant for the state $(n_r,\ell)$ takes the form
\begin{equation}
\label{approximantcuarticD}
    \Psi_{(n_r,\ell)}^{(t)}\ =\
     \frac{r^{\ell}\,P_{n_r}(r^2)}{\left(1\ +\ \tilde{b}_4\,g^2\,r^2\right)^{1/4}
     \left(1\ +\ \sqrt{1\ +\ \tilde{b}_4\, g^2\,r^2}\right)^{D/2}}\,
     \exp\left(-\ \dfrac{\tilde{a}_0\ +\ \tilde{a}_2\,r^2\ +\ \tilde{a}_4\,g^2\,r^4}
     {\sqrt{1\,+\,\tilde{b}_4\,g^2r^2}}\right)\ .
\end{equation}
Here $P_{n_r}(r^2)$ is the polynomial of degree $n_r$ with $n_r$ real {\it positive} roots. In a similar way as in the one-dimensional case, for fixed angular momentum $\ell$ the $(n_r-1)$ free parameters of $P_{n_r}(r^2)$ are found by imposing the orthogonality constraints,
\begin{equation}
\label{constraint}
(\Psi_{(n_r,\ell)}^{(t)},\Psi_{(k_r,\ell)}^{(t)})\ =\ 0\quad ,\quad k_r=0,\ldots,(n_r-1)\ .
\end{equation}
Finally, in order to fix the remaining three free parameters $\tilde{a}_{0,2,4}$ in exponential we use the Approximant, either $\Psi_{(n_r,p)}^{(t)}$ or $\Psi_{(n_r,\ell)}^{(t)}$, as entry in variational calculations.  A description of the computational code we used can be found in I.

The variational energy calculations for four low-lying states with quantum numbers $(0,0)$, $(0,1)$, $(0,2)$, $(1,0)$ for different values of $D>1$ and $g^2$ are presented in Tables \ref{Quartic} - \ref{quartic3}. For some of these states, the variational energy $E_{var} = E_0^{(1)}$, the first correction $E_2$ to it, as well as its corrected value of variational energy $E_0^{(2)} = E_{var} + E_2$ are shown, see (\ref{firsta}) and (\ref{seconda}). Systematically, the variational energy $E_0^{(1)}$  is found with extremely high absolute accuracy: $10^{-8}- 10^{-14}$, which is found by calculating the correction $E_2$. The variational results are compared with numerical ones obtained via LMM, see I for technical details. LMM results are obtained taking 50, 100, 200 mesh points for $g^2=0.1, 1., 10.0$\,, respectively. It allows us to reach, at least, 12 d.d. in the energies of $(0,0)$, $(0,1)$, $(0,2)$ states at $D=1,2,3,6$, see Tables \ref{Quartic} - \ref{quartic2} denoted as $E_0^{(2)}$. Making analysis of numerical results suggests that, when $E_2$ is evaluated, the $E_0^{(2)}$ provides 12 - 13 correct d.d., at least. It implies that, once $E_2$ is taken into account, all digits of $E_0^{(2)}$ printed in Tables \ref{Quartic} - \ref{quartic3} are exact. These highly accurate results are confirmed independently by the calculating the second correction of the variational energy $E_3$ which is always $\leq 10^{-12}$ for all $D$ and $g^2$ that we have studied. It indicates a very fast rate of convergence in the Non-Linearization Procedure when trial function (\ref{approximantcuartic1}), (\ref{approximantcuarticD}) is taken as the zero approximation.
We must mention the hierarchy of eigenstates which holds for any fixed integer $D>1$ and $g^2$: $(0,0)$, $(0,1)$, $(0,2)$, $(1,0)$. Interestingly, it coincides with the hierarchy of the first four eigenstates for the cubic potential established in I.

There is a considerable number of calculations devoted to estimate the energy of the first low-lying states in domain (ii).  Our results, see Tables \ref{Quartic} - \ref{quartic3}, are in complete agreement with \cite{Turbiner2005}, \cite{Turbiner2010} for $D=1$ and superior considerably of those obtained for $D>1$ and different $g^2$, see e.g. \cite{Taseli2dr4}, \cite{WENIGER} and \cite{WITWIT2dr4}.

Deviation of $\Psi_{(n_r,\ell)}^{(t)}$ from the exact (unknown) eigenfunction $\Psi_{(n_r,\ell)}$ can be estimated via the Non-Linearization Procedure. It can be shown that for the ground state function this deviation is extremely small and bounded,
\begin{equation}
\left|\frac{\Psi_{(0,0)}(r)-\Psi_{(0,0)}^{(t)}(r)}{\Psi_{(0,0)}^{(t)}(r)}\right|\lesssim 10^{-6}\ ,
\end{equation}
in the whole range $r \in [0,\infty)$ at any dimension $D$ and any $g^2$ that we considered.
Therefore we can say that our Approximant $\Psi_{(0,0)}^{(t)}$ is a locally accurate approximation of the exact wave function $\Psi_{(0,0)}$ once the optimal parameters are chosen. A similar situation occurs for the Approximants for excited states at different $D$ and $g^2$.

\begin{table}[]
	\caption{Ground state $(0,0)$ energy for the quartic oscillator potential $r^2+g^2r^4$ for
    $D=1,2,3,6$ and $g^2=0.1,1,10$.
    Variational energy $E_0^{(1)}$, the first correction $E_2$ (rounded to 3 s.d.) found with $\Psi_{0,0}^{(t)}$ (see text), the corrected energy $E_0^{(2)}=E_0^{(1)}+E_2$ shown.
    $E_0^{(2)}$ coincides with LMM results (see text) in all displayed 12 d.d.}
\centering
\begin{adjustbox}{max width=\textwidth}
\begin{tabular}{|c|ccc|ccc|}
\hline
\multirow{2}{*}{$\quad g^2\quad$} & \multicolumn{3}{c|}{$D=1$} & \multicolumn{3}{c|}{$D=2$} \\ \cline{2-7}
                       &$\quad\quad\quad E_0^{(1)}\quad\quad\quad$                   &$\quad\quad-E_2\quad\quad$                                                      &$\quad\quad\quad E_0^{(2)}\quad\quad\quad$                  &$\quad\quad\quad E_0^{(1)}\quad\quad\quad$                                                 &$\quad\quad-E_2\quad\quad$                                                                      &$\quad\quad\quad E_0^{(2)}\quad\quad\quad$    \\
\hline
	\rule{0pt}{4ex}0.1   & 1.065\,285\,509\,544  & $3.00\times10^{-14}$  & 1.065\,285\,509\,544
    & 2.168\,597\,211\,269  & $5.28\times10^{-14}$  & 2.168\,597\,211\,269
\\[4pt]
     1.0 & 1.392\,351\,641\,563 & $3.37\times 10^{-11}$  & 1.392\,351\,641\,530 & 2.952\,050\,091\,995                                      & $3.17\times10^{-11}$  & 2.952\,050\,091\,962
\\[4pt]
     10.0 & 2.449\,174\,072\,588 & $4.69\times 10^{-10}$ & 2.449\,174\,072\,118 & 5.349\,352\,819\,751                                      & $3.44\times10^{-10}$  & 5.349\,352\,819\,462
\\[4pt]
\hline
\hline
\multirow{2}{*}{$g^2$} & \multicolumn{3}{c|}{$D=3$} & \multicolumn{3}{c|}{$D=6$} \\
\cline{2-7}
                      & $E_0^{(1)}$ & $-E_2$                                                      &
                        $E_0^{(2)}$ & $E_0^{(1)}$                                                    & $-E_2$      & $E_0^{(2)}$    \\
\hline
	\rule{0pt}{4ex}0.1  & 3.306\,872\,013\,152 & $2.20\times10^{-13}$
      & 3.306\,872\,013\,152  & 6.908\,332\,111\,232 & $9.80\times10^{-14}$                                         & 6.908\,332\,111\,232
\\[4pt]
1.0               & 4.648\,812\,704\,237      & $2.69\times10^{-11}$             & 4.648\,812\,704\,210          & 10.390\,627\,295\,514                                       & $9.68\times10^{-12}$                                       & 10.390\,627\,295\,504
\\[4pt]
10.0              & 8.599\,003\,455\,030      & $2.22\times10^{-10}$             & 8.599\,003\,454\,807            & 19.936\,900\,374\,076                                       & $6.48\times10^{-11}$                                        & 19.936\,900\,374\,011
\\[4pt]
\hline
\end{tabular}
\end{adjustbox}
\label{Quartic}
\end{table}

\begin{table}[]
	\caption{Energy of the 1st excited state in the potential $V=r^2+g^2r^4$ for different
        $D$ and $g^2$ labeled by quantum numbers $(0,1)$. For $D=1$ it corresponds
        to 1st negative parity state $n_r=0$ at $p=1$ in domain (ii) (the whole line, 
        see text).
        Variational energy $E_0^{(1)}$, the first correction $E_2$ to it found with $\Psi_{(0,1)}^{(t)}$, the corrected energy $E_0 ^{(2)}=E_0^{(1)}+E_2$ shown.
        Displayed correction $E_2$ rounded to 3 s.d.; $E_0^{(2)}$ coincides with LMM results (see text) in all 12 displayed d.d.}
		\centering
		\label{quartic1}
		\begin{adjustbox}{max width=\textwidth}
		\begin{tabular}{|c|ccc|ccc|}
			\hline
			\multirow{2}{*}{$\quad  g^2 \quad$} & \multicolumn{3}{c|}{$D=1$} &
                           \multicolumn{3}{c|}{$D=2$} \\
\cline{2-7}
			& $\qquad\qquad E_0^{(1)}\qquad\qquad$ & $\qquad\quad -E_2 \qquad\quad$       &$\quad\quad\quad E_0^{(2)}\quad\quad\quad $      & $\quad\quad\quad E_0^{(1)}\quad\quad\quad$       &$\quad\quad-E_2\quad\quad$       &$\quad\quad\quad E_0^{(2)}\quad\quad\quad$ \\
\hline
			\rule{0pt}{4ex}	0.1 & 3.306\,872\,013\,236  & $8.33\times10^{-11}$ & 3.306\,872\,013\,153      & 4.477\,600\,360\,878 & $1.10 \times 10^{-10}$ & 4.477\,600\,360\,768
\\[4pt]
			1.0 & 4.648\,812\,707\,206 & $2.99 \times 10^{-9}$ & 4.648\,812\,704\,212 & 6.462\,906\,003\,251 & $3.39 \times 10^{-9}$ & 6.462\,905\,999\,864
\\[4pt]
			10.0 & 8.599\,003\,467\,556 & $1.27 \times 10^{-8}$ & 8.599\,003\,454\,810      & 12.138\,224\,752\,729 & $1.38 \times 10^{-8}$ & 12.138\,224\,738\,901
\\[4pt]
\hline\hline
			\multirow{2}{*}{$g^2$} & \multicolumn{3}{c|}{$D=3$} & \multicolumn{3}{c|}{$D=6$} \\ \cline{2-7}
			& $E_0^{(1)}$ & $-E_2$ & $E_0^{(2)}$ & $E_0^{(1)}$ & $-E_2$ & $E_0^{(2)}$
\\
\hline
				\rule{0pt}{4ex}0.1 & 5.678\,682\,663\,377 & $1.33 \times 10^{-10}$ & 5.678\,682\,663\,243 & 9.447\,358\,518\,278 & $1.80 \times 10^{-10}$ & 9.447\,358\,518\,099       \\[4pt]
			1.0 & 8.380\,342\,533\,658 & $3.56 \times 10^{-9}$ & 8.380\,342\,530\,101       & 14.658\,513\,816\,952 & $3.39 \times 10^{-9}$ & 14.658\,513\,813\,563
\\[4pt]
		   10.0 & 15.927\,096\,988\,667 & $1.40 \times 10^{-8}$ & 15.927\,096\,974\,709     & 28.536\,810\,849\,436 & $1.21 \times 10^{-8}$ & 28.536\,810\,837\,360
\\[4pt]
\hline
\end{tabular}
\end{adjustbox}
\end{table}

\begin{table}[]
	\centering
	\caption{Energy of the 2nd excited state in the potential $V=r^2+g^2r^4$ for different 
       $D$ and $g^2$ labeled by quantum numbers $(0,2)$ for $D>1$, as for $D=1$ 
       it corresponds to 1st excitation $n_r=1$ of positive parity $p=0$: $(1,0)$.
       Variational energy $E_0^{(1)}$ found with use of $\Psi_{(1,0)}^{(t)}$ for
       $D = 1$ and $\Psi_{(0,2)}^{(t)}$ for $D > 1$; the first correction $E_2$ and
       the corrected energy $E_0^{(2)}= E_0^{(1)} + E_2$ shown.  Displayed correction $E_2$ rounded to 3 s.d. $E_0^{(2)}$ coincides with LMM results (see text) 
       in all 12 displayed d.d.}
\label{quartic2}
	\begin{adjustbox}{max width=\textwidth}
		\begin{tabular}{|c|ccc|ccc|}
\hline
			\multirow{2}{*}{$\quad g^2\quad$} & \multicolumn{3}{c|}{$D=1$} & \multicolumn{3}{c|}{$D=2$} \\
\cline{2-7}
			& \multicolumn{3}{c|}{$E_0^{(1)}$}    &$\qquad\qquad E_0^{(1)}\qquad\qquad$       &$\quad\quad-E_2\quad\quad$         &$\qquad\qquad E_0^{(2)}\qquad\qquad$
\\ \hline
			\rule{0pt}{4ex}	0.1  & \multicolumn{3}{c|}{5.747\,959\,269\,942} & 6.908\,332\,112\,167 & $9.35 \times 10^{-10}$ & 6.908\,332\,111\,232
\\[4pt]
			1.0 & \multicolumn{3}{c|}{8.655\,049\,995\,062 } & 10.390\,627\,321\,799 & $2.63 \times 10^{-8}$ & 10.390\,627\,295\,506
\\[4pt]
			10 & \multicolumn{3}{c|}{16.635\,921\,650\,401} & 19.936\,900\,479\,247        & $1.05 \times 10^{-7}$ & 19.936\,900\,374\,040
\\
\hline\hline
			\multirow{2}{*}{$g^2$} & \multicolumn{3}{c|}{$D=3$} & \multicolumn{3}{c|}{$D=6$}
\\ \cline{2-7}
			& $E_0^{(1)}$ & $\qquad\qquad -E_2 \qquad\qquad$ & $E_0^{(2)}$ & $E_0^{(1)}$ & $-E_2$         & $E_0^{(2)}$
\\
\hline
				\rule{0pt}{4ex}0.1 & 8.165\,006\,438\,494 & $1.00 \times 10^{-9}$ & 8.165\,006\,437\,493 & 12.084\,471\,853\,886 & $1.11 \times 10^{-9}$ & 12.084\,471\,852\,776      \\[4pt]
			1.0 & 12.485\,556\,075\,670 & $2.47 \times 10^{-8}$ & 12.485\,556\,051\,000 & 19.217\,523\,515\,555 & $1.97 \times 10^{-8}$ & 19.217\,523\,495\,879
\\[4pt]
		   10.0 & 24.145\,857\,689\,623 & $9.48 \times 10^{-8}$ & 24.145\,857\,594\,824 & 37.811\,402\,320\,699 & $6.90 \times 10^{-8}$ & 37.811\,402\,251\,702
\\
\hline
\end{tabular}
\end{adjustbox}
\end{table}

\begin{table}[]
	\caption{Energy $E$ of the 3rd excited state $(1,0)$ for $D=2,3,6$ - 
             the first radial excitation - for the potential $V=r^2 + g^2 r^4$ at 
             $g^2=0.1, 1.0, 10.0$ and its node $r_0$, calculated in LMM with 50,100,200 
             mesh points, respectively.
             Variational energy $E_0^{(1)}$ and its node position $r_0^{(0)}$ correspond
             to underlined digits, both found with $\Psi_{(1,0)}^{(t)}$. First
             corrections $E_2$ and $r_0^{(1)}$ to variational results (not shown) 
             contribute to the first non-underline figure (and next ones).}
	\label{quartic3}
\resizebox{\textwidth}{!}{\begin{tabular}{|c|cc|cc|cc|}
\cline{1-7}
\multicolumn{1}{|c|}{\multirow{2}{*}{$\quad\,g^2\quad\,$}} & \multicolumn{2}{c|}{$D=2$} & \multicolumn{2}{c|}{$D=3$} & \multicolumn{2}{c|}{$D=6$} \\ \cline{2-7}
\multicolumn{1}{|c|}{}                  &    $\quad \qquad E_0^{(1)}\quad \qquad $  &
$\quad \qquad r_0^{(0)}\quad \qquad $   &    $\quad \qquad E_0^{(1)}\quad \qquad $  &
$\quad \qquad r_0^{(0)}\quad \qquad $   &    $\quad \qquad E_0^{(1)}\quad \qquad $  &
$\quad \qquad r_0^{(0)}\quad \qquad $
\\
\hline
	\rule{0pt}{5ex}0.1 &\underline{7.039\,707\,58}4  &\underline{0.918\,78}3\,458  &\underline{8.352\,677\,82}5   &\underline{1.111\,521}\,078  &\underline{12.415\,256\,1}77  & \underline{1.522\,966}\,591
\\
 1.0 &\underline{10.882\,435}\,576  &\underline{0.733\,72}4\,778 &\underline{13.156\,803\,9}22  &\underline{0.875\,5}67\,486  &\underline{20.293\,829}\,707  & \underline{1.166\,7}53\,149
\\
 10.0 &\underline{21.175\,135}\,370  &\underline{0.524\,08}3\,057  &\underline{25.806\,276}\,215  &\underline{0.621\,79}5\,290  & \underline{40.388\,142\,9}70  & \underline{0.820\,068}\,428
\\[8pt]
\hline
\end{tabular}}
\end{table}

\begin{figure}[]
	\includegraphics[width=0.99\textwidth]{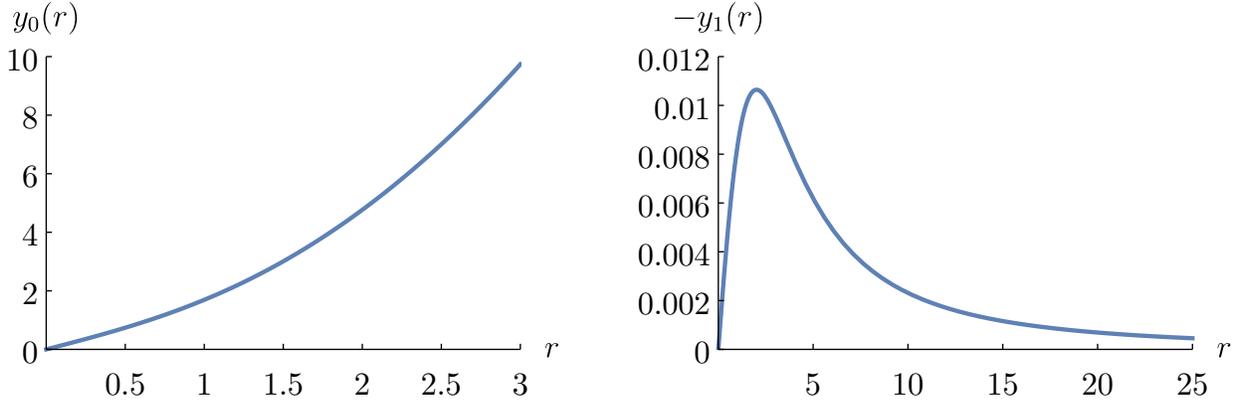}
	\caption{Quartic oscillator at $D=1$: function $y_0=(\Phi_t)'$  (on left) and its
     first correction $y_1$  (on right) {\it vs} $r$ for $g^2=1$. }
\label{fig:D=1q}
\end{figure}

\begin{figure}[]
	\includegraphics[width=0.99\textwidth]{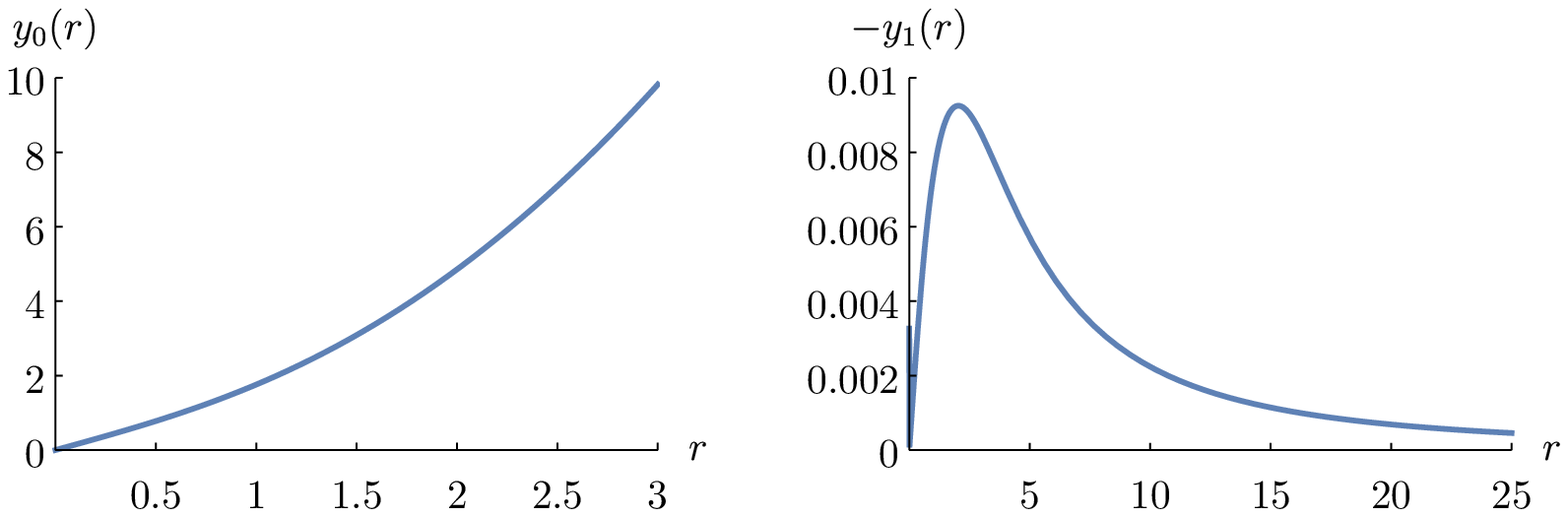}
	\caption{Quartic oscillator at $D=2$: function $y_0=(\Phi_t)'$ (on left) and its
     first correction $y_1$ (on right) {\it vs} $r$ for $g^2=1$.}
\label{fig:D=2q}
\end{figure}

\begin{figure}[]
	\includegraphics[width=0.99\textwidth]{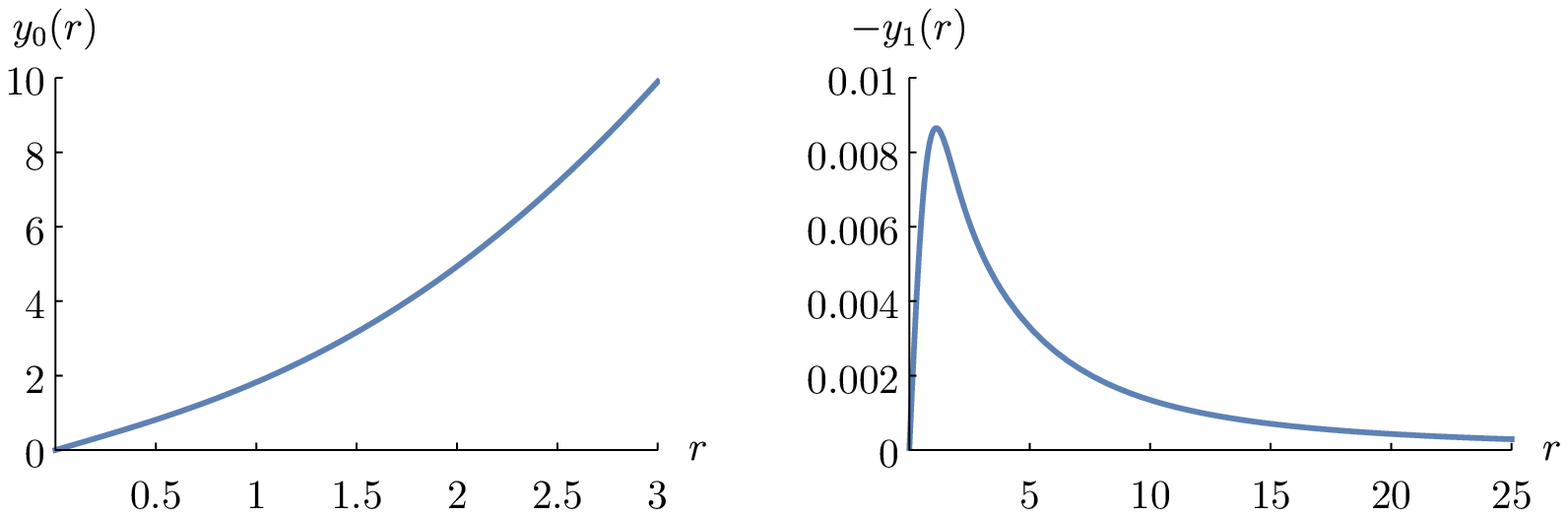}
	\caption{Quartic oscillator at $D=3$: function $y_0=(\Phi_t)'$ (on left) and its
     first correction $y_1$ (on right) {\it vs} $r$ for $g^2=1$.}
\label{fig:D=3q}
\end{figure}

In all cases the first order correction $y_1$ to the logarithmic derivative of the ground state is a bounded function at different $D$ and $g^2$. For example, for $g^2=1$ the first correction $y_1$ has the upper bound
\begin{equation}
\label{cases-4}
|y_1|_{max} \sim
\begin{cases}
 0.0106\ ,\qquad D=1 \\
 0.0092\ ,\qquad D=2 \\
 0.0086\ ,\qquad D=3 \\
 0.0072\ ,\qquad D=6 \\
\end{cases}
\end{equation}
It is the consequence of the fact that by construction the derivative of $\Phi_t$ reproduces exactly all growing terms at large $r$ in expansion (\ref{qexpansion}). ``Boundness" of $y_1$ and its small value of the maximum implies that we deal with smartly designed zero order-approximation $\Psi_{0,0}^{(t)}$ that leads, in framework of the Non-Linearization Procedure, to a fastly convergent series for the energy and wave function. In Figs. \ref{fig:D=1q} - \ref{fig:D=3q} $y_0$ and $y_1$ {\it vs} $r$ are presented for $g^2=1$ in physics dimensions $D=1,2,3$. Let us emphasize that all curves in these figures are slow-changing {\it vs} $D$. Therefore, it is not a surprise that similar plots should appear for $D=6$ (not shown) as well as for other values of $g$. An analysis of these plots indicates that $|y_1|$ is extremely small function in comparison with $|y_0|$ in the domain $0 \leq r \lesssim 1.7$, thus, in domain which provides the dominant contribution in variational integrals. It is the consequence of minimization of the energy functional, see (\ref{evar}). It is the real reason why the energy correction $E_2$ is extremely small being of order $\sim 10^{-8}$, or sometimes even smaller, $\sim 10^{-10}$. Similar situation occurs for the phase (and its derivative) of the Approximants for the excited states.

The Approximant  $\Psi_{(n_r,\ell)}^{(t)}$ (\ref{approximantcuarticD}) also allows us to get an accurate estimate of the position of the radial nodes of the exact wave function. For example, in the state $(1,0)$ where is a single positive node, $r_0 > 0$, the trial function (\ref{approximantcuarticD}) provides the zero order estimate of the radial node $r_0^{(0)}$  coming directly from the orthogonality constraint (\ref{constraint}). Results are presented in Table \ref{quartic3}. A comparison of these numerical results with those coming from the LMM indicates that the Approximant  $\Psi_{1,0}^{(t)}$ defines the node with not less than 5 d.d. From Table \ref{quartic3} it can be noted that the radial node is an increasing function of $D$ at fixed $g^2$, but decreasing with the increase of $g^2$ at fixed $D$.

\subsection{The Strong Coupling Expansion}

In this Section assuming $2M=\hbar=1$ we will focus on finding the first two terms of the strong coupling expansion (\ref{Scoupling}) of the ground state energy for the quartic anharmonic oscillator (\ref{potquartic}),
\begin{equation}
\label{stq}
  E\ \equiv \ g^{2/3}\, {\tilde \veps}\ =\ g^{2/3}(\tilde{\veps}_0\ +\ \tilde{\veps}_2g^{-4/3}\ +\ \tilde{\veps}_4g^{-8/3}\ +\
  \ldots) \ .
\end{equation}
In contrast to the weak coupling expansion, see (\ref{eps-in-la-4}) at $\la=g$, the expansion (\ref{stq}) has a finite radius of convergence in $1/g$. This expansion corresponds to PT in powers of $\hat{\la}$ for the potential
\begin{equation}
\label{qST}
    V(w)\ =\ w^4\ +\ \hat{\la}^2\,w^2\ ,\quad \hat{\la}^2\ =\ g^{-4/3} \ ,
\end{equation}
in the Schr\"odinger equation defined in $w \in [0,\infty)$. The transformed RB equation suitable to develop such a PT has the form, see Eq.(III.8) in I and (\ref{riccati-bloch-4-tilde}),
\begin{equation}
\label{riccatiST}
   2{\rm w}\pa_{\rm w}{\mathcal{Y}}({\rm w})\ -\ {\mathcal{Y}}({\rm w})
   \left({\rm w}{\mathcal{Y}}({\rm w})-{D}\right)\ =\ \tilde{\veps}(\hat{\la})\ -\ {\hat{\la}^2}\,{\rm w} - {\rm w}^2 \quad ,\quad \pa_{\rm w}\equiv\frac{d}{d{\rm w}}\ ,\ {\rm w}\equiv w^2\ ,
\end{equation}
c.f. (\ref{riccati-bloch}) with different r.h.s., where $\hat{\la}$ plays a role of effective coupling constant, and $\tilde{\veps}(\tilde{\la})$ plays a role of energy.

In order to calculate the first two terms $\tilde{\veps}_0$ and $\tilde{\veps}_2$ of the strong coupling expansion (\ref{stq}) we use the Approximant (\ref{ApproximantQuartic}).  In Table \ref{table:stcuartic1} for different $D$  the leading coefficient  $\tilde{\veps}_0$ and the second perturbative correction $\hat{\veps}_2$ as well as  $\tilde{\veps}_0^{(2)}=\tilde{\veps}_0^{(1)}+\hat{\veps}_2$, calculated via the Non-Linearization Procedure, are presented. Numerical results for $\tilde{\veps}_0$, based on the LMM and obtained with 12 d.d., indicate that $\tilde{\veps}_0^{(2)}$ found in Non-Linearization procedure with the Approximant (\ref{ApproximantQuartic}) reproduce not less that 10 d.d. This accuracy is verified independently by calculating the next correction $\hat{\veps}_3$ which results in the order of $\hat{\veps}_3 \sim 10^{-2}\hat{\veps}_2$. In turn, Table \ref{table:stcuartic2} contains the results of the first two approximations for the coefficient $\tilde{\veps}_2$ in (\ref{stq}).
It should be mentioned that our final results for the coefficient $\tilde{\veps}_0$ reproduce and sometimes exceed the best results available in literature so far for $D=1$, see e.g. \cite{TurbinerST}, \cite{StrongFernandez}, \cite{WENIGST}.
\begin{table}[]
\centering
	\caption{Ground state $(0,0)$ energy $\tilde{\veps}_0$ for the potential
     $W=r^4$ (see (\ref{qST})) for $D=1,2,3,6$ found in PT based on the Approximant $\Psi_{(0,0)}^{(t)}$: $\tilde{\veps}_0^{(1)}$ corresponds to the variational energy, $\hat{\veps}_2$ is
     the second PT correction, $\tilde{\veps}_0^{(2)}=\tilde{\veps}_0^{(1)}+\hat{\veps}_2$
     is the corrected variational energy. 10 d.d. in $\tilde{\veps}_0^{(2)}$
     confirmed independently in LMM.}
\label{table:stcuartic1}
\begin{adjustbox}{max width=\textwidth}
\begin{tabular}{|ccc|ccc|}
\hline
		\multicolumn{3}{|c|}{$D=1$} & \multicolumn{3}{c|}{$D=2$} \\ \hline
			$\quad\quad\quad \tilde{\veps}_0^{(1)}\quad\quad\quad$      &$\quad\quad\quad-\hat{\veps}_2\quad\quad\quad$       &$\quad\quad\quad \tilde{\veps}_0^{(2)}\quad\quad\quad$       & $\quad\quad\quad \tilde{\veps}_0^{(1)}\quad\quad\quad$      &$\quad\quad\quad-\hat{\veps}_2\quad\quad\quad$       &$\quad\quad\quad \tilde{\veps}_0^{(2)}\quad\quad\quad$       \\[4pt]
		1.060\,362\,090\,491 & $7.02 \times 10^{-12}$ & 1.060\,362\,090\,484 & 2.344\,829\,072\,753       & $9.27 \times 10^{-12}$ & 2.344\,829\,072\,744
\\[4pt] \hline \hline
		\multicolumn{3}{|c|}{$D=3$} & \multicolumn{3}{c|}{$D=6$} \\ \hline
			$\tilde{\veps}_0^{(1)}$      &$-\hat{\veps}_2$       &$\tilde{\veps}_0^{(2)}$       & $\tilde{\veps}_0^{(1)}$      &$-\hat{\veps}_2$       &$\tilde{\veps}_0^{(2)}$       \\
		3.799\,673\,029\,810 & $9.27 \times 10^{-12}$ & 3.799\,673\,029\,801 & 8.928\,082\,199\,890       & $4.07 \times 10^{-11}$ & 8.928082199850
\\[4pt]  \hline
\end{tabular}
\end{adjustbox}
\end{table}

\begin{table}[h]
\centering
  \caption{Subdominant coefficient $\tilde{\veps}_2$ in the strong coupling  expansion (\ref{stq}) for the ground state  $(0,0)$ energy for the quartic radial anharmonic
  potential (\ref{qST}) for different  $D=1,2,3,6$. First order correction $\tilde{\veps}_{2,1}$ in PT, see text, included. 10 d.d. in $\tilde{\veps}_2^{(2)}$
  confirmed independently in LMM.}
\label{table:stcuartic2}
\begin{adjustbox}{max width=\textwidth}
\begin{tabular}{|ccc|ccc|}
\hline
		\multicolumn{3}{|c|}{$D=1$} & \multicolumn{3}{c|}{$D=2$} \\ \hline
			$\quad\quad\quad \tilde{\veps}_2^{(1)}\quad\quad\quad$& $\quad\quad\quad \tilde{\veps}_{2,1}\quad\quad\quad$  &      $\quad\quad\quad \tilde{\veps}_2^{(2)}\quad\quad\quad$    &$\quad\quad\quad \tilde{\veps}_2^{(1)}\quad\quad\quad $    &    $\quad\quad\quad \tilde{\veps}_{2,1}\quad\quad\quad$     &$\quad\quad\quad \tilde{\veps}_2^{(2)}\quad\quad\quad$             \\
		0.362\,022\,648\,388& $3.96 \times 10^{-10}$ & 0.362\,022\,648\,784 & 0.651\,477\,773\,845           & $4.38 \times 10^{-10}$ & 0.651\,477\,774\,283
\\[4pt]
\hline \hline
		\multicolumn{3}{|c|}{$D=3$} & \multicolumn{3}{c|}{$D=6$}
\\
\hline
			$\tilde{\veps}_2^{(1)}$ & $\tilde{\veps}_{2,1}$ &  $\tilde{\veps}_2^{(2)}$
         & $\tilde{\veps}_2^{(1)}$  & $\tilde{\veps}_{2,1}$ &  $\tilde{\veps}_2^{(2)}$ \\
		  0.901\,605\,894\,682 & $2.03 \times 10^{-9}$ & 0.901\,605\,896\,709 & 1.526\,804\,282\,772
         & $-3.06 \times 10^{-8}$ & 1.526\,804\,252\,175
\\[4pt]
\hline	
\end{tabular}
\end{adjustbox}
\end{table}

\subsection{Quartic Radial Anharmonic Oscillator: conclusions}

It is shown that the 2-parametric Approximants (\ref{ApproximantQuartic}), (\ref{approximantcuartic1}), (\ref{approximantcuarticD}) taken as variational trial functions for the first four states $(0,0), (0,1), (0,2), (1,0)$ of the quartic radial $D$-dimensional anharmonic oscillator with the potential (\ref{potquartic}) provide extremely high relative accuracy in energy ranging from $\sim 10^{-14}$ to  $\sim 10^{-8}$ for different coupling constants $g$ and dimension $D$. Variational parameters depend on $g$ and $D$ in a smooth manner and can be easily interpolated.
For $D=1$ the Approximant (\ref{approximantcuartic1}) appears as a slight generalization
of the trial functions proposed in \cite{Turbiner2005,Turbiner2010}: they differ in a form of the pre-exponential factors. 
If variationally optimized Approximants are taken as zero approximation in Non-Linearization (iteration) procedure, they lead to a fastly convergent scheme
with rate of convergence $\sim 10^{-4}$. For the ground state it was calculated
the relative deviation of the logarithmic derivative of variationally optimized Approximant from the exact one {\it vs} radial coordinate $r$ for different $g$ and $D$. It was always smaller than $\sim 10^{-6}$. It implies that the Approximants with interpolated parameters ${\tilde a}_{4,0}$ {\it vs} $g$ and $D$ provide highly accurate uniform approximation of the eigenfunctions of the Quartic Radial Anharmonic Oscillator while the respectful eigenvalues are given by ratio of two integrals with integrands proportional to Approximants.


\section{Sextic Anharmonic Oscillator}

In this Section it will be considered the sextic anharmonic radial oscillator two-term potential
\begin{equation}
\label{potsextic}
  V(r)\ = \ r^2\ +\ g^4\,r^6\ ,
\end{equation}
see (\ref{potential}) at $m=6$ and $a_3=a_4=a_5=0, a_6=1$, cf. (\ref{potquartic}).

\subsection{PT in the Weak Coupling Regime}

In the Weak Coupling Regime the perturbative expansion for $\veps$ and $\mathcal{Y}(v)$, developed in RB equation (\ref{riccati-bloch}),
\begin{equation}
\label{riccati-bloch-6}
 \pa_v\mathcal{Y}\ -\ \mathcal{Y}\left(\mathcal{Y} - \frac{D-1}{v}\right)\ =\
 \veps\left(\la\right)\ -\ v^2\ \ -\ \la^4\,v^6 \quad ,
 \quad \pa_v \equiv \frac{d}{dv}\ ,
\end{equation}
where $v$ and $\la$ are defined in (\ref{change-v}) and (\ref{effective}), correspondingly,
are of the form
\begin{equation}
\label{encorrection-6}
   \veps\ =\ \veps_0\ +\ \veps_4\,\la^4\ +\ \veps_8\,\la^8\ +\ \ldots\quad , \qquad \veps_0\ =\ D\ ,
\end{equation}
and
\begin{equation}
\label{Yncorrection-6}
   \mathcal{Y}(v)\ =\ \mathcal{Y}_0\ +\ \mathcal{Y}_4\,\la^4\ +\ \mathcal{Y}_8\,\la^8\ +\ \ldots
   \quad , \quad \mathcal{Y}_0\ =\ v \ ,
\end{equation}
respectively. All coefficients in front of terms $\la^{4n+1}$, $\la^{4n+2}$, $\la^{4n+3}$, $n=0,1,2,\ldots$ are equal to zero. Since the potential (\ref{potsextic}) is even, PT can be constructed by algebraic means. The first non-vanishing corrections are
\begin{equation}
\label{correction2-6}
\veps_4\ =\ \frac{1}{8}\,D\,(D+2)\,(D+4)\quad ,\quad \mathcal{Y}_4(v)\ =\ \frac{1}{2}\,v^5\ +\ \frac{1}{4}(D+4)\,v^3\ +\ \frac{1}{8}\,(D+2)\,(D+4)\,v\ ,
\end{equation}
while the next two corrections $\veps_{8,12}$ and $\mathcal{Y}_{8,12}(v)$ are presented in Appendix B.
It can be shown that the correction $\mathcal{Y}_{4n}(v)$ has the form of odd polynomial in $v$,
\begin{equation}
\mathcal{Y}_{4n}(v)\ =\ v\,\sum_{k=0}^{2n}c_{2k}^{(4n)}v^{2(2n-k)}\ ,
\label{Yncorrection-sex}
\end{equation}
with coefficients $c_{2k}^{(4n)}$ being polynomial in $D$ of degree $k$,
\begin{equation}
c^{(4n)}_{2k}\ =\ P^{(4n)}_{k}(D)\quad ,\qquad c_{4n}^{(4n)}\ =\ \frac{\veps_{4n}}{D}\ ,
\label{propertiessextic}
\end{equation}
cf. (\ref{Y2n}), (\ref{Y2n-c}).
The correction $\veps_{4n}$  has the factorization property
\begin{equation}
\veps_{4n}(D)\ =\ D\,(D+2)\,(D+4)\,R_{2n-2}(D)\ ,
\label{factorizations}
\end{equation}
where $R_{2n-2}(D)$ is a polynomial in $D$ of degree $(2n-2)$, cf. (\ref{factorizationq}), in particular, $R_0=\frac{1}{8}$.

Due to invariance $v \rar -v$ of the original equation (\ref{riccati-bloch-6}) it is convenient to simplify it by introducing a new unknown function and changing $v$-variable to its square,
\[
    \mathcal{Y}\ =\ v\, \mathcal{\tilde Y}\quad \mbox{and}\quad {\rm v}\ =\ v^2\ .
\]
As a result, (\ref{riccati-bloch-6}) becomes
\begin{equation}
\label{riccati-bloch-6-tilde}
 2 {\rm v} \pa_{\rm v} \mathcal{\tilde Y}\ -\ \mathcal{\tilde Y}\left({\rm v} \mathcal{\tilde Y} - D\right)\ =\
 \veps\left(\la\right)\ -\ {\rm v}\ \ -\ \la^4\,{\rm v}^3 \quad ,
 \quad \pa_v \equiv \frac{d}{dv}\ .
\end{equation}
This is a convenient form of the RB equation to carry out the PT consideration. It particular, the first correction (\ref{correction2-6}) in expansion of $\mathcal{\tilde Y}$ becomes a second degree polynomial in ${\rm v}$,
\[
  \mathcal{\tilde Y}_4({\rm v})\ =\ \frac{1}{2}\,{\rm v}^2\ +\ \frac{D+4}{4}\,{\rm v}\ +\ \frac{(D+2)\,(D+4)}{8}\,\ ,
\]
and, in general, $\mathcal{\tilde Y}_{4n}({\rm v})$ is a polynomial in ${\rm v}$ of degree $(2n)$, see (\ref{Y2n}).
Corrections $\mathcal{\tilde Y}_{4,6}$ are presented in Appendix B.

From (\ref{propertiessextic}) one can see that all corrections $\veps_{4n}$ vanish at $D=0, -2, -4$, hence, their formal sum results in $\veps=0, -2, -4$, respectively. In the case $D=0$ the radial  Schr\"odinger equation takes the form
\begin{equation}
\label{D=0-s}
  -\frac{\hbar^2}{2M}\left(\frac{d^2\Psi(r)}{dr^2}\ -\ \frac{1}{r}\frac{d\Psi(r)}{dr}\right)\ +\
  (r^2\ +\ g^4\,r^6)\,\Psi(r)\ =\ 0\ ,
\end{equation}
cf.(\ref{D=0-q}). Its formal solution, cf. \cite{DOLGOVPOPOV1979}, is given in terms of the parabolic cylinder functions \cite{abramowitz+stegun} (also known as Weber functions), it reads
\begin{equation}
\label{D=0-s-psi}
  \Psi\ =\ C_1\,D_{\nu_{-}}(\la v^2)\  +\ C_2\,D_{\nu_{-}}\left(i\,\la v^2\right)\ ,\quad\quad\quad \nu_{\pm}\ =\ -\frac{1}{2}\ \pm\ \frac{1}{4\la^2}\ ,
\end{equation}
if written in $v$ and $\la$, see (\ref{change-v}) and (\ref{effective}), respectively, cf. (\ref{D=0-q-psi}). It has the meaning of the zero mode of the Schr\"odinger operator at $D=0$.
The function (\ref{D=0-s-psi}) cannot be made normalizable by any choice of constants  $C_1$ and $C_2$.
Hence, the Schr\"odinger operator at $D=0$ for sextic potential (\ref{potsextic}) has no zero mode in the Hilbert space. It complements the similar statement made for quartic potential (\ref{potquartic}).
One can guess that zero mode in the Hilbert space is absent for the Schr\"odinger operator with anharmonicity $r^{2m}$ at $D=0$.

Like in the quartic oscillator case, the assumption that $E(D=0)=0$ is incorrect: non perturbative contribution in $\la$ (or $g$) to energy should be present at $D=0$. Note even though at $D=-2$ and $D=-4$ all corrections $\veps_{4n}$ vanish and we formally have $\veps=-2$ and $\veps=-4$, respectively, no exact solutions have been found for the corresponding radial Schr\"odinger equation. In Appendix \ref{appendix:C} some general results are presented for the two-term potentials.

\subsection{Generating Functions}

For the sextic anharmonic oscillator using the GB equation (\ref{Bloch}),
\begin{equation}
 \la^2\,\pa_u\mathcal{Z}\ -\ \mathcal{Z}\left(\mathcal{Z} - \frac{\la^2(D-1)}{u}\right)\ =\ \la^2\,\veps(\la)\ -\ u^2 - u^6 \quad , \quad \pa_u\equiv\frac{d}{du}\ ,
\label{Bloch-6}
\end{equation}
the expansion of $\mathcal{Z}(u)$ in generating functions
\begin{equation}
 \label{expansionZs}
 \mathcal{Z}(u)\ =\ \mathcal{Z}_0(u)\ +\ \mathcal{Z}_2(u)\,\la^2\ +\ \mathcal{Z}_4(u)\,\la^4\ +\ \ldots\ ,
\end{equation}
can be constructed, where the reduced energy expansion is given by (\ref{encorrection-6}),
\[
   \veps\ =\ \veps_0\ +\ \veps_4\,\la^4\ +\ \veps_8\,\la^8\ +\ \ldots\quad , \qquad \veps_0\ =\ D\ ,
\]
Interestingly (\ref{expansionZs}) has the same structure  as the expansion for the quartic anharmonic case: all generating functions $\mathcal{Z}_{2k+1}(u)$, $k=1,2,...$ of odd orders $\la^{2k+1}$ are absent in expansion, cf. (\ref{expansionZq}). It contrasts with the expansion of $\mathcal{Y}(v; \la)$ and $\veps(\la)$ in which the powers $\la^{4n}, n=0,1,2, \ldots$ are present only. In fact, for any even radial  anharmonic potential, $V(r)=V(-r)$,  $\mathcal{Z}(u)$  is written  in terms of generating functions as an expansion in powers of $\la^2$. As the result there are two different families of generating  functions,
\begin{equation}
\mathcal{Z}_{4k}(u)\ =\ u\,\sum_{n=k}^{\infty}c_{4k}^{(4n)}u^{4(n-k)}\ ,
\end{equation}
and
\begin{equation}
\mathcal{Z}_{4k+2}(u)\ =\ u\,\sum_{n=k+1}^{\infty}c_{4k+2}^{(4n)}u^{4(n-k)-2}\ ,
\end{equation}
those occur in correspondence to $\veps_{4k+2}=0$ and $\veps_{4k} \neq 0$.
Following (\ref{Yncorrection-sex}) and  (\ref{propertiessextic}) it is easy to see that for both families the generating function $\mathcal{Z}_{2p}(u)$ is a polynomial in $D$ of degree $p$,
\begin{equation}
 \mathcal{Z}_{2p}(u)\ =\ u\sum_{n=0}^{p}f^{(p)}_n(u^2)\,D^n\ ,
\end{equation}
where $f_n^{(p)}(u^2)$, $n=0,1,...,k$ are some real functions. The first two terms in expansion (\ref{expansionZs}) are
\begin{align}
 \label{1sextic}
 \mathcal{Z}_0(u)&\ =\   u\,\sqrt{1+u^4}\ ,\\
 \mathcal{Z}_2(u)&\ =\ \frac{2 u^4+D\left( 1+u^4- \sqrt{1+u^4}\right)}{2 u \left(1+u^4\right)}\ .
\label{2sextic}
\end{align}

Due to invariance $u \rar -u$ one can simplify the non-linear equation (\ref{Bloch-6}) by introducing
\[
    \mathcal{Z}\ =\ u \mathcal{\tilde Z}\quad \mbox{and}\quad {\rm u}\ =\ u^2\ .
\]
Finally, (\ref{Bloch-6}) is reduced to
\begin{equation}
 2 \la^2\,{\rm u}\,\pa_{\rm u} \mathcal{\tilde Z}\ -\ \mathcal{\tilde Z}\left({\rm u}
 \mathcal{\tilde Z} - {\la^2\,D}\right)\ =\ \la^2\,\veps(\la)\ -\ {\rm u} \ -\ {\rm u}^3 \quad ,
 \quad \pa_u\equiv\frac{d}{du}\ .
\label{Bloch-6-tilde}
\end{equation}
In this case the first two terms in expansion (\ref{expansionZs}) are simplified,
\begin{align}
 \label{1sextic-tilde}
 \mathcal{\tilde Z}_0({\rm u})&\ =\   \sqrt{1+{\rm u}^2}\ ,\\
 \mathcal{\tilde Z}_2({\rm u})&\ =\ \frac{2 {\rm u}^2+D\left( 1+{\rm u}^2- \sqrt{1+{\rm u}^2}\right)}{2 {\rm u} \left(1+{\rm u}^2\right)}\ ,
\label{2sextic-tilde}
\end{align}
cf. (\ref{1sextic}), (\ref{2sextic}).

The asymptotic behavior at large $u$ of $\mathcal{Z}_{2p}(u)$ is related with the expansion of $y$ at large $r$.  For the sextic anharmonic oscillator, the expansion of $y$ at large $r$ and fixed (effective) coupling constant $g(\lambda)$ is rewritten conveniently in variable $v$, see (\ref{change-v}),
\begin{equation}
\label{asymptoticsextic}
   y\ =\ (2M\hbar^2)^{\frac{1}{4}}\left(\la^2 v^3\ +\ \frac{(D+2)\la^2+1}{2\la^2}v^{-1} -\ \frac{\veps}{2\la^2}v^{-3}\ +\  \ldots\right)\ ,\quad\quad v\rar\infty\ .
\end{equation}
The first two terms of this expansion are $\veps$-independent, while the first term is also $D$-independent.
Following an analogous procedure to that  used for the quartic oscillator, we can transform the expansion of $\mathcal{Z}_{2p}(u)$ at large $u$ into an expansion  at large $v$ via the connection between the classical and quantum coordinate (\ref{u vs v}). The first two terms in  (\ref{expansionZs}) expanded at  large $v$ are
 \begin{equation}
\left(\frac{2M}{g^2}\right)^{1/2} \mathcal{Z}_0(\la v)\ =\ (2M\hbar^2)^{\frac{1}{4}}\left(\la^2 v^3\ +\ \frac{1}{2\la^2}v^{-1}\ -\ \frac{1}{8\la^6}v^{-5}\ +\ \ldots \right)\ , \quad\quad v\rar\infty\ ,
 \end{equation}
 and
  \begin{equation}
\left(\frac{2M}{g^2}\right)^{1/2} \la^2 \mathcal{Z}_2(\la v)\ =\ (2M\hbar^2)^{\frac{1}{4}}\left(\frac{D+2}{2}v^{-1}\ -\ \frac{D}{2\la^2}v^{-3}\  -\ \frac{1}{\la^{4}}v^{-5}\ +\  \ldots \right)\ ,\quad\quad v\rar\infty\ ,
 \end{equation}
 see (\ref{change-to-Z}).
We  explicitly observe that the sum   $(\frac{2M}{g^2})^{1/2}(\mathcal{Z}_0(\la v)+\la^2 \mathcal{Z}_2(\la v))$ at large $v$ reproduces exactly the first two terms in the expansion (\ref{asymptoticsextic}). However, all higher generating functions, $\mathcal{Z}_4(\la v)$, $\mathcal{Z}_6(\la v), ...$  contribute to the same order $O(v^{-3})$ for large $v$,
\begin{equation}
  \left(\frac{2M}{g^2}\right)^{1/2} \la^{2p} \mathcal{Z}_{2p}(\la v)\ =\ (2M\hbar^2)^{\frac{1}{4}}\left(-\frac{\veps_{4p-4}\la^{4p-6}}{2}v^{-3}\ +\ \ldots\right)\ ,
  \quad v \rar \infty \ ,
\end{equation}
here $\veps_{4p-4}$ is the energy correction of order $\la^{4p-4}$.  Therefore, it does not matter how many generating functions are considered in the expansion $(\frac{2M}{g^2})^{1/2}(\mathcal{Z}_0(\la v)+\la^2 \mathcal{Z}_2(\la v)+...)$ for large $v$, the $\veps$-dependent coefficient in front of term of order $O(v^{-3})$ is never reproduced exactly. Similar situation occurred in the quartic case.

\subsection{The Approximant and Variational Calculations}

From now on we set again $\hbar=1$ and $M=1/2$, consequently, $v=r$ and $\veps=E$.
Following formulas (\ref{1sextic}) and (\ref{2sextic}), the first two terms of expansion of the phase (\ref{phase}) can be calculated,
\begin{align}
   G_0(r;g) & \ =\ \frac{r^2}{4g^2}\, \sqrt{1+g^4r^4}\ +\ \frac{1}{4g^2}\log\left[g^2r^2+\sqrt{1+g^4r^4}\right]\ ,
\label{firstr6}
\end{align}
and
\begin{align}
 g^2\,G_2(r;g)&\ =\ \frac{1}{4}\log\left[1+g^4r^4\right]\ +\ \frac{D}{4}\log\left[1+\sqrt{1+g^4r^4}\right]\ .
\label{secondr6}
\end{align}
cf. (\ref{firstr4}).
The next two generating functions $G_4(r;g)$ and $G_6(r;g)$ can be also calculated, see Appendix \ref{appendix:C}. Keeping the expressions for $G_0(r;g)$ and $G_2(r;g)$ in mind
we can proceed to construct the Approximant $\Psi_{0,0}^{(t)}$.

\begin{figure}[]
	\centering
	\begin{subfigure}[t]{0.47\textwidth}
		\centering
		\includegraphics[width=\linewidth]{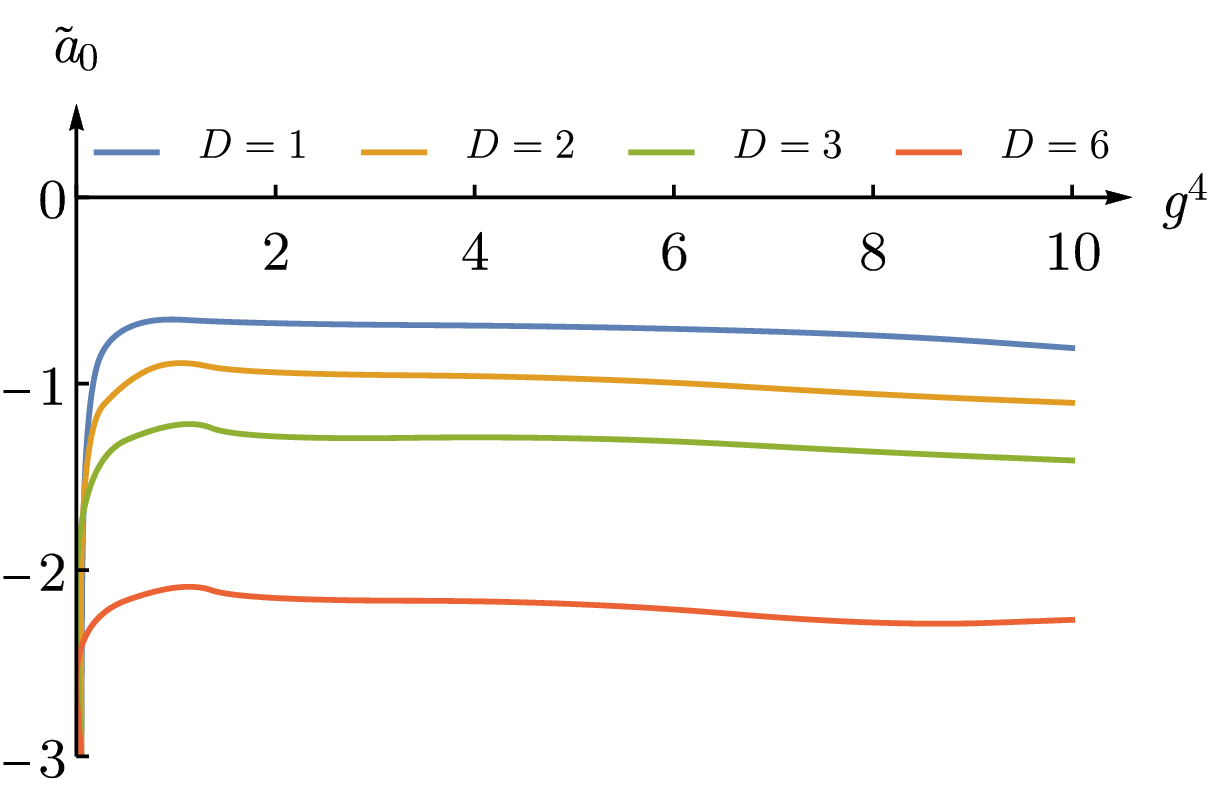}
		\caption{} \label{fig:a0s}
	\end{subfigure}
	\hfill
	\begin{subfigure}[t]{0.47\textwidth}
		\centering
		\includegraphics[width=\linewidth]{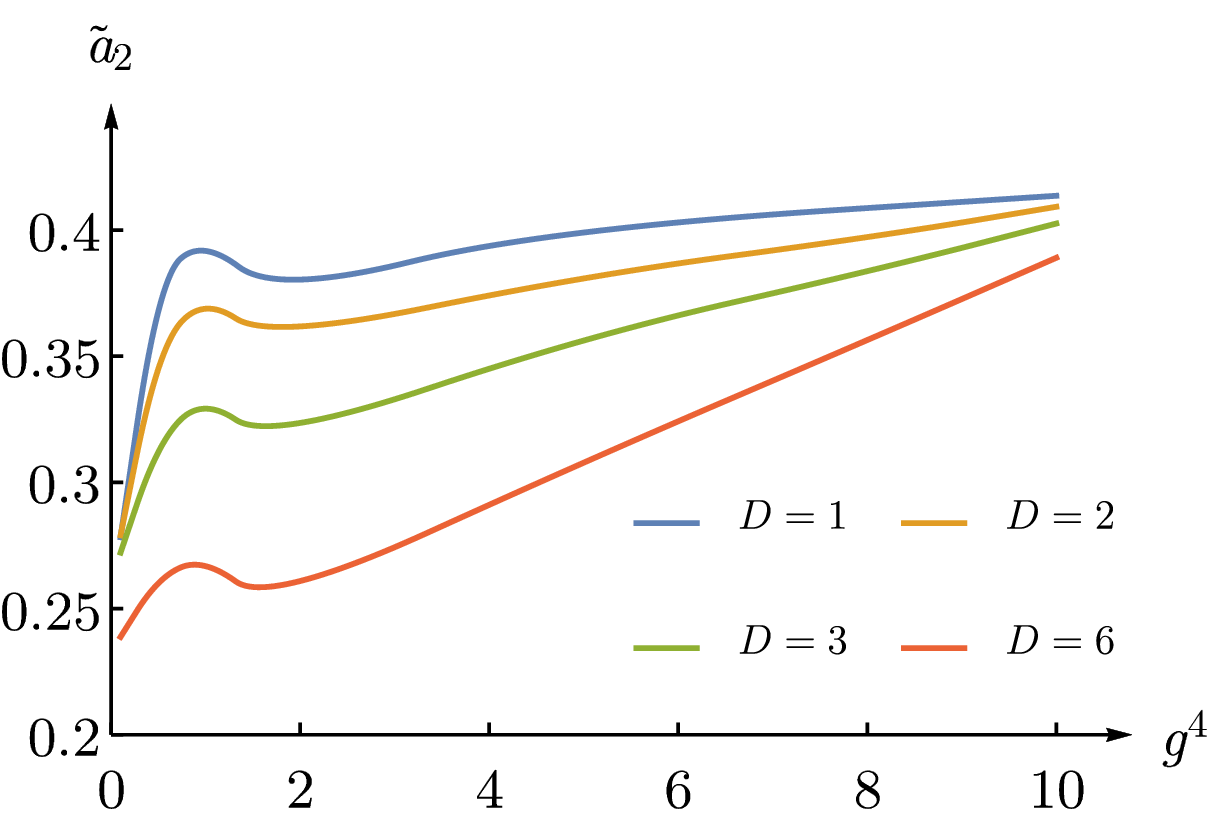}
		\caption{} \label{fig:a2s}
	\end{subfigure}
	\vspace{1cm}
	\begin{subfigure}[t]{0.47\textwidth}
		\centering
		\includegraphics[width=\linewidth]{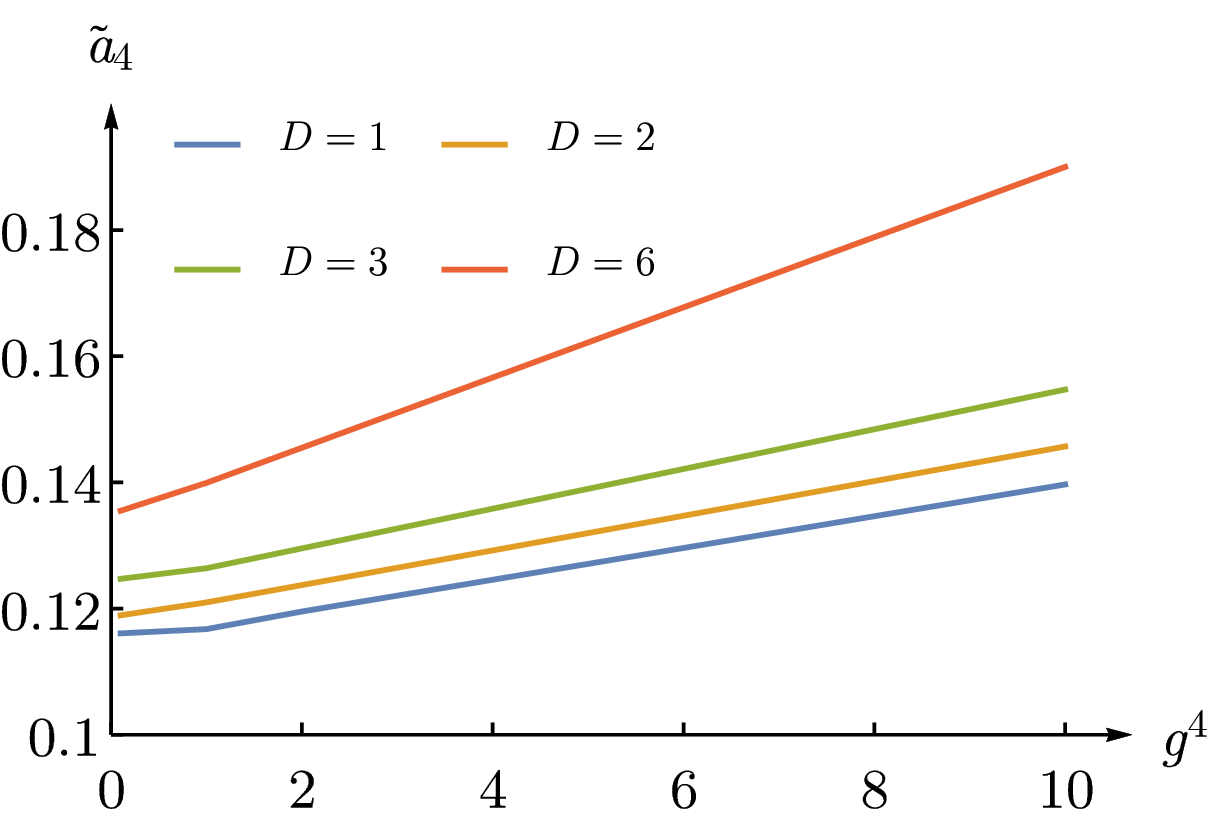}
		\caption{} \label{fig:a4s}
	\end{subfigure}
	\hfill
	\begin{subfigure}[t]{0.47\textwidth}
		\centering
		\includegraphics[width=\linewidth]{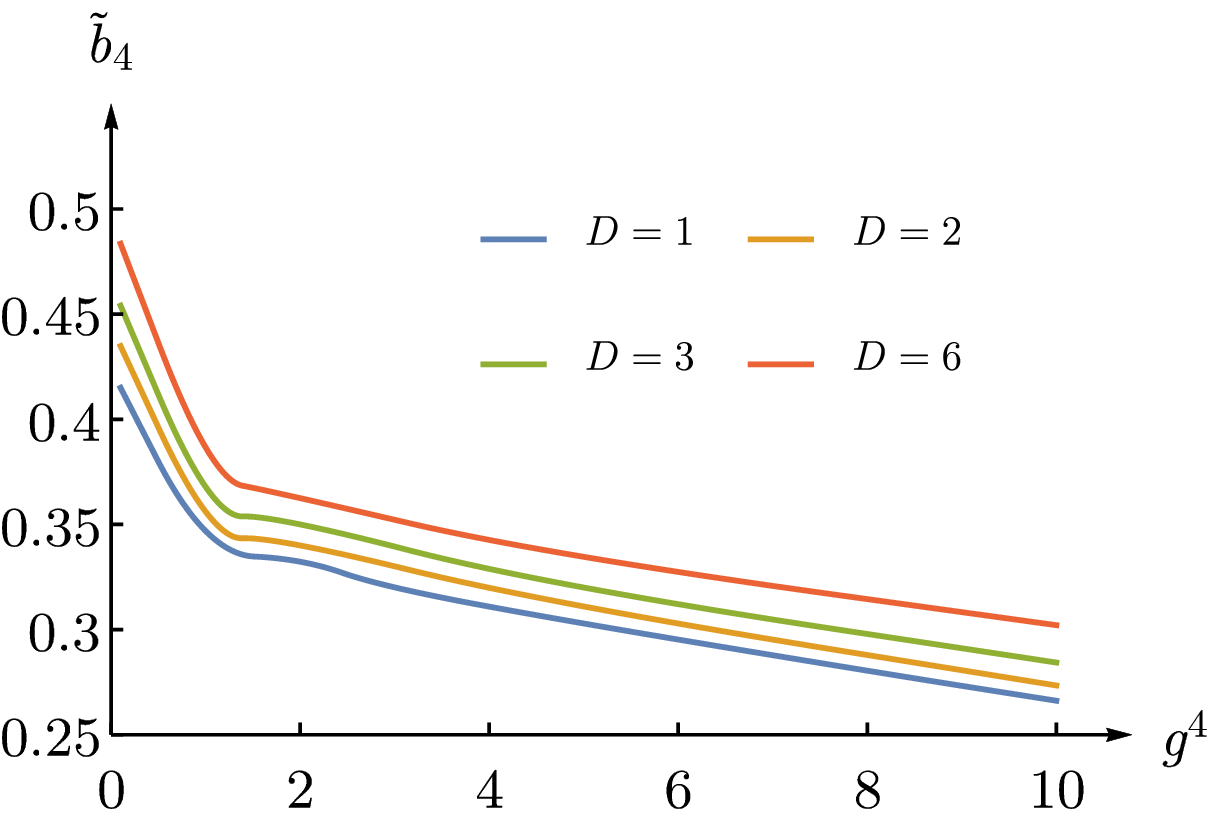}
		\caption{} \label{fig:b4s}	
	\end{subfigure}
	\vspace{1cm}
	\begin{subfigure}[t]{0.47\textwidth}
		\centering
		\includegraphics[width=\linewidth]{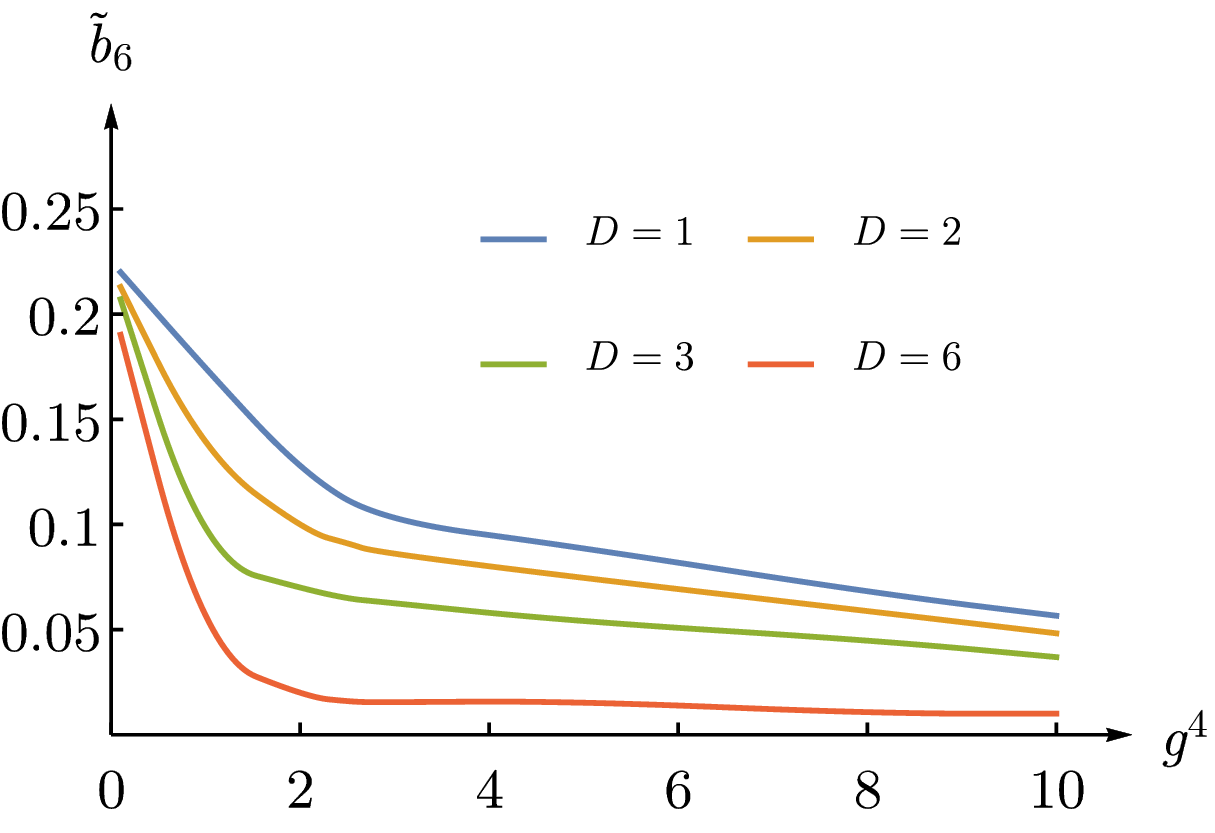}
		\caption{} \label{fig:b6s}
	\end{subfigure}
	\hfill
	\begin{subfigure}[t]{0.47\textwidth}
		\centering
		\includegraphics[width=\linewidth]{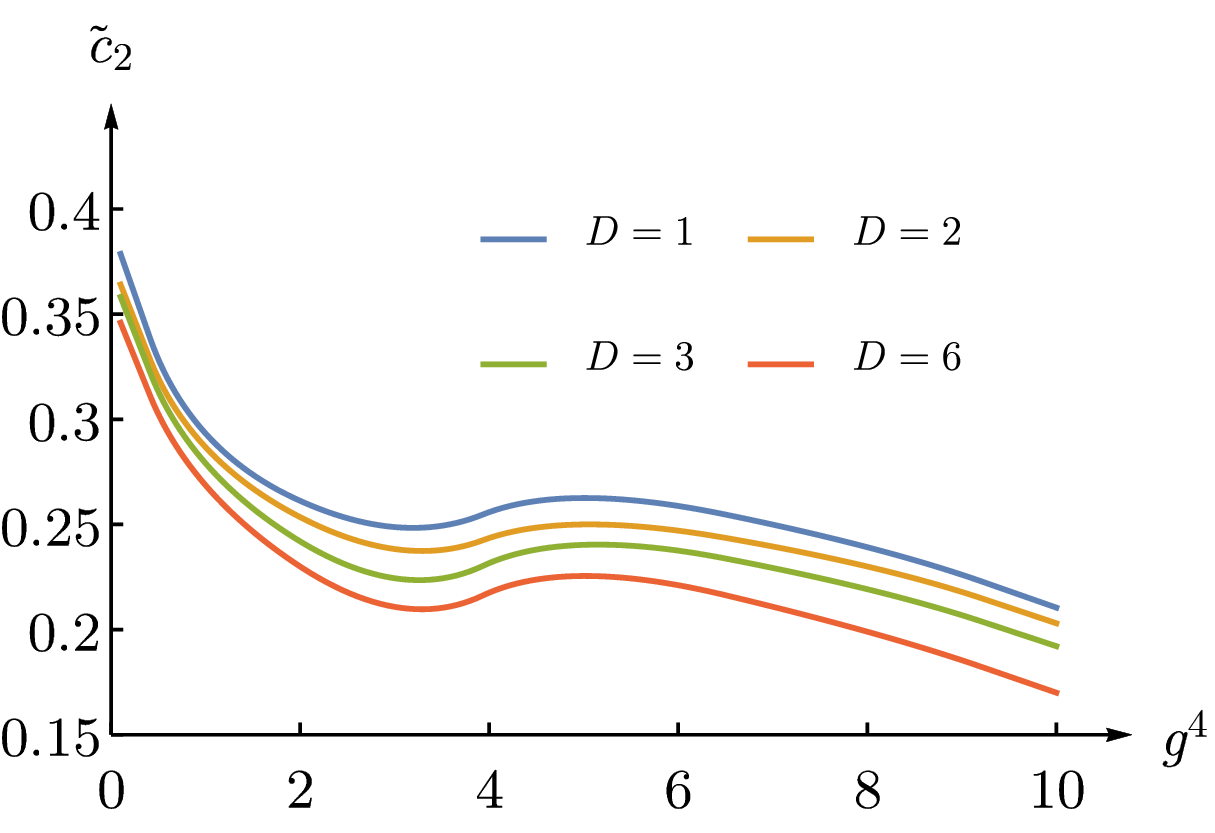}
		\caption{} \label{fig:c2s}	
	\end{subfigure}
	\caption{Ground state $(0,0)$: Variational parameters
     ${\tilde a}_0\ (a)$, ${\tilde a}_2\ (b)$, ${\tilde a}_4\ (c)$,
     ${\tilde b}_4\ (d)$, ${\tilde b}_6\ (e)$, ${\tilde c}_2\ (f)$ {\it vs} the coupling constant
     $g^4$ in domain $g^4 \in [0, 10]$ for $D=1, 2, 3,6$ }
	\label{fig:varparfixeds}
\end{figure}

\begin{figure}[]
	\centering
	\begin{subfigure}[t]{0.47\textwidth}
		\centering
		\includegraphics[width=\linewidth]{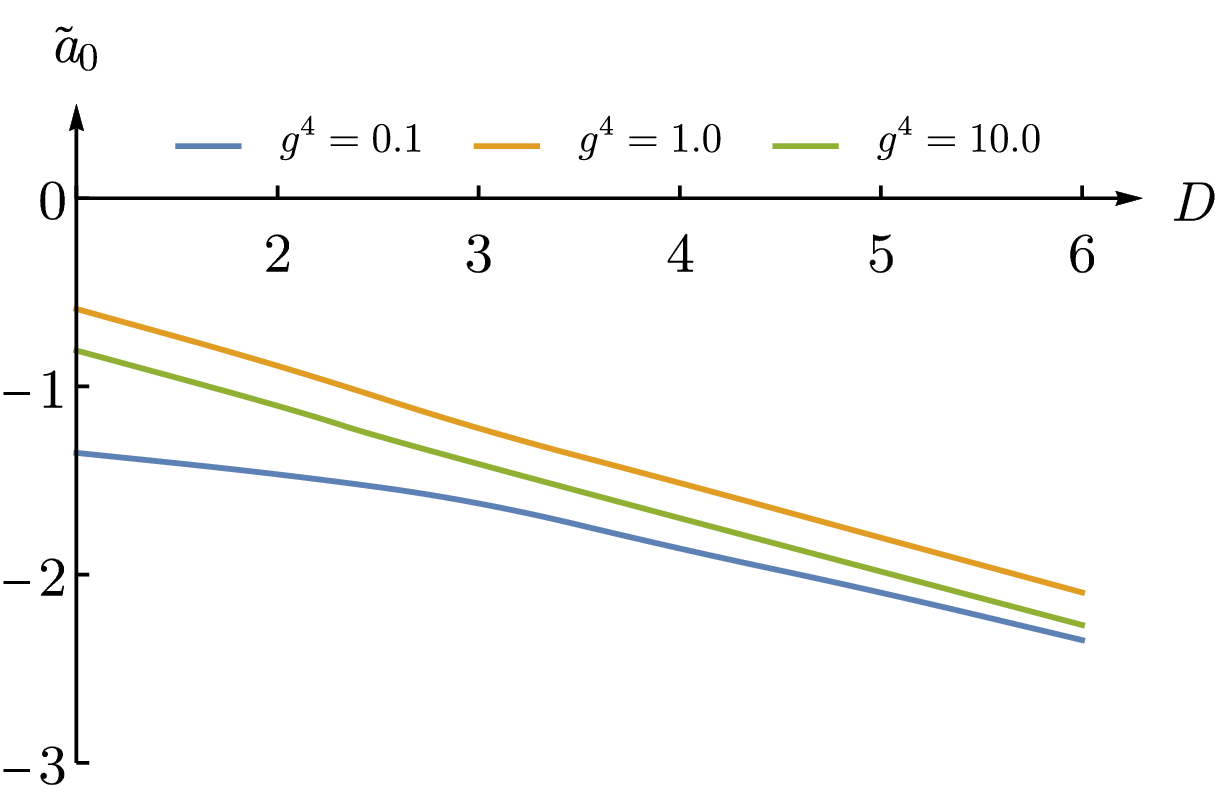}
		\caption{}
		\label{fig:a0fixeds}
	\end{subfigure}
	\hfill
	\begin{subfigure}[t]{0.47\textwidth}
		\centering
		\includegraphics[width=\linewidth]{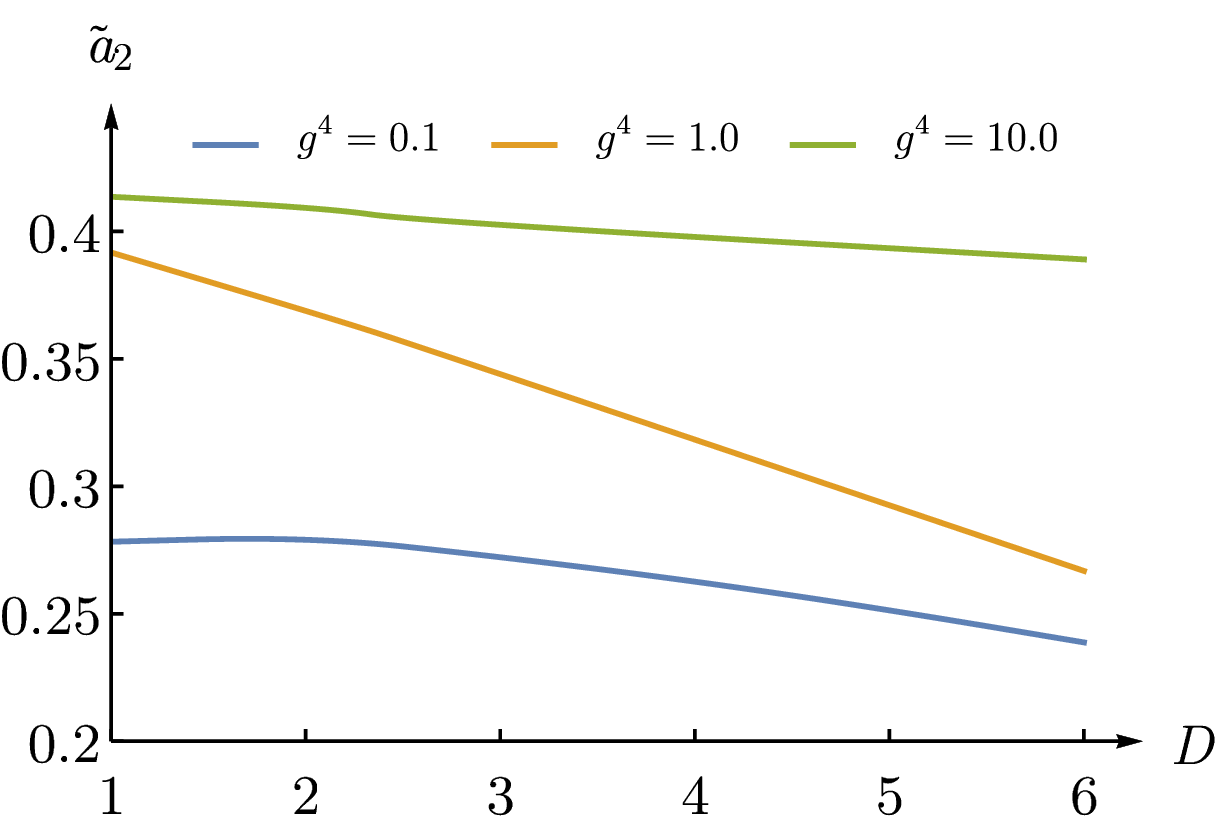}
		\caption{}
		\label{fig:a2fixeds}
	\end{subfigure}
	\vspace{1cm}
	\begin{subfigure}[t]{0.47\textwidth}
		\centering
		\includegraphics[width=\linewidth]{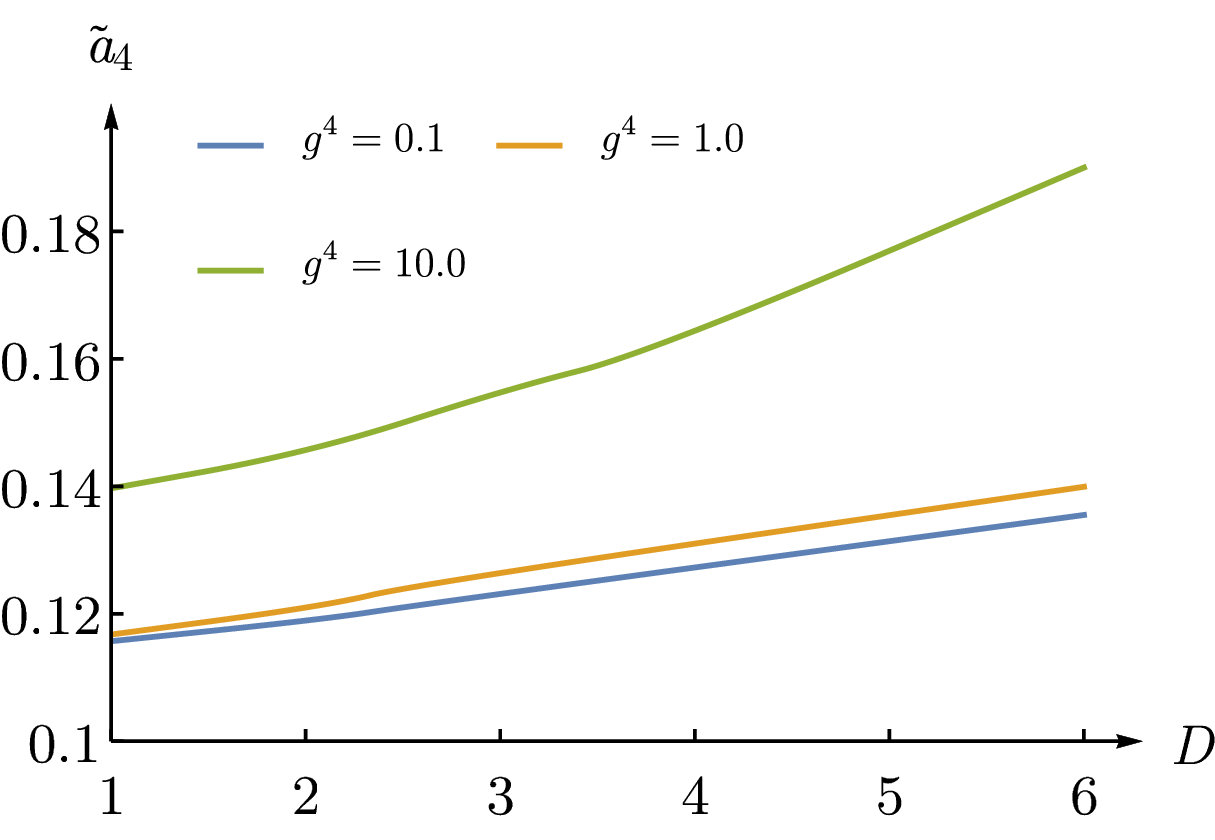}
		\caption{}
		\label{fig:a4fixeds}
	\end{subfigure}
	\hfill
	\begin{subfigure}[t]{0.47\textwidth}
		\centering
		\includegraphics[width=\linewidth]{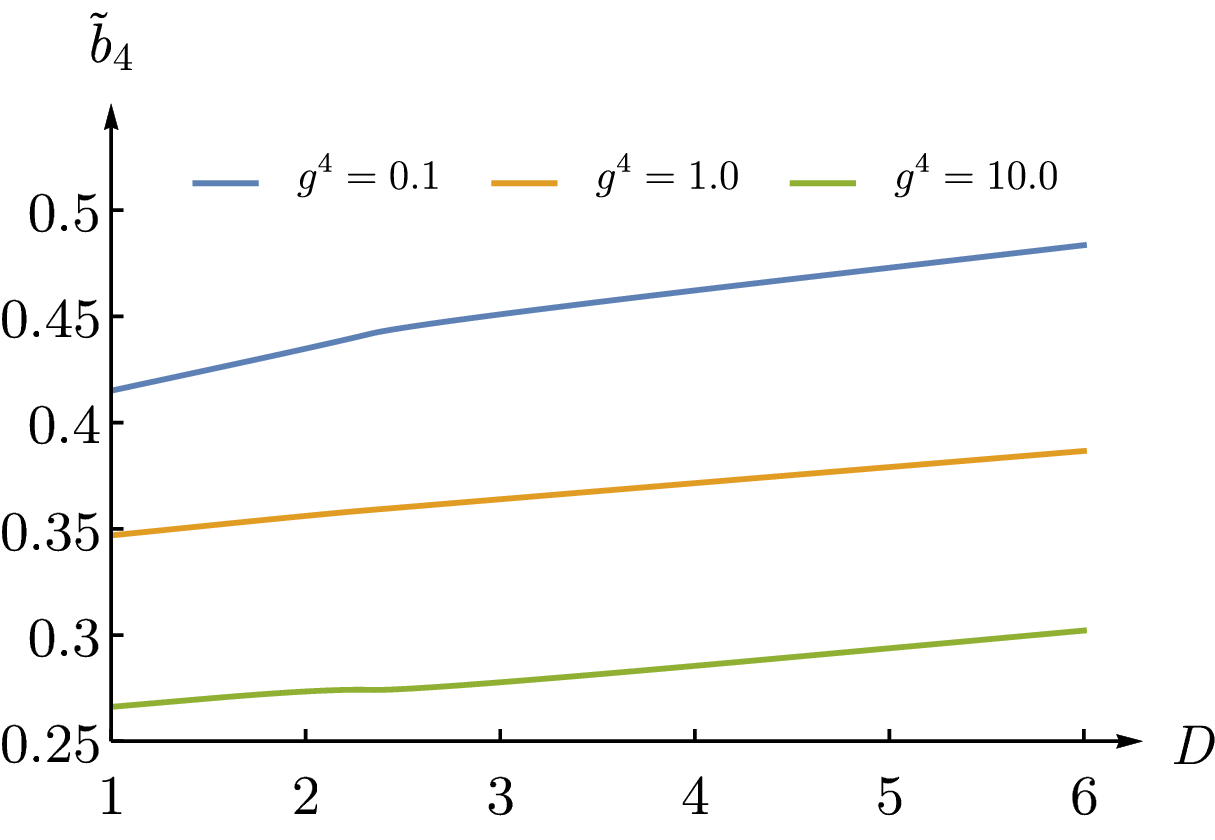}
		\caption{}
		\label{fig:b4fixeds}
	\end{subfigure}
	\vspace{1cm}
	\begin{subfigure}[t]{0.47\textwidth}
		\centering
		\includegraphics[width=\linewidth]{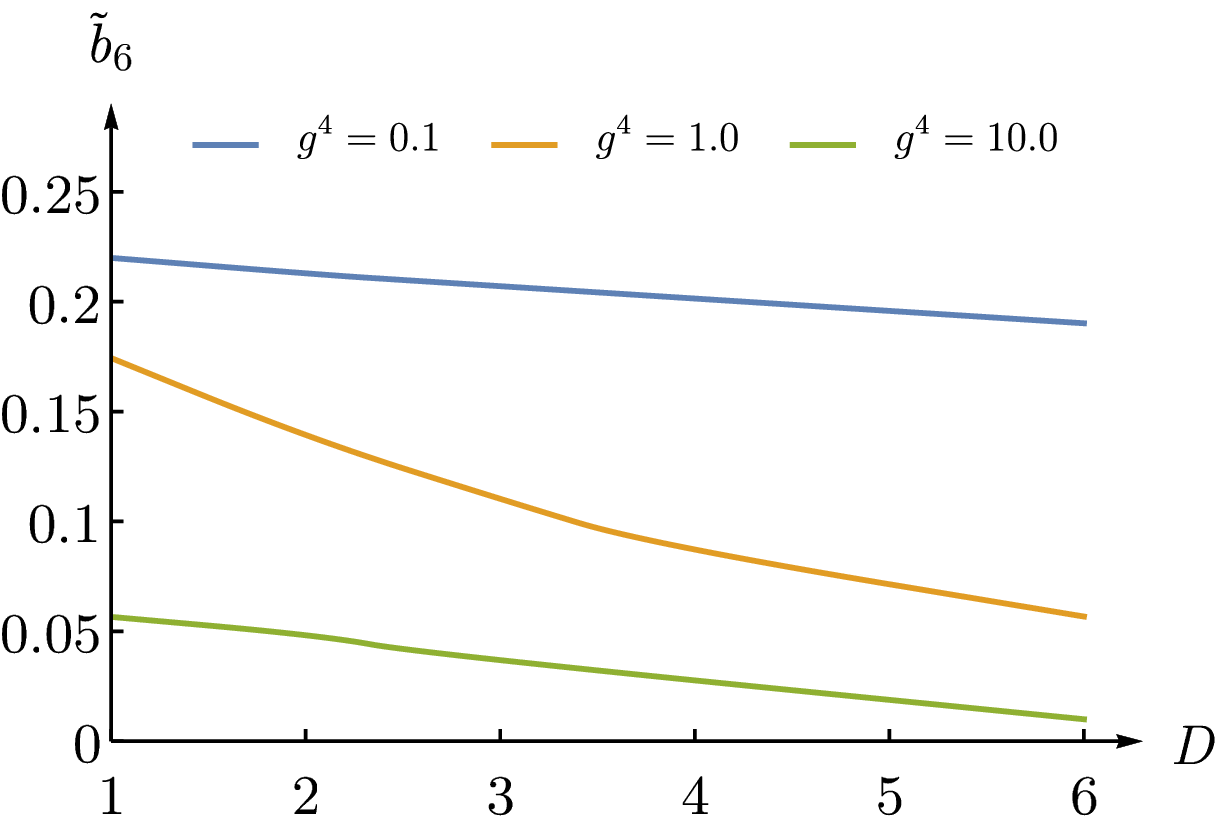}
		\caption{}
		\label{fig:b6fixeds}
	\end{subfigure}
	\hfill
	\begin{subfigure}[t]{0.47\textwidth}
		\centering
		\includegraphics[width=\linewidth]{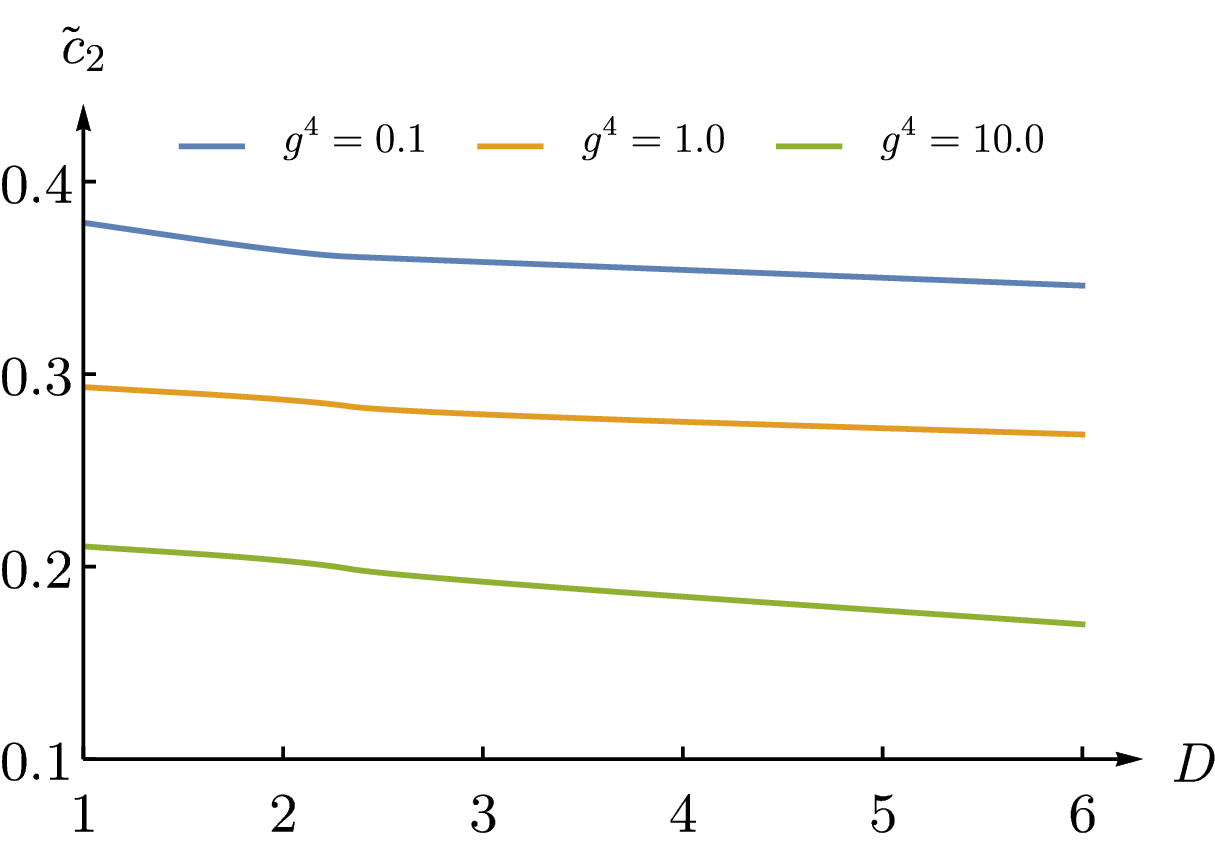}
		\caption{}
		\label{fig:c2fixeds}
	\end{subfigure}
	\caption{Ground state $(0,0)$: Variational parameters ${\tilde a}_0\ (a)$, ${\tilde a}_2\ (b)$, ${\tilde a}_4\ (c)$, ${\tilde b}_4\ (d)$, ${\tilde b}_6\ (e)$, ${\tilde c}_2\ (f)$ {\it vs} dimension $D$ for $g^4=0.1, 1.0, 10.0$ }
	\label{fig:varparfixedsD}
\end{figure}

Following the prescription (\ref{generalrecipe}), the exponential phase of the Approximant (the Phase Approximant) should have the form
\[
   \Phi_t\ =\ \dfrac{\tilde{a}_0\ +\ \tilde{a}_2\,r^2\ +\  \tilde{a}_4\,g^2\,r^4\  +\
         \tilde{a}_6\,g^4\,r^6}
   {\sqrt{1\ +\  \tilde{b}_4\,g^2\,r^2\ +\   \tilde{b}_6\,g^4\,r^4}}\ +
\]
\[
 \frac{1}{4g^2}\,\log\left[\tilde{c}_2\, g^2\,r^2\ +\ \sqrt{1\ +\   \tilde{b}_4\,g^2\,r^2\ +\  \tilde{b}_6\,g^4\,r^4}\right]\ +
\]
\begin{equation}
   \dfrac{1}{4}\log\left[1\ +\   \tilde{b}_4\,g^2\,r^2\ +\   \tilde{b}_6\,g^4 r^4\right]\ +\ \dfrac{D}{4}\log\left[1+\sqrt{1\ +\   \tilde{b}_4\,g^2\,r^2\ +\  \tilde{b}_6g^4r^4}\right]\ ,
\label{sextictrial}
\end{equation}
where $\tilde{a}_{0,2,4,6}$, $\tilde{b}_{4,6},\ {\tilde c}_2$ are seven free parameters. All three logarithmic terms inserted in (\ref{sextictrial}) appear as a \textit{minimal} modification of those that come from generating functions $G_{0,2}(r;g)$. For arbitrary $D>1$ the Phase Approximant (\ref{sextictrial}) can be transformed to the Approximant of the ground state function,
\[
   \Psi_{(0,0)}^{(t)}\ =\
   \frac{1}{\left(1\ +\  \tilde{b}_4\,g^2\,r^2\ +\  \tilde{b}_6\,g^4r^4\right)^{1/4}
   \left(1\ +\ \sqrt{1\ +\   \tilde{b}_4\,g^2\,r^2\ +\  \tilde{b}_6 g^4\,r^4}\right)^{D/4}}\ \times
\]
\[
\frac{1}{
	\left(\tilde{c}_2 g^2r^2\ +\ \sqrt{1\ +\   \tilde{b}_4\,g^2\,r^2\ +\
          \tilde{b}_6\,g^4\,r^4}\right)^{1/{4g^2}}}\ \times
\]
\begin{equation}
\label{approximantsextic}
\exp \left(-\dfrac{\tilde{a}_0\ +\ \tilde{a}_2\,r^2\ +\  \tilde{a}_4\,g^2\,r^4\ +\ \tilde{a}_6\,g^4\,r^6}
{\sqrt{1\ +\   \tilde{b}_4\,g^2\,r^2\ +\  \tilde{b}_6\,g^4\,r^4}}\right)\ .
\end{equation}
This is the central formula for this Section. Later it will be shown that it leads to a highly accurate uniform approximation of the exact ground state eigenfunction and also to a highly accurate variational energy for the ground state. Following the general prescription, the constraint
\begin{equation}
\label{a6}
  \tilde{b}_6\ =\  16\,\tilde{a}_6^{2}\ ,
\end{equation}
cf. (\ref{a4}), guarantees that the approximate phase $\Phi_{t}$ (\ref{sextictrial}) reproduces exactly the dominant term $\sim r^4$ in expansion (\ref{asymptoticsextic}). It is worth mentioning that the relaxing this constraint by keeping $\tilde{a}_6$ and $\tilde{b}_6$ free demonstrates that as the result of minimization this constraint is restored with very high accuracy. Furthermore, there exists another constraint,
\begin{equation}
\label{a4-6}
  \tilde{b}_4\ =\  32\,\tilde{a}_6\,\tilde{a}_4\ ,
\end{equation}
cf. (\ref{a2}), which corresponds to absence of subdominant term $\sim r^2$ in expansion of the trial phase $\Phi_{t}$ (\ref{sextictrial}) at $r \rar \infty$, see (\ref{asymptoticsextic}). Again relaxing the condition (\ref{a4-6}), keeping the parameters $\{\tilde{a}_6,\tilde{a}_4, \tilde{b}_4\}$ free and making minimization of the energy functional one can see that resulting parameters obey the condition (\ref{a4-6}) with very high accuracy.
Thus, the Approximant (\ref{approximantsextic})  contains, finally, 5 free parameters,   $\{\tilde{a}_0,\tilde{a}_2,\tilde{a}_4,\tilde{a}_6,\tilde{c}_2\}$. We must emphasize that the choice of parameters
\begin{equation}
\label{reproduction2}
     \tilde{a}_0\ =\ 0\ ,\qquad\tilde{a}_2\ =\ \frac{1}{4}\ ,
     \qquad\tilde{a}_4\ =\ 0\ ,\qquad\tilde{b}_4\ =0\ ,\qquad \tilde{a}_6\ =\ \frac{1}{4}\ ,\qquad \tilde{c}_2\ =\ 1\ ,
\end{equation}
in the (phase) Approximant $\Phi_t$ (\ref{sextictrial}) allows us to reproduce exactly the first two terms in expansion (\ref{expansionZs}). Note that contrary to the quartic radial anharmonic oscillator case (\ref{potquartic}), no single parameter in (\ref{reproduction2}) has explicit dependence on the coupling constant $g$, see (\ref{reproduction1}). However,
this choice of parameters  (\ref{reproduction2}) is not optimal from the point of the variational calculations of the energy. Performing minimization of the expectation value of 
the radial Scr\"odinger operator (\ref{radialop}), saying differently, of the variational energy, for different values of $g^4$ and $D$ one can see that all five parameters $\{\tilde{a}_0,\tilde{a}_2,\tilde{a}_4,\tilde{a}_6,\tilde{c}_2\}$ are smooth, slow-changing functions {\it vs} $g^4$ and $D$. Plots of variational parameters for the ground state,  as functions of $g^4$ for fixed $D$, are shown in Fig. \ref{fig:varparfixeds}. In turn, Fig. \ref{fig:varparfixedsD} presents the plots of the parameters as functions of $D$ for fixed $g^4$.
Similar behavior of parameters occurs for the excited states $(0,1), (0,2), (1,0)$.

As was indicated in I, the Approximant of the ground state function $\Psi_{(0,0)}^{(t)}$ is a building block to construct the Approximants of excited states. In particular, for $D=1$ in the case of the $n_r$-th excited state (at $r \geq 0$, see below), the Approximant has the form
\[
\Psi_{(n_r, p)}^{(t)}\ =\
   \frac{r^p\,P_{n_r}(r^2)}{\left(1\ +\  \tilde{b}_4\,g^2\,r^2\ +\  \tilde{b}_6\,g^4r^4\right)^{1/4}
    \left(1\ +\ \sqrt{1\ +\   \tilde{b}_4\,g^2\,r^2\ +\  \tilde{b}_6 g^4\,r^4}\right)^{D/4}}\ \times
\]
\[
   \frac{1}{
   \left(\tilde{c}_2 g^2r^2\ +\ \sqrt{1\ +\   \tilde{b}_4\,g^2\,r^2\ +\  \tilde{b}_6\,g^4\,r^4}\right)^{1/{4g^2}}}\ \times
\]
\begin{equation}
\label{1dsextic}
  \exp \left(-\dfrac{\tilde{a}_0\ +\ \tilde{a}_2\,r^2\ +\  \tilde{a}_4\,g^2\,r^4\ +\ \tilde{a}_6\,g^4\,r^6}
  {\sqrt{1\ +\   \tilde{b}_4\,g^2\,r^2\ +\  \tilde{b}_6\,g^4\,r^4}}\right)\ ,
\end{equation}
where $P_{n_r}(r^2)$ is a polynomial of degree $n_r$ in $r^2$ with real coefficients and all real roots, and with $P_{n}(0)=1$ chosen for normalization, here $p=0,1$ and $(-)^p=\pm$ has the meaning of parity w.r.t. reflection $(r \rar -r)$. Similarly to the quartic case at $D=1$ there are two possible domains for the Schr\"odinger operator: $r \in [0, \infty)$ (i) and $r \in (-\infty, +\infty)$ (ii). For the first domain (i) there exist the states of positive parity $p=0$ only, we denote them $(n_r, 0)$, thus negative nodes in (\ref{1dsextic})
at $r < 0$ are ignored. Hence the state $(n_r, 0)$ is the $n_r$-th excited state. As for the second domain (ii) there exist the states of both positive and negative parity, we denote the state as $(n_r, p)$. The state $(n_r, p)$ corresponds to $(2n_r+p)$-th excited state. It is evident that the energy of the state $(n_r, 0)$ for the first domain (i) coincide with energy of the state $(n_r, 0)$ for the second domain (ii). Like in the quartic radial anharmonic case it is easily demonstrated that energies $E_{(n_r, p)}$ obey the following inequality
\[
   E_{(n_r, 0)} < E_{(n_r, 1)} < E_{(n_r+1, 0)} \ ,
\]
for any coupling constant.
For fixed $n_r$, the $(n_r-1)$ free parameters of $P_{n_r}(r^2)$ are found by imposing the orthogonality constraints
\begin{equation}
\label{constraint6}
(\Psi_{(n_r,p)}^{(t)},\Psi_{(k_r,p)}^{(t)})\ =\ 0\quad ,\qquad k_r=0,\ldots,(n_r-1)\ .
\end{equation}
see previous Section I for details, cf. (\ref{constraint1}).

As for $D>1$, the Approximant of the state $(n_r,\ell)$ reads
\[
\Psi_{(n_r,\ell)}^{(t)}\ =\
   \frac{r^{\ell}P_{n_r}(r^2)}{\left(1\ +\   \tilde{b}_4\,g^2\,r^2\ +\  \tilde{b}_6\,g^4r^4\right)^{1/4} \left(1\ +\ \sqrt{1\ +\   \tilde{b}_4\,g^2\,r^2\ +\  \tilde{b}_6 g^4\,r^4}\right)^{D/4}}\ \times
\]
\[
   \frac{1}{
   \left(\tilde{c}_2 g^2r^2\ +\ \sqrt{1\ +\   \tilde{b}_4\,g^2\,r^2\ +\  \tilde{b}_6\,g^4\,r^4}\right)^{1/{4g^2}}}\ \times
\]
\begin{equation}
\label{appsextic}
  \exp \left(-\dfrac{\tilde{a}_0\ +\ \tilde{a}_2\,r^2\ +\  \tilde{a}_4\,g^2\,r^4\ +\ \tilde{a}_6\,g^4\,r^6}{\sqrt{1\ +\   \tilde{b}_4\,g^2\,r^2\ +\   \tilde{b}_6\,g^4\,r^4}}\right)\ .
\end{equation}
Like for the quartic anharmonic radial oscillator case, $P_{n_r}(r^2)$ is a polynomial of degree $n_r$ with $n_r$ positive roots. Needless to say, the procedure to fix the value of its coefficients is identical to that followed for the  quartic case via orthogonality constraints.
Once they are imposed, any Approximant  depends only on five free non-linear parameters $\{\tilde{a}_0,\tilde{a}_2,\tilde{a}_4,\tilde{a}_6,\tilde{c}_2\}$ regardless quantum numbers or dimension. Their values are fixed when we take either $\Psi_{(n_r,p)}^{(t)}$ or $\Psi_{(n_r,\ell)}^{(t)}$ as entry in variational calculations.

In Tables \ref{Sextic1} - \ref{Sextic4}  we present the calculations of  variational energy  for the first four low-lying states with quantum numbers $(0,0), (1,0), (0,1)$ and $(0,2)$ for different values of $D>1$ and $g^4$. As for $D=1$ the states $(0,0), (1,0), (0,1)$ are studied 
for the operator defined in the domain {ii}. The variational energy $E_0^{(1)}$, the correction $E_2$, and the corrected value $E_0^{(2)}$ are presented for some of these states, see (\ref{firsta}) and (\ref{seconda}). The variational energy $E_0^{(1)}$  provides  absolute accuracy of order $10^{-9} - 10^{-13}$, as it is confirmed by calculating the order of the correction $E_2$.  Using the LMM, we evaluated independently the energy for these states. The same number of mesh points that we used for the quartic oscillator case was used for the sextic case, i.e. 50, 100 and 200 for $g^4=0.1, 1.0, 10.0$\,, respectively. As for energy the LMM allows us to reach not less than 12 d.d. which coincide with $E_0^{(2)}$. In other words, all digits for $E_0^{(2)}$  printed in Tables \ref{Sextic1} - \ref{Sextic4} are exact. The next correction to variational energy $E_3$ confirmed it: systematically $E_3 \lesssim 10^{-12}$ for any $D$ and $g^4$ considered. It indicates an extremely fast rate of convergence of (\ref{nth}) as $n \rar \infty$ when trial function (\ref{1dsextic}) or (\ref{appsextic}) is taken as zero approximation.  The hierarchy of eigenstates for fixed integer $D > 1$ and positive $g^4$ is the same as the cubic and quartic potentials: $(0,0)$, $(0,1)$, $(0,2)$, $(1,0)$. We guess that the same hierarchy should hold for any two-term potential.

In contrast with the quartic oscillator case, following the available literature the estimates of the energy of the low-lying states for the sextic radial anharmonic oscillator are limited. Most of calculations are made for the one-dimensional case. In general, our results reproduce all known numerical ones that can be found for $D=1,2,3$, see e.g.  \cite{WENIGER}, \cite{Taseli2dr4} \cite{Meissner} and  \cite{WITWIT2dr6}. At $D=6$ the calculations are carried out for the first time.

The relative deviation of $\Psi_{(0,0)}^{(t)}$ from the exact (unknown) ground state eigenfunction $\Psi_{(0,0)}$ is estimated with the Non-Linearization procedure, it is bounded and very small,
\begin{equation}
\label{deviationsextic}
    \left|\frac{\Psi_{(0,0)}(r)-\Psi_{(0,0)}^{(t)}(r)}{\Psi_{(0,0)}^{(t)}(r)}\right|
    \lesssim 10^{-6}\ ,
\end{equation}
in the whole range of $r\in(0,\infty]$ at any integer $D$ and at any coupling constant $g^4$ which we considered.
Thus, our Approximant leads to a locally accurate approximation of the exact wave function once optimal parameters are chosen. A similar situation occurs for excited states for different $D$ and $g^4$.

For all $D$ and $g^4$, the correction $|y_1|$ to the logarithmic derivative of the ground state is very small function in comparison with $|y_0|$ in the domain
$0 \lesssim r \lesssim 1.7$ in which the dominant contribution of integrals - required by the variational method - occurs. In all cases the first order correction $(-y_1)$ to the logarithmic derivative of the ground state function $y$ is positive and bounded function in abome-mentioned domain in $r$ for all $D$ and $g^4$ we have studied. For example, for $g^4=1$ and $D=1,2,3,6$ the first correction $y_1$ has the upper bound
\begin{equation}
\label{cases-6}
|y_1|_{max} \sim
\begin{cases}
 0.0078\ ,\qquad D=1 \\
 0.0065\ ,\qquad D=2 \\
 0.0048\ ,\qquad D=3 \\
 0.0031\ ,\qquad D=6 \\
\end{cases}
\end{equation}
cf. (\ref{cases-4}). It is the consequence of the fact that by construction in the derivative $y_t=(\Phi_t)^{\prime}$ the first two terms at large $v$ in expansion (\ref{asymptoticsextic}) are reproduced exactly: $\sim~v^3$ and $\sim~1/v$~. Moreover, the minimization of the variational energy leads to the approximate fulfillment of the condition (\ref{a4-6}), hence, the coefficient in front of $\sim~v$ - this term is absent in the expansion (\ref{asymptoticsextic}) - is very small. Correspondingly, $|y_1|$ grows at large $v$ with a very small rate.

\begin{figure}[]
	\includegraphics[width=0.99\textwidth]{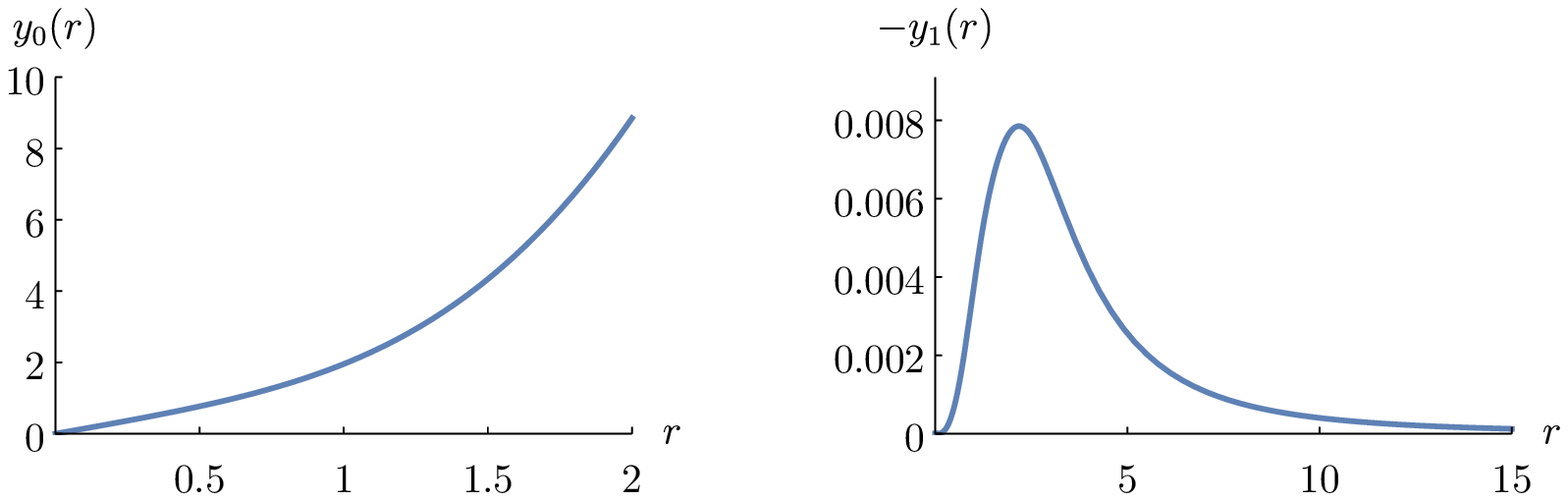}
	\caption{Sextic oscillator at $D=1$: function $y_0=(\Phi_t)'$ (on left) and its
     first correction $y_1$ (on right) {\it vs} $r$ for $g^2=1$.}
\label{fig:D=1s}
\end{figure}

\begin{figure}[]
	\includegraphics[width=0.99\textwidth]{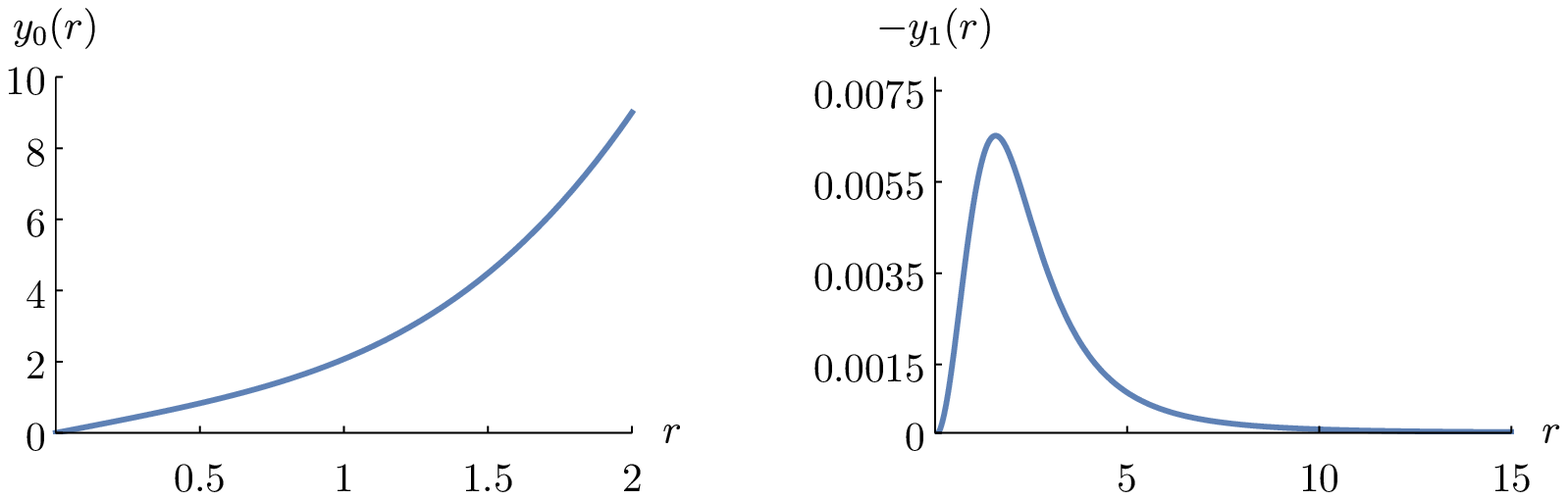}
	\caption{Sextic oscillator at $D=2$: function $y_0=(\Phi_t)'$ (on left) and its
     first correction $y_1$ (on right) {\it vs} $r$ for $g^2=1$.}
\label{fig:D=2s}
\end{figure}

\begin{figure}[]
	\includegraphics[width=0.99\textwidth]{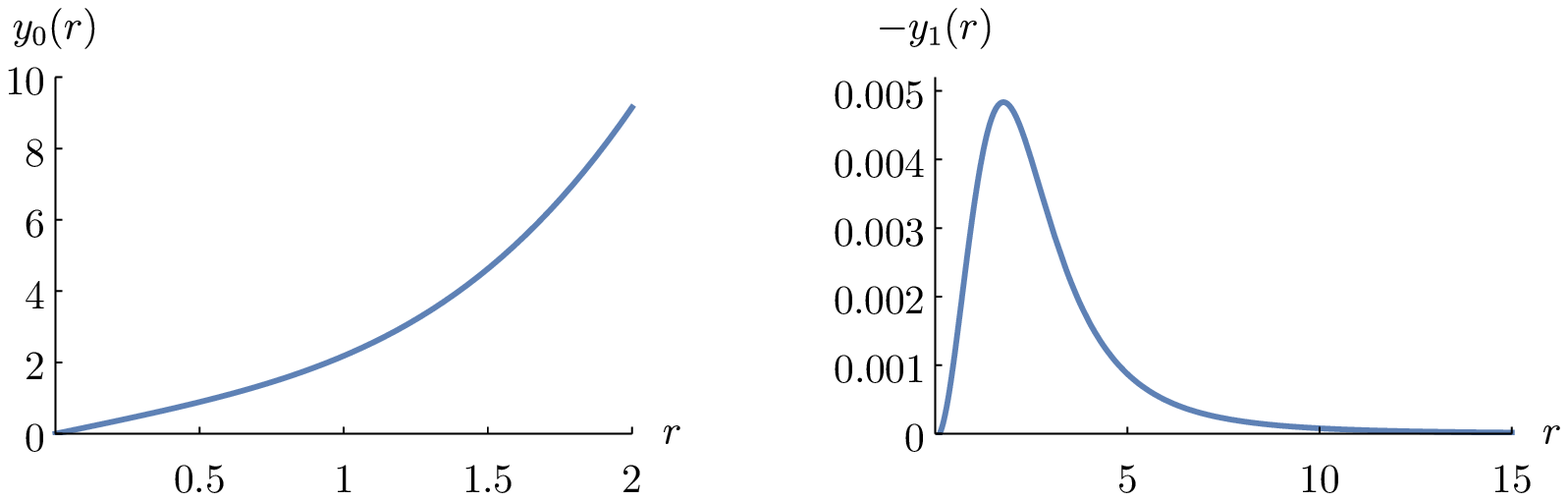}
	\caption{Sextic oscillator at $D=3$: function $y_0=(\Phi_t)'$ (on left) and its
     first correction $y_1$ (on right) {\it vs} $r$ for $g^2=1$.}
\label{fig:D=3s}
\end{figure}

``Boundness" of $y_1(r)$ at $r \geq 2$, which gives essential contribution to the variational integrals, and its small value of the maximum, and slow growth at large $r$ implies that we deal with a smartly designed zero order-approximation $\Psi_{(0,0)}^{(t)}$. It leads, in framework of the Non-Linearization Procedure, to a fast convergent iteration procedure for the energy and wave function. In Figs. \ref{fig:D=1s} - \ref{fig:D=3s}, $y_0$ and $y_1$ {\it vs} $r$ are presented for $g^4=1$ in physics dimensions $D=1,2,3$. We emphasize that all curves in these figures are slow-changing {\it vs} $D$. Therefore, it is not a surprise that similar plots should appear for $D=6$ (not shown) as well as for other values of $g > 0$. An analysis of these plots indicates that $(-y_1)^2$ is extremely small function in comparison with $y_0$ in the domain $0 \leq r \lesssim 1.7$, thus, in domain which provides the dominant contribution in variational integrals. It is the real reason why the energy correction $E_2$ is extremely small being of order $\sim 10^{-8}$, or sometimes even smaller, $\sim 10^{-10}$. Similar situation occurs for the phase (and its derivative) of the Approximants for the excited states.
\begin{table}[]
	\centering
	\caption{Ground state $(0,0)$ energy for the  sextic potential $V=r^2+g^4r^6$  for
            $D=1,2,3,6$ and $g^4=0.1,1.0,10.0$. Variational energy $E_0^{(1)}$, the first correction $E_2$ (rounded to 3 s.d.) found with $\Psi_{(0,0)}^{(t)}$, the corrected energy $E_0^{(2)}=E_0^{(1)}+E_2$  shown. $E_0^{(2)}$ coincides with LMM results (see text) in 12 d.d., thus, in all printed digits}
	\label{Sextic1}
	\begin{adjustbox}{max width=\textwidth}
		{\setlength{\tabcolsep}{0.15cm}	
			\begin{tabular}{|c|ccc|ccc|}
\hline
				\multirow{2}{*}{$g^4$} & \multicolumn{3}{c|}{$D=1$} & \multicolumn{3}{c|}{$D=2$}
\\
\cline{2-7}
				& $E_0^{(1)}$ &        $-E_2$       & $E_0^{(2)}$  &  $E_0^{(1)}$  &       $-E_2$       &  $E_0^{(2)}$ \\
				\hline
				\rule{0pt}{4ex}
				0.1&1.109\,087\,078\,465&$1.20 \times 10^{-13}$&1.109\,087\,078\,465 &
                2.307\,218\,600\,932&$7.04 \times 10^{-13}$&2.307\,218\,600\,931
\\[4pt]
				1.0&1.435\,624\,619\,003&$3.22 \times 10^{-13}$&1.435\,624\,619\,003 &
                3.121\,935\,474\,246&$9.81 \times 10^{-13}$&3.121\,935\,474\,246
\\[4pt]
				10.0&2.205\,723\,269\,598&$3.22 \times 10^{-12}$&2.205\,723\,269\,595 &
                4.936\,774\,524\,584&$1.72 \times 10^{-12}$&4.936\,774\,524\,582
\\[4pt]
				\hline
				\hline
				\multirow{2}{*}{$g^4$}
				& \multicolumn{3}{c|}{$D=3$} & \multicolumn{3}{c|}{$D=6$}
\\
				\cline{2-7}
				& $E_0^{(1)}$ & $-E_2$ & $E_0^{(2)}$ & $E_0^{(1)}$ & $-E_2$ & $E_0^{(2)}$
\\
\hline
				\rule{0pt}{4ex}
				0.1 & 3.596\,036\,921\,222 & $1.76 \times 10^{-12}$ &
                            3.596\,036\,921\,220 &
                7.987\,905\,269\,800 & $7.11 \times 10^{-13}$ &
                            7.987\,905\,269\,799
\\[4pt]
                1.0 & 5.033\,395\,937\,721 & $5.21 \times 10^{-12}$ & 5.033\,395\,937\,720 & 11.937\,202\,695\,862 & $9.62 \times 10^{-13}$ & 11.937\,202\,695\,862
\\[4pt]
               10.0 & 8.114\,843\,118\,826 & $7.60 \times 10^{-12}$ &
                8.114\,843\,118\,819 &
                19.880\,256\,604\,739 & $3.12 \times 10^{-12}$ & 19.880\,256\,604\,736
\\[4pt]
\hline
		\end{tabular}}
	\end{adjustbox}
\end{table}

\vskip -1.2cm

In a similar way as it was discussed for the quartic radial anharmonic oscillator, the Approximant for the first radial excited state $\Psi_{(1,0)}^{(t)}$ provides an estimate of the position of the radial node by imposing the orthogonality constraint to the Approximant of the ground state. For sextic anharmonic potential the zero order approximation $r_0^{(0)}$  gives not less than 5 d.d. of accuracy.
The first order correction $r_0^{(1)}$ gives a contribution to the 6 d.d.
Variational results are presented in Table \ref{Sextic4}, they compared with ones obtained in LMM with 50 mesh points for $g^4=0.1$, with 100 mesh points for $g^4=1$\,, with 200 mesh points for $g^4=10$\,. It can be noted that the radial node $r_0$ grows with an increase of $D$ at fixed $g^4$, but decreases with the increase of $g^4$ at fixed $D$.

\begin{table}[]
	\caption{The 1st excited  state energy for the sextic potential $V=r^2+g^4r^6$ for
     different $D$ and $g^4$ labeled by quantum numbers (0,1). 
     For $D=1$ it corresponds to 1st negative parity state $n_r=0$ at $p=1$ in domain (ii) (the whole line, see text). Variational energy $E_0^{(1)}$, the first correction $E_2$  found with use of $\Psi_{0,1}^{(t)}$; the corrected energy $E_0 ^{(2)}=E_0^{(1)}+E_2$  shown. Displayed  correction $E_2$ rounded to 3 s.d. $E_0^{(2)}$ coincides with LMM results (see text) in all 12 displayed d.d.}
	\centering
	\label{Sextic2}
	\begin{adjustbox}{max width=\textwidth}
		{\setlength{\tabcolsep}{0.15cm}	
			\begin{tabular}{|c|ccc|ccc|}
				\hline
				\multirow{2}{*}{$g^4$}
				& \multicolumn{3}{c|}{$D=1$} & \multicolumn{3}{c|}{$D=2$}               \\
				\cline{2-7}
				& $E_0^{(1)}$ & $-E_2$ & $E_0^{(2)}$ & $E_0^{(1)}$ & $-E_2$ &
                  $E_0^{(2)}$
\\
\hline
				\rule{0pt}{4ex}
				0.1 & 3.596\,036\,921\,295 & $7.50 \times 10^{-11}$ &
                  3.596\,036\,921\,220 & 4.974\,197\,493\,807 &
                  $9.01 \times 10^{-11}$ & 4.974\,197\,493\,717
\\[4pt]
				1.0 & 5.033\,395\,937\,795 & $7.52 \times 10^{-11}$ &
                  5.033\,395\,937\,720 & 7.149\,928\,601\,496 &
                  $5.84 \times 10^{-11}$ & 7.149\,928\,601\,438
\\[4pt]
               10.0 & 8.114\,843\,118\,966 & $1.48 \times 10^{-10}$ &
                  8.114\,843\,118\,818 & 11.688\,236\,0345\,77 &
                  $1.81 \times 10^{-10}$ & 11.688\,236\,034\,396
\\[4pt]
\hline
\hline
				\multirow{2}{*}{$g^4$}
				& \multicolumn{3}{c|}{$D=3$} & \multicolumn{3}{c|}{$D=6$}                \\
				\cline{2-7}
				& $E_0^{(1)}$ & $-E_2$ & $E_0^{(2)}$ & $E_0^{(1)}$ & $-E_2$ &
                  $E_0^{(2)}$
\\
\hline
				\rule{0pt}{4ex}
				0.1 & 6.439\,143\,322\,388 & $6.64 \times 10^{-11}$ &
                6.439\,143\,322\,321 & 11.324\,899\,788\,818 &
                $8.15 \times 10^{-10}$ & 11.324\,899\,788\,004
\\[4pt]
				1.0 & 9.455\,535\,276\,950 & $1.09 \times 10^{-10}$ &
                9.455\,535\,276\,841 & 17.387\,207\,808\,723 &
                $1.26 \times 10^{-9}$ & 17.387\,207\,807\,460
\\[4pt]
               10.0 & 15.619\,579\,279\,334 & $5.05 \times 10^{-10}$ &
                15.619\,579\,278\,830 & 29.302\,506\,554\,618 &
                $1.22 \times 10^{-9 }$ & 29.302\,506\,553\,402
\\[4pt]
\hline
		\end{tabular}}
	\end{adjustbox}
\end{table}
\begin{table}[]
	\centering
	\caption{The second excited state energy for the sextic potential $V=r^2+g^4r^6$ for
       different $D$ and $g^4$ labeled by quantum numbers $(0,2)$ for $D>1$, as for $D=1$
       it corresponds to 1st excitation $n_r=1$ of positive parity $p=0$: $(1,0)$. 
       Variational energy $E_0^{(1)}$ found with $\Psi_{(1,0)}^{(t)}$ for $D = 1$ and        $\Psi_{(0,2)}^{(t)}$ for $D > 1$, the first correction $E_2$ to it and the corrected energy $E_0^{(2)}= E_0^{(1)} + E_2$  shown. Displayed correction
       $E_2$ rounded to 3 s.d. $E_0^{2}$ coincides with LMM results (see text) 
       in all 12 displayed d.d.}
\label{Sextic3}
\begin{adjustbox}{max width=\textwidth}
		{\setlength{\tabcolsep}{0.15cm}
\begin{tabular}{|c|ccc|ccc|}
\hline
				\multirow{2}{*}{$ g^4$}
				& \multicolumn{3}{c|}{$D=1$} & \multicolumn{3}{c|}{$D=2$}
\\
				\cline{2-7}
				& \multicolumn{3}{c|}{$E_0^{(1)}$} & $E_0^{(1)}$ & $-E_2$ & $E_0^{(2)}$
\\
\hline
				\rule{0pt}{4ex}	
				0.1 & \multicolumn{3}{c|}{6.644\,391\,710\,782} &
                7.987\,905\,270\,111 & $3.12 \times 10^{-10}$ & 7.987\,905\,269\,799
\\[4pt]
				1.0 & \multicolumn{3}{c|}{9.966\,622\,004\,356} &
                11.937\,202\,696\,127 & $2.66 \times 10^{-10}$ & 11.937\,202\,695\,862
\\[4pt]				
               10.0 & \multicolumn{3}{c|}{16.641\,218\,168\,076} &
                19.880\,256\,605\,756 & $1.02 \times 10^{-9 }$ & 19.880\,256\,604\,742
\\[4pt]
\hline
\hline
				\multirow{2}{*}{$g^4$}
				& \multicolumn{3}{c|}{$D=3$} & \multicolumn{3}{c|}{$D=6$}
\\
				\cline{2-7}
				& $E_0^{(1)}$ & $-E_2$ & $E_0^{(2)}$ & $E_0^{(1)}$ & $-E_2$ &
                  $E_0^{(2)}$
\\
\hline
				\rule{0pt}{4ex}
				0.1 & 9.617\,462\,285\,440 &
                $1.50 \times 10^{-10}$ & 9.617\,462\,285\,290 & 14.962\,630\,328\,506 &
                $1.60 \times 10^{-10}$ & 14.962\,630\,328\,346
\\[4pt]
				1.0 & 14.584\,132\,948\,883 & $3.15 \times 10^{-9}$ &
                14.584\,132\,945\,729 & 23.431\,551\,835\,405 & $2.19
                \times 10^{-9}$ & 23.431\,551\,833\,215
\\[4pt]
			   10.0 & 24.447\,468\,037\,325 & $5.42 \times 10^{-9}$ &
                24.447\,468\,031\,906 & 39.815\,551\,142\,800 & $7.05
                \times 10^{-9}$ & 39.815\,551\,135\,750
\\[4pt]
\hline
\end{tabular}}
\end{adjustbox}
\end{table}
\begin{table}[]
 	\caption{Energy $E$ of the 3rd excited state $(1,0)$ - the first radial excitation -
        for the potential $V=r^2 + g^4\, r^6$ for $D=2,3,6$ at $g^4=0.1, 1.0, 10.0$ and its radial node $r_0$ calculated in LMM with 9 correct d.d.. Variational energy $E_0^{(1)}$ and its radial node $r_0^{(0)}$ given by underlined digits, both found with $\Psi_{(1,0)}^{(t)}$.}
\label{Sextic4}
\begin{adjustbox}{max width=\textwidth}
 		{\setlength{\tabcolsep}{0.15cm} 	
\begin{tabular}{|c|cc|cc|cc|}
 				\cline{1-7}
 				\multicolumn{1}{|c|}{\multirow{2}{*}{$g^4$}} &
                \multicolumn{2}{c|}{$D=2$} & \multicolumn{2}{c|}{$D=3$} & \multicolumn{2}{c|}{$D=6$}
\\
 				\cline{2-7}
 				\multicolumn{1}{|c|}{}
 				& $E$ & $r_0$ & $E$ & $ r_0$ & $E$ & $r_0$ \\
\hline
 				\rule{0pt}{5ex}
 				0.1 & \underline{8.402\,580\,46}2 & \underline{0.837\,31}0\,052 &
                     \underline{10.237\,873\,7}21 & \underline{0.995\,78}7\,872 & \underline{16.154\,260\,6}10 & \underline{1.308\,54}3\,484
\\
 				1.0 & \underline{12.914\,938\,79}3 & \underline{0.671\,82}1\,606 &
                      \underline{15.989\,440\,78}7 & \underline{0.790\,36}4\,964 & \underline{25.938\,441\,0}37 & \underline{1.019\,16}6\,200
\\
 			   10.0 & \underline{21.792\,578\,2}51 & \underline{0.515\,91}4\,526 &
                      \underline{27.155\,085}\,604 & \underline{0.604\,32}2\,682 & \underline{44.521\,781}\,513 & \underline{0.773\,860}\,964
\\[8pt] 				
\hline
\end{tabular}}
\end{adjustbox}
\end{table}

\subsection{The Strong Coupling Expansion}

In this section we calculate the first two terms in the strong coupling expansion (\ref{Scoupling}). For  the sextic anharmonic oscillator potential (\ref{potsextic}) this (convergent) expansion has the form
\begin{equation}
\label{sexticST}
 \veps\ =\ g (\tilde{\veps}_0\ +\ \tilde{\veps}_6\,g^{-3}\ +\ \tilde{\veps}_{12}\,g^{-6}\ +\
  \ldots)\ ,
\end{equation}
as long as $2M=\hbar=1$. Evidently, this expansion corresponds to PT in power of $\hat{\la}$ for the potential
\begin{equation}
\label{sexticSTW}
 V(w)\ =\ w^6\ +\ \hat{\la}\,w^2\ ,\quad \hat{\la}\ =\ g^{-3}
\end{equation}
in the Schr\"odinger equation defined in $w \in [0,\infty)$.

We use the Approximant (\ref{appsextic}) and develop the perturbation procedure already described in I and also in the previous Section to calculate $\tilde{\veps}_0$ and $\tilde{\veps}_6$. In Table \ref{sexticst1} we present for different $D$ the variational estimate for $\tilde{\veps}_0$ (denoted by $\tilde{\veps}_0^{(1)}$), the correction $\hat{\veps}_2$  and the corrected value $\tilde{\veps}_0^{(2)}$ calculated via the Non-Linearization Procedure. Using the LMM it was verified that $\tilde{\veps}_0^{(2)}$ provides not less than 12 exact d.d. at every $D$ considered. Hence, the first 10 d.d. in variational energy $\veps_0^{(1)}$ printed in Table \ref{sexticst1} are exact. In this case, the $\tilde{\veps}_3 \sim 10^{-2}\veps_2$, it indicates a high rate of convergence in PT. Finally, in Table \ref{sexticst2} we present the first two approximations of the coefficient in front of subdominant term $\tilde{\veps}_6$,
see (\ref{sexticST}). Comparing our results for $\tilde{\veps}_0$  and $\tilde{\veps}_6$ with those available in literature, see e.g. \cite{Taseli2dr4} and \cite{Meissner}, one can see that we usually reproduce, although often exceed them.
\begin{table}[htb]
\centering
	\caption{Ground state $(0,0)$ energy $\tilde{\veps}_0$ for the potential $W=r^6$
     (see (\ref{sexticSTW})) for $D=1,2,3,6$ found in PT, based on the Approximant $\Psi_{(0,0)}^{(t)}$ : $\tilde{\veps}_0^{(1)}$ corresponds to the variational
     energy, $\hat{\veps}_2$ is the second PT correction,  $\tilde{\veps}_0^{(2)}=\tilde{\veps}_0^{(1)}+\hat{\veps}_2$ is the corrected variational energy. 10 d.d. in $\tilde{\veps}_0^{(2)}$ confirmed independently
     in LMM, see text.}
\label{sexticst1}
	\begin{adjustbox}{max width=\textwidth}
		{\setlength{\tabcolsep}{0.15cm}
			\begin{tabular}{|ccc|ccc|}
\hline
				\multicolumn{3}{|c|}{$D=1$} & \multicolumn{3}{c|}{$D=2$}                    \\
\hline
				$\tilde{\veps}_0^{(1)}$ & $-\hat{\veps}_2$ &
                $\tilde{\veps}_0^{(2)}$&$\tilde{\veps}_0^{(1)}$ & $-\hat{\veps}_2$ & $\tilde{\veps}_0^{(2)}$
\\
				1.144\,802\,453\,800 & $3.21\times10^{-12}$ & 1.144\,802\,453\,797 &
                2.609\,388\,463\,259     &$5.72\times10^{-12}$&2.609\,388\,463\,253
\\[4pt]
\hline \hline
				\multicolumn{3}{|c|}{$D=3$} & \multicolumn{3}{c|}{$D=6$}                    \\
\hline
				$\tilde{\veps}_0^{(1)}$ & $-\hat{\veps}_2$  & $\tilde{\veps}_0^{(2)}$ &
                $\tilde{\veps}_0^{(1)}$ & $-\hat{\veps}_2$ & $\tilde{\veps}_0^{(2)}$
\\
				4.338\,598\,711\,518 & $4.73\times10^{-12}$ & 4.338\,598\,711\,513 &
                10.821\,985\,609\,895 & $7.21\times10^{-12}$ & 10.821\,985\,609\,888
\\[4pt]
\hline
		\end{tabular}}
	\end{adjustbox}
\end{table}
\begin{table}[htb]
	\caption{Subdominant coefficient $\tilde{\veps}_6$ in the strong coupling
      expansion (\ref{sexticST}) of the ground state $(0,0)$ energy for the sextic radial anharmonic potential (\ref{potsextic}) for $D=1,2,3,6$.
      First order correction $\tilde{\veps}_{6,1}$ in PT, see text, included.
      10 d.d. in $\tilde{\veps}_6$ confirmed independently in LMM.}
\label{sexticst2}
\center
\begin{adjustbox}{max width=\textwidth}
		{\setlength{\tabcolsep}{0.15cm}			
\begin{tabular}{|ccc|ccc|}
\hline
				\multicolumn{3}{|c|}{$D=1$} & \multicolumn{3}{c|}{$D=2$}
\\
\hline
		$\tilde{\veps}_6^{(1)}$ &
        $-\tilde{\veps}_{6,1}$ & $\tilde{\veps}_6^{(2)}$ & $\tilde{\veps}_6^{(1)}$ & $-\tilde{\veps}_{6,1}$ & $\tilde{\veps}_{6}^{(2)}$
\\
		0.307\,920\,304\,114 & $3.83 \times 10^{-10}$ & 0.307\,920\,303\,731 &
        0.534\,591\,069\,789 & $2.85 \times 10^{-10}$ & 0.534\,591\,069\,504
\\[4pt]
\hline \hline
				\multicolumn{3}{|c|}{$D=3$} & \multicolumn{3}{c|}{$D=6$}
\\
\hline			
        $\tilde{\veps}_6^{(1)}$ & $-\tilde{\veps}_{6,1}$ & $\tilde{\veps}_6^{(2)}$ & $\tilde{\veps}_6^{(1)}$ & -$\tilde{\veps}_{6,1}$ & $\tilde{\veps}_6^{(2)}$
\\
		0.718\,220\,134\,970 & $1.55 \times 10^{-9 }$ & 0.718\,220\,133\,425 &
        1.137\,762\,108\,070 & $2.68 \times 10^{-10}$ & 1.137\,762\,107\,802
\\[4pt]
\hline
\end{tabular}}
\end{adjustbox}
\end{table}


\subsection{Sextic Radial Anharmonic Oscillator: Conclusions}

It is shown that the 5-parametric Approximants (\ref{approximantsextic}), (\ref{1dsextic}), (\ref{appsextic}) taken as variational trial functions for the first four states $(0,0), (0,1), (0,2), (1,0)$ of the sextic radial $D$-dimensional anharmonic oscillator with the potential (\ref{potsextic}) provide extremely high relative accuracy in energy ranging from $\sim 10^{-14}$ to  $\sim 10^{-8}$ for different coupling constants $g$ and dimension $D$. Variational parameters depend on $g$ and $D$ in a smooth manner and can be easily interpolated. If variationally optimized Approximants are taken as zero approximation in Non-Linearization (iteration) procedure, they lead to fast convergent procedure with rate of convergence $\sim 10^{-4}$. For the ground state it was calculated the relative deviation of the logarithmic derivative of variationally optimized Approximant from the exact one {\it vs} radial coordinate $r$ for different $g$ and $D$. It was always smaller than $\sim 10^{-6}$ at any $r \in [0, \infty)$. It implies that the Approximants with interpolated parameters $\{\tilde{a}_0,\tilde{a}_2,\tilde{a}_4,\tilde{a}_6,\tilde{c}_2\}$ {\it vs}
$g$ and $D$ provide highly accurate approximation of the eigenfunctions of the Sextic Radial Anharmonic Oscillator while the respectful eigenvalues are given by ratio of two integrals.

\section{Conclusions}

Based on the analysis of asymptotic behavior of the phase $\Phi(r)$ of the wavefunction for the state $(n_r, \ell)$ defined in the form
\[
       \Psi(r)\ =\ r^{\ell} Q_{n_r}(r)\, e^{-\Phi(r)}\ ,
\]
at large and small $r$, its weak and strong coupling expansions in $r-$ and $(g\,r)-$ spaces we constructed the interpolating expression for the phase in extremely compact form (\ref{generalrecipe}), which should be valid for the general radial polynomial potential and its {\it any} eigenstate. This expression was called the (Phase) Approximant and denoted $\Phi_t(r)$. Here $Q_{n_r}(r)$ is a polynomial with real, positive roots. Intuitively, it is clear that the Approximant is very close to the exact phase thus we have no intention to make rigorous analysis of it for the general polynomial potential in order to answer the question of how close is it. It was checked successfully for a few, particular, physically important cases. The (Phase) Approximant depends on a number of free parameters which can be fixed variationally by taking the function $\Psi_t(r) = r^{\ell} Q_{n_r}(r)\, e^{-\Phi_t(r)}$ as trial function in the energy functional and then either imposing orthogonality conditions to previous functions with $k_r < n_r$, or simply requiring the roots of $Q_{n_r}(r)$ to be positive.
It turns out that the optimal variational parameters allow us to reproduce with high accuracy all coefficients in growing terms of the semiclassical expansion at large $r$, thus, $\Psi_t(r)$ is close to the exact wavefunction $\Psi(r)$ even in classically-prohibited domain where it is exponentially-small! This phenomenon allows to draw two conclusions:
(i) $\Psi_t(r)$ is of very high ``quality": after minimization it becomes close to exact eigenfunction even in domain where it is exponentially-small, thus, where it gives exponentially small contribution to the variational integrals and
(ii) without loosing much of accuracy we can impose constraints on free parameters $\Phi_t(r)$ by requiring the exact reproduction of all growing terms of the phase in semiclassical domain $r \gg 1$. Now we proceed to discuss three particular cases which were studied in detail.

In concrete, for three, two-term radial anharmonic oscillators in $D$-dimensional space with the potential
\[
     V_p(r)\ =\ r^2\ +\ g^p\,r^{p+2}\ ,\quad r \geq 0\ ,
\]
at $p=1,2,4$ we constructed highly-accurate-locally, compact, few-parametric, uniform approximation,
\[
       \Psi(r)\ =\ r^{\ell} P_{n_r}(r^2)\, e^{-\Phi_p(r)}\ ,
\]
for the eigenfunction of the state $(n_r, \ell)$ (with radial quantum number $n_r$ and angular momentum $\ell$), where $P_{n_r}(r^2)$ is a polynomial in $r^2$ of degree $n_r$ with positive roots. Phases $\Phi_p(r)$ had appeared of the form:

\bigskip
\noindent
{\bf (I)}\ Cubic anharmonic oscillator, $p=1$,
\[
   \Phi_1(r)\ =\ \frac{\tilde{a}_0\ +\ \tilde{a}_1\,gr+\tilde{a}_2\,r^2\ +\ \tilde{a}_3\,g\,r^3}{\sqrt{1\ +\ \tilde{b}_3\, g\, r}}\ +\ \frac{1}{4}\,\log[1\ + \tilde{b}_3\, g\,r] + D \log \left[1\ +\ \sqrt{1\ +\ \tilde{b}_3\, g\,r}\right]\ ,
\]
see Eq.(V.13) in I. If all parameters are chosen to be optimal in variational calculation with $\Phi_t(r)$ as trial function \footnote{Note that as the result of minimization the ratio of optimal parameters $\frac{\tilde{b}_3}{\tilde{a}^2_3}$ is equal to $\frac{25}{4}$ with accuracy $\sim 10^{-5}$ for all $D$ and $g$ we had studied.} , the variational energy is obtained with absolute accuracy $10^{-8}$ (8 s.d.) for $D=1, 2, 3, 6$ and $g=0.1, 1.0, 10.0$\,, see I, for the four lowest states: $(0,0), (0,1), (0,2), (1,0)$. All parameters are smooth slow-changing functions, see Fig.4 in paper I for the ground state $(0,0)$. Undoubtedly the same accuracy would be reached for other dimensions and coupling constants, as well as for excited states. The first correction $E_2$ to the variational energy $E_0^{(1)}$ is always of the order $10^{-8}$, while the rate of convergence of energy in Non-Linearization procedure is $\frac{E_3}{E_2} \sim 10^{-4}$. Relative deviation of $\Phi_t(r)$ from the exact eigenfunction is bounded function, it does not exceed $\sim 10^{-4}$.

\begin{figure}[]
	\includegraphics[width=0.7\textwidth]{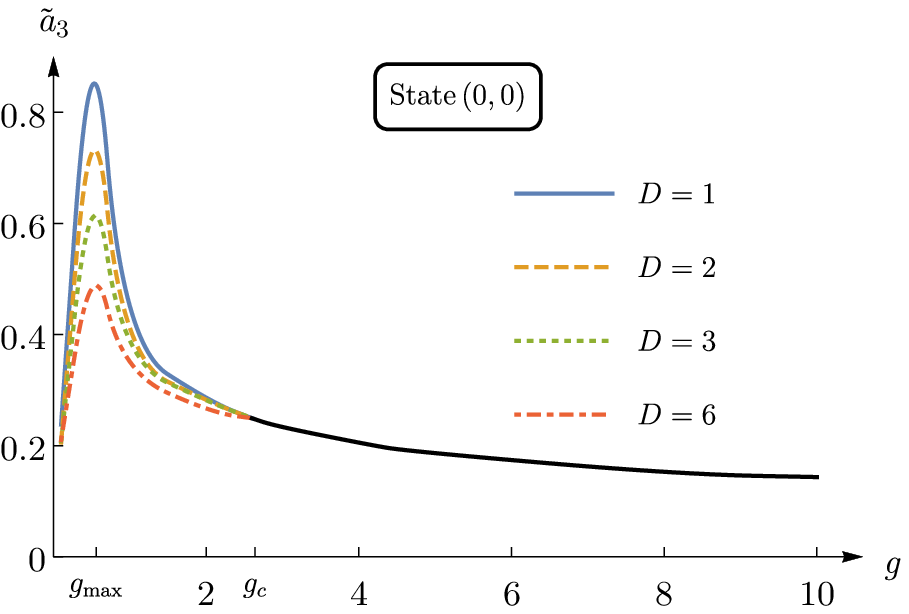}
	\caption{Cubic oscillator: parameter $\tilde{a}_3$ {\it vs} $g$ at $D=1,2,3,6$
             for the ground state $(0,{0})$. Positions of maximum for different curves about to coincide $g_{max} \approx 0.56$, and for $g \gtrsim g_c \approx 2.6$ the curves become non-distinguishable.}
\label{(0_0)}
\end{figure}

\begin{figure}[]
	\includegraphics[width=0.7\textwidth]{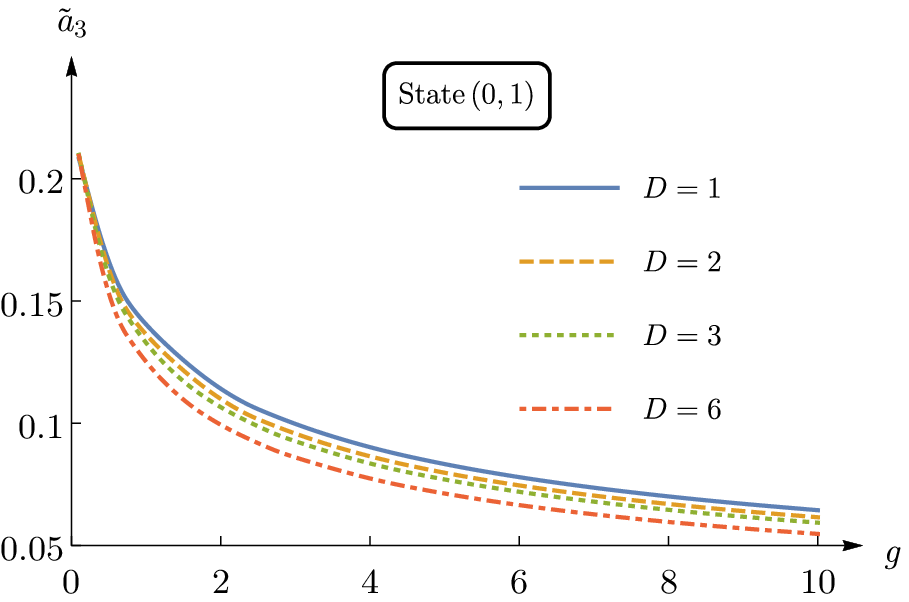}
	\caption{Cubic oscillator: parameter $\tilde{a}_3$ {\it vs} $g$ at $D=1,2,3,6$
             for the 1st excited state $(0,{1})$.}
\label{(0_1)}
\end{figure}

\begin{figure}[]
	\includegraphics[width=0.7\textwidth]{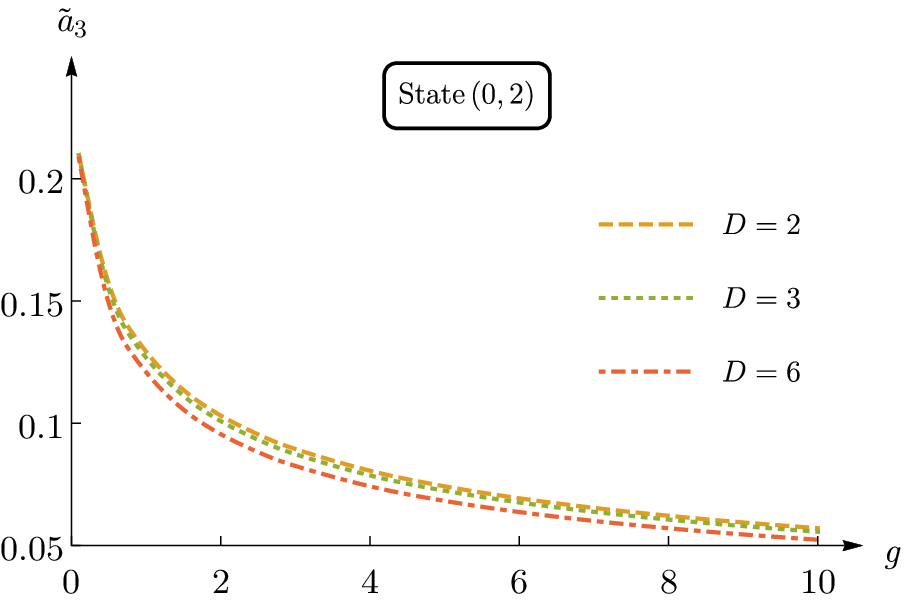}
	\caption{Cubic oscillator: parameter $\tilde{a}_3$ {\it vs} $g$ at $D=1,2,3,6$
             for the 2nd excited state $(0,{2})$.}
\label{(0_2)}
\end{figure}

\begin{figure}[]
	\includegraphics[width=0.7\textwidth]{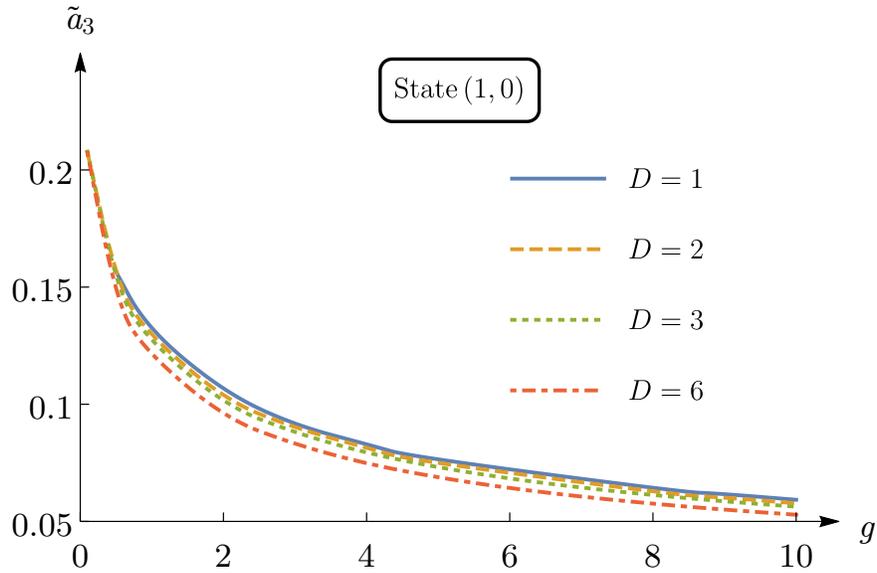}
	\caption{Cubic oscillator: parameter $\tilde{a}_3$ {\it vs} $g$ at $D=1,2,3,6$
             for the 3rd excited state $(1,{0})$.}
\label{(1_0)}
\end{figure}

\begin{figure}[]
	\includegraphics[width=0.99\textwidth]{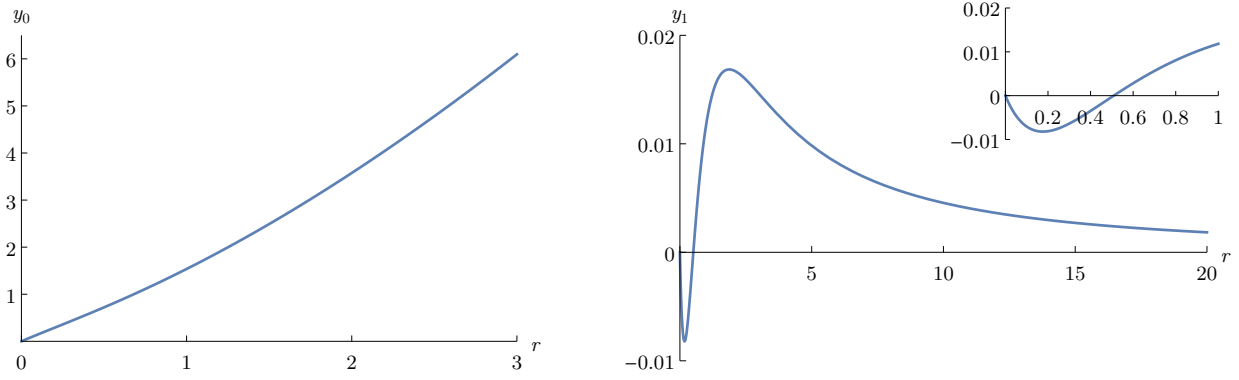}
	\caption{Cubic oscillator at $D=1$: function $y_0=(\Phi_t)'$  (on left) and its
     first correction $y_1$ (on right, with amplification at small $r$ at sub-figure) {\it vs} $r$ for $g=1$. }
\label{D=1-y01_c}
\end{figure}

\begin{figure}[]
	\includegraphics[width=0.99\textwidth]{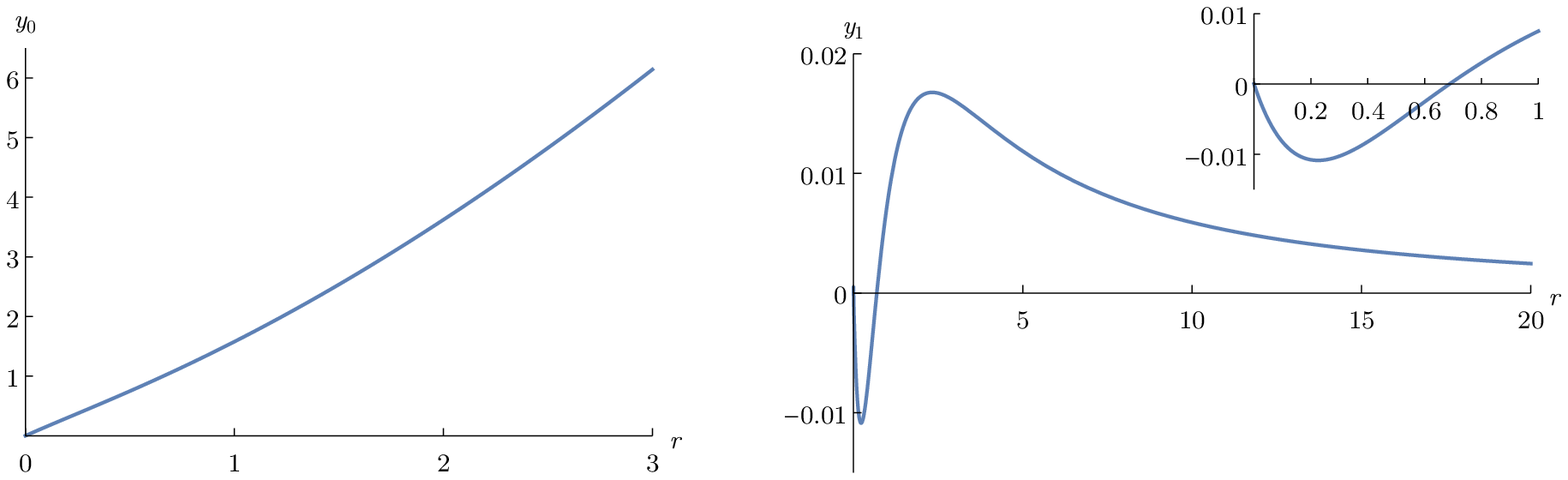}
	\caption{Cubic oscillator at $D=2$: function $y_0=(\Phi_t)'$ (on left) and its
     first correction $y_1$ (on right, with amplification at small $r$ at sub-figure) {\it vs} $r$ for $g=1$.}
\label{D=2-y01_c}
\end{figure}

\begin{figure}[]
	\includegraphics[width=0.99\textwidth]{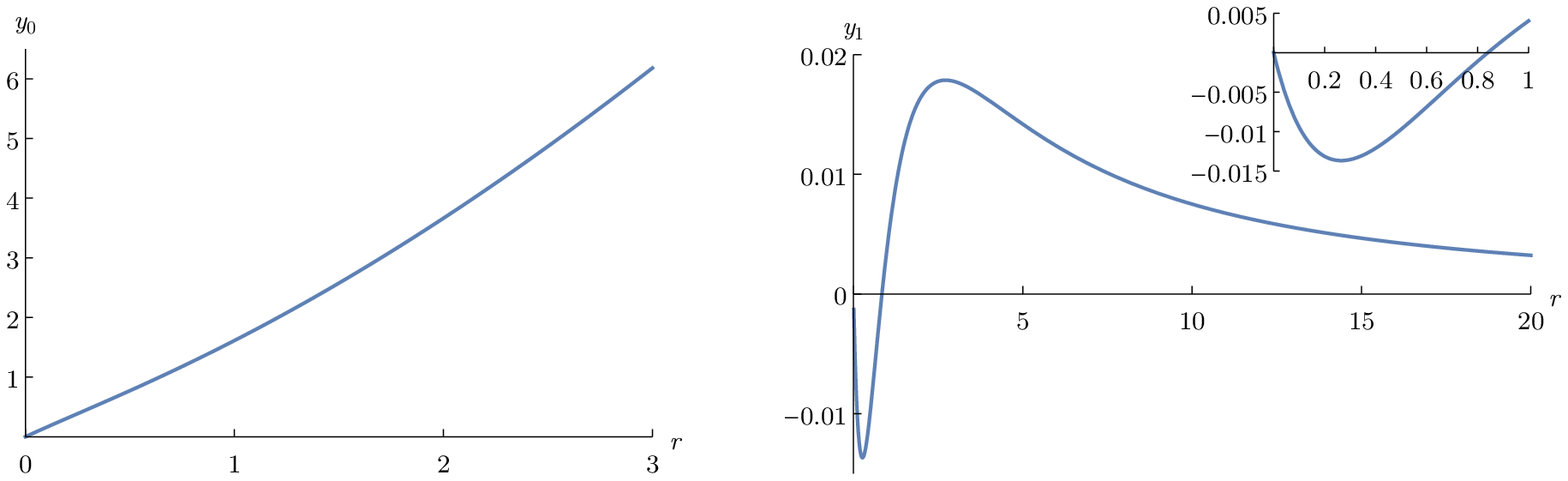}
	\caption{Cubic oscillator at $D=3$: function $y_0=(\Phi_t)'$ (on left) and its
     first correction $y_1$ (on right, with amplification at small $r$ at sub-figure) {\it vs} $r$ for $g=1$.}
\label{D=3-y01_c}
\end{figure}

Now we fix in $\Phi_1(r)$ the parameters $\tilde{b}_3, \tilde{a}_2, \tilde{a}_1, \tilde{a}_0$ to be the function of the parameter $\tilde{a}_3$ following the constraint Eq.(V.14) in paper I
\[
 \tilde{b}_3\ =\ \frac{25}{4}\,\tilde{a}^2_3\ ,
\]
and two more constraints
\[
 \tilde{a}_2\ =\ \frac{125 \tilde{a}^2_3 + 12}{150 \tilde{a}_3}\quad ,\quad
 -g^2 \tilde{a}_1\ =\ \frac{9375 \tilde{a}^4_3 - 1000 \tilde{a}^2_3 + 48}{15000 \tilde{a}^3_3} \ ,
\]
- it allows to reproduce {\it exactly} all four growing terms in the phase at semiclassical domain $r \rar \infty$: $r^{5/2}, r^{3/2}, r^{1/2}, \log r$ -
and also putting
\[
 \tilde{a}_0 \ =\ \frac{2\,\tilde{a}_1}{\tilde{b}_3}\ +\ \frac{D+1}{2}\ ,
\]
to keep $y(r=0)={\Phi^{\prime}_1}|_{r=0}=0$ - no linear term in $r$ in the expansion of phase at $r \rar 0$ occurs. Interestingly, those four constraints occur approximately for optimal parameters in variational calculus. Hence, the trial phase becomes one-parametric, it depends on the parameter $\tilde{a}_3$ only. If now we remake the minimization of the variational energy w.r.t. $\tilde{a}_3$ only, see plots on Figs.\ref{(0_0)} - \ref{(1_0)}, it allows us to reproduce not less than 5 s.d. correctly in energy for $D=1,2,3,6$ and $g = 0.1, 1.0, 10.0$ for all four low-lying states: $(0,0), (0,1), (0,2), (1,0)$. This result seems unprecedented: we are not aware about a situation when one-parametric trial function had led to such an accuracy. The rate of convergence of energy in Non-Linearization procedure reduces to $\frac{E_3}{E_2} \sim 10^{-2}$ in comparison to $\sim 10^{-4}$ as for non-constraint function Eq.(V.13) in paper I.
Note that in this case the first correction $y_1$ to derivative of the trial phase $y_0= {\Phi^{\prime}_1}$ is bounded function with maximal deviation of the order of $10^{-2}$, for illustration see Figs.\ref{D=1-y01_c}-\ref{D=3-y01_c} at $g=1$ (cf. Figs.5-7 in paper I),
\[
|y_1|_{max} \sim
\begin{cases}
 0.0167\ ,\qquad D=1 \\
 0.0168\ ,\qquad D=2 \\
 0.0178\ ,\qquad D=3 \\
 0.0224\ ,\qquad D=6 \\
\end{cases}
\]
cf. (\ref{cases-4}), (\ref{cases-6}). It is evident that the first correction to the phase itself $\Phi_1$ is also bounded function for all $r \in [0, \infty)$.

\bigskip

\noindent
{\bf (II)}\ Quartic anharmonic oscillator, $p=2$
\[
   \Phi_2(r)\ =\ \dfrac{\tilde{a}_0\ +\ \tilde{a}_2\, r^2\ +\
 \tilde{a}_4\, g^2\, r^4}{\sqrt{1\ +\ \tilde{b}_4\,g^2\,r^2}}\ +\
 \dfrac{1}{4}\log\left[1\ +\ \tilde{b}_4\,g^2\, r^2\right]\ +\
 \dfrac{D}{2}\log\left[1\ +\ \sqrt{1\ +\ \tilde{b}_4\,g^2\,r^2}\right] \ ,
\]
see (\ref{quartictrialg}), where $\tilde{a}_0,\tilde{a}_4$ are two {\it free} non-linear parameters only, see Figs.1-2 (a),(c); the parameters $\tilde{b}_4, \tilde{a}_2$ are fixed following the exact constraints (\ref{a4}), (\ref{a2}),
\[
  \tilde{b}_4\ =\ 9\,\tilde{a}_4^2\quad ,\quad \tilde{a}_2\ =\
  \frac{ 27\,\tilde{a}_4^2+1}{18 \tilde{a}_4}\ .
\]
It allows to reproduce {\it exactly} all growing terms at $r \rar \infty$ in the phase: $r^3, r, \log r$. In this case the variational energy coincides with exact one with not less than 8 s.d. for the states $(0,0), (0,1), (0,2), (1,0)$ for $D=1,2,3,6$ and $g^2=0.1, 1.0, 10.0$\,. However, there are no doubts that it will be the same for any integer $D$ and $g^2 > 0$, and for any excited state. The rate of convergence of energy in Non-Linearization procedure remains $\frac{E_3}{E_2} \sim 10^{-4}$ as in non-constraint function (\ref{quartictrialg}).  Note that in this case the first correction $y_1$ to derivative of the trial phase $y_0 = {\Phi^{\prime}_2}$ is bounded function with maximal deviation of the order of $10^{-2}$. The same is true for the phase itself ${\Phi_2}$.

\bigskip

\noindent
{\bf (III)}\ Sextic anharmonic oscillator, $p=4$
\[
   \Phi_4(r)\ =\ \dfrac{\tilde{a}_0\ +\ \tilde{a}_2\,r^2\ +\ \tilde{a}_4\,g^2\,r^4\
    +\ \tilde{a}_6\,g^4\,r^6}{\sqrt{1\ +\ \tilde{b}_4\,g^2\,r^2\ +\   \tilde{b}_6\,g^4\,r^4}}\ +\ \frac{1}{4g^2}\log\left[\tilde{c}_2\, g^2\,r^2\ +\ \sqrt{1\ +\   \tilde{b}_4\,g^2\,r^2\ +\  \tilde{b}_6\,g^4\,r^4}\right] \ +\
\]
\[
  \dfrac{1}{4}\log\left[1\ +\ \tilde{b}_4\,g^2\,r^2\ +\ \tilde{b}_6\,g^4 r^4\right]\ +\ \dfrac{D}{4}\log\left[1+\sqrt{1\ +\ \tilde{b}_4\,g^2\,r^2\ +\ \tilde{b}_6g^4r^4}\right]\ ,
\]
see (\ref{sextictrial}), where $\{\tilde{a}_0,\tilde{a}_2,\tilde{a}_4,\tilde{a}_6,\tilde{c}_2\}$ are five {\it free} non-linear parameters only, see Figs.6-7; the parameters $\tilde{b}_4,\tilde{b}_6$ are fixed following the constraints (\ref{a6}), (\ref{a4-6}),
\[
    \tilde{b}_6\ =\  16\,\tilde{a}_6^{2}\quad ,\quad \tilde{b}_4\ =\  32\,\tilde{a}_6\,\tilde{a}_4\ ,
\]
in order to reproduce exactly all growing terms at $r \rar \infty$ in the phase: $r^4, r^2, \log r$. It is worth emphasizing that optimal parameters in variational calculus obey the above constraints with high accuracy.
In this case the variational energy coincides with exact one with not less than 8 s.d. for the states $(0,0), (0,1), (0,2), (1,0)$ for $D=1,2,3,6$ and $g^4=0.1, 1.0, 10.0$. However, there are no doubts that it will be the same for any integer $D$ and
$g^4 > 0$, and for any excited state. The rate of convergence of energy in Non-Linearization procedure remains $\frac{E_3}{E_2} \sim 10^{-4}$ as in non-constraint function (\ref{sextictrial}).  Note that in this case the first correction $y_1$ to derivative of the trial phase $y_0= {\Phi^{\prime}_4}$ is bounded function with maximal deviation of the order of $10^{-2}$. The same is true for the phase itself ${\Phi_4}$.

In order to have benchmark results for energies we developed LMM for all three potentials at $p=1,2,4$ for $D=1,2,3,6$ and three values of the coupling constant $g^p=0.1, 1.0, 10.0$\,. For those values of coupling constants LMM was used with 50, 100, 200 mesh points, respectively, independently on $D$. These calculations had allowed us to find eigenenergies with 12 figures.

Concluding we state that the constrained (phase) Approximant (\ref{generalrecipe})
written for the general radial polynomial potential and arbitrary eigenstate,
which guarantees the absence of linear in $r$ term at small distances and reproduces exactly all growing, energy-independent terms at large $r$, provides the uniform approximation of the exact phase in the whole domain $r \in [0, \infty)$. In all studied particular cases the absolute deviation of the derivative of phase turned to be less or order $10^{-2}$. The variational energy, found with constrained (phase) Approximant taken as the phase of trial function, provides accuracy not less than 5 significant digits for any integer dimension $D$ and any coupling constant $g > 0$.
From physics point of view the constrained (phase) Approximant corresponds to a classical action for a potential which coincides with original potential at small and large distances and differs (slightly) for intermediate distances of order one.
Bohr-Sommerfeld quantization condition for this potential will be studied elsewhere.

It is straightforward to write the Riccati-Bloch (cf.(7)) and the generalized Bloch (cf.(11)) equations for perturbed two-body Coulomb problem or, saying differently, radially perturbed Hydrogen atom,
\begin{equation*}
\label{Coulomb}
V_c(r)\ =\ g\,\tilde{V}(gr) = -\frac{b_0}{r} + b_1 g^2 r + b_2 g^3 r^2 + \ldots \ ,\ b_0=1\ ,\quad r \in [0,\infty)\ ,
\end{equation*}
where $b_i, i=1,2,\ldots$ are parameters, and perform a perturbation theory analysis in powers of $g$ and construct the expansion in generating functions, see (13)-(15). It will be done elsewhere. In a similar way as was done in the paper I and the present paper the constrained (Phase) Approximant can be constructed. On intuitive level, it was already realized a long ago for the quarkonium (funnel-type) potential,
\[
   V_c\ =\ -\frac{1}{r}\ +\ b_1 g^2 r \ ,
\]
in \cite{Turbiner:1987}, and, recently, for the Yukawa potential,
\[
   V_c\ =\ -\frac{b_0}{r}\,e^{- g r} \ ,
\]
see \cite{delValle-Nader:2018}, which had led to highly accurate variational
energies for both cases.

\section*{Acknowledgments}

The authors thank J.C.~L\'opez~Vieyra and H.~Olivares~Pil\'on for their interest to the work and useful remarks, and especially for help with numerical and computer calculations. J.C.~del\,V. is supported by CONACyT PhD Grant No.570617 (Mexico). This work is partially supported by CONACyT grant A1-S-17364 and DGAPA grant IN113819 (Mexico).


\bibliography{references2}

\newpage

\appendix
\section{First PT Corrections and Generating Functions $G_4,6$ for the Quartic Anharmonic Oscillator}
\label{appendix:A}

We present the corrections $\veps_{4,6}$ and $\mathcal{Y}_{4,6}$ in explicit form for the quartic anharmonic potential (\ref{potquartic}) in the expansion (\ref{Y2n}) and (\ref{Y2n-c}),
\[
\veps_4 \ =\ -\frac{1}{16}\,D(D+2)(2D+5)\ ,
\]
\[
\veps_6 \ =\ \frac{1}{64}\,D(D+2)(8D^2+43D+60)\ ,
\]
\[
\mathcal{Y}_4(v)\ =\ -\frac{1}{8}v^5\ -\ \frac{1}{16}(3D+8)v^3\ +\ \frac{\veps_4}{D}\,v\ ,
\]
\[
\mathcal{\tilde Y}_4({\rm v})\ =\ -\frac{1}{16}\left[2{\rm v}^2\ +\ (3D+8){\rm v}\ +\ (D+2)(2D+5)\right]\ ,
\]

\[
\mathcal{Y}_6(v)\ =\ \dfrac{1}{16}v^7\ +\ \frac{1}{32}(5D+16)v^5\ +\ \frac{1}{16}(3D^2+17D+25)v^3\ +\ \frac{\veps_6}{D}\,v\ .
\]
\[
\mathcal{\tilde Y}_6({\rm v})\ =\ \dfrac{1}{64}\left[4{\rm v}^3\ +\ {2}(5D+16){\rm v}^2\ +\ 4 (3D^2+17D+25){\rm v}\ +\ (D+2)(8D^2+43D+60)\right]\ .
\]

We also present two generating functions for the phase, see expansion (\ref{phase}) and (\ref{genexpphi}),
\[
  G_4(r;g)\ =\ \frac{5}{24 g^2 w^3}\ +\ \frac{D \left(1+w+w^2\right)}{4 g^2  (w+1)w^2}\ +\
  \frac{D^2}{8 g^2 w}\ ,
\]
\[
  G_6(r;g)\ =\ -\frac{5}{16 g^2 w^6}\ -\ \frac{D \left(15+30 w+20 w^2+16 w^3+20 w^4+30 w^5+15 w^6\right)}{48 g^2  (w+1)^2w^5}
\]
\[
 -\frac{D^2 \left(4+8 w+8 w^2+12 w^3+18 w^4+9 w^5\right)}{32 g^2  (w+1)^2w^4}\ -\
 \frac{D^3 \left(1+3 w^2\right)}{48 g^2 w^3}\ ,
\]
where $w=\sqrt{1+g^2v^2}$. Two remarks in a row: $(i)$ in the variable $w$ all generating functions are rational functions, $(ii)$ the polynomial structure in $D$ of generating functions becomes evident.

\section{First PT Corrections and Generating Functions $G_{4,6} $ for the Sextic Anharmonic Oscillator}
\label{appendix:B}

We present explicitly the  first corrections $\veps_{8,12}$ and $\mathcal{Y}_{8,12}$ for the sextic anharmonic oscillator (\ref{potsextic}), see (\ref{Yncorrection-sex}) and (\ref{factorizations}),
\[
\veps_8\ =\ -\frac{1}{128}\,D\,(D+2)\,(D+4)\,(9 D^2+72 D+152)\ ,
\]
\[
\veps_{12}\ =\ \dfrac{1}{1024}\,D\,(D+2)\,(D+4)\,(81 D^4+1404 D^3+9624 D^2+31152 D+40384)\ .
\]

\[
\mathcal{Y}_8(v)\ =\ -\frac{1}{8}v^9-\frac{1}{16}(3D+16)v^7-\frac{1}{16}(3D^2+27D+64)v^5
 -\frac{1}{32}(4D^3+49D^2+204D+288)v^3+\frac{\veps_{8}}{D}\,v
\]
\[
  \mathcal{\tilde Y}_8({\rm v})\ =\ -\frac{1}{8}{\rm v}^4-\frac{1}{16}(3D+16){\rm v}^3-\frac{1}{16}(3D^2+27D+64){\rm v}^2
 -\frac{1}{32}(4D^3+49D^2+204D+288){\rm v}
\]
\[
-\frac{1}{128}\,(D+2)\,(D+4)\,(9 D^2+72 D+152)\ ,
\]
\[
\mathcal{Y}_{12}(v)\ =\ \frac{1}{16}v^{13}\ +\ \frac{1}{32} (5 D+32) v^{11}\ +\ \frac{5}{64} (3 D^2+34 D+104) v^9
\]
\[
+\ \frac{1}{32} (8 D^3+125 D^2+688 D+1344) v^7
 +\ \frac{1}{256} (55 D^4+1038 D^3+7708 D^2+26784 D+36800) v^5
\]
\[
+\ \frac{1}{256} (36 D^5+783 D^4+7040 D^3+32768 D^2+78912 D+78336) v^3
\]
\[
+\ \frac{\veps_{12}}{D}\,v\ .
\]
\[
\mathcal{\tilde Y}_{12}({\rm v})\ =\ \frac{1}{16}\,{\rm v}^{6}\ +\ \frac{1}{32}\,(5 D+32)\,{\rm v}^{5}
   \ +\ \frac{5}{64}\,(3 D^2+34 D+104)\,{\rm v}^4
\]
\[
+\ \frac{1}{32}\,(8 D^3+125 D^2+688 D+1344)\,{\rm v}^3
 +\ \frac{1}{256}\,(55 D^4+1038 D^3+7708 D^2+26784 D+36800)\,{\rm v}^2
\]
\[
+\ \frac{1}{256}\,(36 D^5+783 D^4+7040 D^3+32768 D^2+78912 D+78336)\,{\rm v}
\]
\[
+\ \dfrac{1}{1024}\,(D+2)\,(D+4)\,(81 D^4+1404 D^3+9624 D^2+31152 D+40384)\ .
\]

Now we present explicitly two generating functions $G_4(r), G_6(r)$ in the expansion (\ref{phase})
for sextic oscillator,
\[
  G_4(r)\ =\   \frac{r^2}{4w}\bigg(\frac{\left(5 + w^2\right)}{3 w^2}\ +\ \frac{D  \left(1+w+w^2\right)}{w (w+1)}\ +\ \frac{D^2}{4}\bigg)\ ,
\]
\[
  4 g^2\,G_6(r)\ =\ \frac{5-3w^2}{w^6}\ +\ \frac{D \left(15+15 w-4 w^2+2 w^3+6 w^4+6 w^5\right)}
  {6 (w+1)w^5}
\]
\[
  \ +\ \frac{D^2 \left(2+2 w+w^2+3 w^3+3 w^4\right)}{4 (w+1)w^4}
  +\ \frac{D^3 \left(1+3 w^2\right)}{24 w^3}\ ,
\]
where $w=\sqrt{1+g^4r^4}$. Two remarks in a row: $(i)$ in the variable $w$, all generating functions are rational function; $(ii)$ The polynomial structure in $D$ of these generating functions is evident.

\section{General Aspects of the  Two-Term Radial Anharmonic Oscillator  }
\label{appendix:C}
Let us consider the general two-term radial anharmonic oscillator potential,
\begin{equation}
\label{twoterm}
V(r)\ =\ r^2\ +\ g^{m-2}r^{m}\ ,\quad m\ >\ 2\  .
\end{equation}
cf. (\ref{potential}), where $a_2=a_m=1$ and $a_3,a_4,...,a_{m-1}=0$.
Expansion of $\veps$ in powers of $\la$ has the form
\begin{equation}
\label{generalseries}
    \veps(\la)\ =\ \veps_0\ +\ \la^{m-2}\veps_{m-2}\ +\ \la^{2(m-2)}\veps_{2(m-2)}\ +\ ...\ ,
\end{equation}
see  (\ref{eps-in-la}). Not surprisingly, for even potentials ($m=2p$, $p=1,2,...$) the PT coefficients $\veps_{2n(p-1)}$, $n=1,2,...$, are polynomials in $D$ factorized as follows
\begin{equation}
\veps_{2n(p-1)}(D)\ =\ D(D+2)(D+4) \ldots (D+2p-2)R_{(n-1)(p-1)}(D)
\end{equation}
where $R_{(n-1)(p-1)}(D)$ is a polynomial in $D$ of degree $(n-1)(p-1)$, see \cite{ELETSKY}. We emphasize that, in the framework of the Non-Linearization Procedure, any PT coefficient $\veps_{2n(p-1)}(D)$ is calculated by linear algebra means. Series (\ref{generalseries}) is divergent due to a Dyson instability argument. It is evident that for larger $D$ and $m$, the index of divergence is larger, see e.g. \cite{IVANOV}.

For potential (\ref{twoterm}) generating functions $G_0(r;g)$ and $G_2(r;g)$ can be written in closed form,
\begin{equation}
 \frac{G_0(r)}{(2M)^{1/2}}\ =\ \frac{2}{m+2}r^2\left\{\sqrt{1+(gr)^{m-2}}+\frac{m-2}{4}\, {}_2F_1\left(\frac{1}{2},\frac{2}{m-2};\frac{m}{m-2};-(gr)^{m-2}\right)\right\}\ ,
\end{equation}

\begin{equation}
  \frac{g^2\,G_2(r)}{(2M)^{1/2}}\ =\ \frac{1}{4}\,\log\left[1+(gr)^{m-2}\right]\ +\ \frac{D}{m+2}\,\log\left[1+\sqrt{1+(gr)^{m-2}}\right]\ ,
\end{equation}
where ${}_2 F_1$ is the hypergeometric function, see I.

\end{document}